\documentclass{article}
\usepackage{graphicx}
\def\users{us}
\def\users{world}
\usepackage{epsfig}
\usepackage{amsfonts}
\usepackage{amssymb}
\usepackage{amsmath}
\usepackage{upgreek} % for upshape greek letters
\usepackage{times}\usepackage{mathptm}
\allowdisplaybreaks
\usepackage{mathrsfs} % nicer caligraphic fonts ``mathscr''
\usepackage{color}
\usepackage[normalem]{ulem}
\usepackage{ifthen}
\definecolor{brown}{rgb}{0.5,0,0}
\ifthenelse{\equal{\users}{world}}{

    \newcommand{\DELETE}[1]{}
    \newcommand{\COMMENT}[1]{}

    \newcommand{\COL}[1]{#1}}
{
 
 \newcommand{\DELETE}[1]{{\color{brown}\sout{#1}\color{black}}}
 \newcommand{\COMMENT}[1]{{\color{blue}\uuline{#1}\color{black}}}

 \newcommand{\COL}[1]{{\color{blue}{#1}\color{black}}}

\usepackage[notcite,notref]{showkeys}
}

\newcommand{\R}{{\mathbb R}}
\newcommand{\N}{{\mathbb N}}

\renewcommand{\d}{{\rm d}}

\newcommand\DT[1]{\mathchoice
                 {{\buildrel{\hspace*{.1em}\text{\LARGE.}}\over{#1}}}
                 {{\buildrel{\hspace*{.1em}\text{\Large.}}\over{#1}}}
                 {{\buildrel{\hspace*{.1em}\text{\large.}}\over{#1}}}
                 {{\buildrel{\hspace*{.1em}\text{\large.}}\over{#1}}}}
\newcommand\DDT[1]{\mathchoice
   {{\buildrel{\hspace*{.1em}\text{\LARGE.\hspace*{-.1em}.}}\over{#1}}}
   {{\buildrel{\hspace*{.1em}\text{\Large.\hspace*{-.1em}.}}\over{#1}}}
   {{\buildrel{\hspace*{.1em}\text{\large.\hspace*{-.1em}.}}\over{#1}}}
   {{\buildrel{\hspace*{.1em}\text{\large.\hspace*{-.1em}.}}\over{#1}}}}

\newcommand\eps{\varepsilon}

\newcommand{\DDD}[3]{\begin{array}[t]{c}#1\vspace*{-1em}\\_{#2}\vspace*{-.3em}\\_{#3}\end{array}}
\newcommand{\ddd}[3]{\DDD{\begin{array}[t]{c}\underbrace{#1}\vspace*{.6em}\end{array}}{\text{\footnotesize #2}}{\text{\footnotesize #3}}}

\newcommand{\OF}{\Omega_{_{\rm F}}}
\newcommand{\OS}{\Omega_{_{\rm S}}}

\usepackage{psfrag}
\usepackage{epsfig}
\newcounter{myfigure}
\newenvironment{my-picture}[3]{\refstepcounter{myfigure}\label{#3}\setlength{\unitlength}{\textwidth}\begin{picture}(#1,#2)}{\end{picture}}
\usepackage{cite}
\newcommand{\lineunder}[2]{\LU{\begin{array}[t]{c}\underbrace{#1}\vspace*{.5em}\end{array}}{\mbox{\footnotesize\rm #2}}}

\newcommand{\LU}[2]{\begin{array}[t]{c}#1\vspace*{-1em}\\_{#2}\end{array}}
\newcommand{\bbb}{\relax} %{\color{blue}}
\newcommand{\eee}{\relax} %{\color{black}}

\begin{document}
\begin{sloppypar}
%\allowdisplaybreaks
  
\global\def\refname{\centerline{\large\sc References}}
\baselineskip12pt
%\title[A monolithic model for seismic sources and waves]{A monolithic 
%model for seismic sources and seismic waves}
%\title
\begin{center}{\LARGE\bf A monolithic model for phase-field fracture and
    waves\\in solid-fluid
    media towards earthquakes\footnote{The support from the grants
%16-03823S 
%``Homogenization and multi-scale computational modelling of flow and 
%nonlinear interactions in porous smart structures'' and
    17-04301S (as far as dissipative evolutionary aspects concerns)
%``Advanced mathematical methods for dissipative evolutionary systems''
and 19-04956S (as far as modelling of dynamic and nonlinear behaviour concerns)
%``Dynamic and nonlinear behaviour of smart structures; modelling and
%optimization''.
of the Czech Sci.\ Foundation and 
VEGA 1/0078/16 of the Ministry of Education, Science,
Research and Sport of the Slovak Republic,
and the institutional support RVO:61388998 (\v CR) 
are acknowledged.}}

  \bigskip
  
    {\large\sc Tom\'a\v s Roub\'{\i}\v cek\footnote{Institute of 
  Thermomechanics, Czech Academy of Sciences,
  Dolej\v skova 5, CZ-182 00 Praha 8, Czech Republic.} %, and\\\hspace*{2em}
 \footnote{Mathematical Institute, Charles University,
  Sokolovsk\'a 83, CZ-186~75~Praha~8, Czech Republic} and Roman Vodi\v cka\footnote{Civil Engineering Faculty,
   Technical University of Ko\v sice,
   Vysoko\v skolsk\' a 4, SK-042 00 Ko\v sice, Slovakia}.}

\end{center}

\author{Tom\'a\v s Roub\'{\i}\v cek \and Roman Vodi\v cka}

%~~~~~~~~~~~~~~~~~~~~~~~~~~~~~~~~~~~~~~~~~~~~~~~~~~~~~~~~~~~~~~~~~~~~~~~~~~~~~~

%\author[T. Roub\'{\i}\v cek and R. Vodi\v cka]{T. Roub\'{\i}\v cek$^{1,2}$ and  R. Vodi\v cka$^{3}$
%\\
%$^1$Institute of Thermomechanics, Czech Academy of Sciences, Dolej\v skova 5, 
%CZ-182 00 Praha 8, Czech Republic
%\\$^2$Mathematical Institute, Charles University, Sokolovsk\'a 83, 
%CZ-186~75~Praha~8, Czech Republic
%\\$^3$Civil Engineering Faculty, 
%Technical University of Ko\v sice,
%Vysoko\v skolsk\' a 4, SK-042 00 Ko\v sice, Slovakia}

%\author{\hspace*{4em}{\sc T. Roub\'{\i}\v cek}\footnote{Institute of 
%Thermomechanics, Czech Academy of Sciences,
%Dolej\v skova 5, CZ-182 00 Praha 8, Czech Republic.}
%\footnote{Mathematical Institute, Charles University, Sokolovsk\'a 83, 
%CZ-186~75~Praha~8, Czech Republic.}\,
%\ \ and\ \  {\sc R. Vodi\v cka}\footnote{Civil Engineering Faculty, 
%Technical University of Ko\v sice,
%Vysoko\v skolsk\' a 4, SK-042 00 Ko\v sice, Slovakia}
%\hfill\hspace*{1em}\medskip\baselineskip10pt
%\\\small\sl Mathematical Institute, Charles University, Praha \hfill\hspace*{1em}
%\\[-.5em]\small\sl
%and\hfill\hspace*{1em}
%\\[-.5em]\small\sl Institute of Thermomechanics of the ASCR, Praha,     
%Czech Republic\hfill\hspace*{1em}
%\\\\\\\\
%\hspace*{-2mm}\begin{minipage}[b]{1\textwidth}\baselineskip10pt
%{\small\sf \tableofcontents }\end{minipage}
%}

%\date{Received: Nov. 5, 2018 / Revised: May 10, 2019 / Accepted: .............}

%\maketitle

\begin{abstract}
  %\begin{minipage}[t]{50em}
  Coupling of rupture processes in solids with waves also propagating  in fluids is a prominent
  phenomenon arising during tectonic earthquakes. It is executed here in a
  single `monolithic' model which can asymptotically capture both damageable
  solids (rocks) and (visco-)elastic fluids (outer core or oceans). Both
  ruptures on pre-existing 
lithospheric faults and a birth of new faults in compact rocks are covered 
by this model, together with emission and propagation of seismic 
waves, including, e.g., reflection of S-waves and refraction of P-waves 
on the solid-fluid interfaces. A robust, energy conserving, 
and convergent staggered FEM discretisation is devised. Using a rather 
simplified variant of such models for rupture, three computational 
experiments documenting the applicability of this approach are presented. 
Some extensions of the model towards more realistic geophysical 
modelling are outlined, too.

%\keywords

\medskip


\medskip

%\PACS

    \medskip

    %\subclass
  \end{abstract}
  %\end{minipage}

%Keywords selected from:
%https://academic.oup.com/DocumentLibrary/GJI/gji%20keyword%20list%20updated2017.pdf

%\begin{keywords}
%Earthquake dynamics,
%elasticity and anelasticity,
%fault zone rheology,
%body waves,
%numerical modelling,
%non-linear differential equations. 
%%Dynamic rupture, shear/pressure waves, damage, global seismicity,
%%tectonic earthquakes, solid-fluid interaction, staggered finite-element schemes,
%%computational modelling.
%\end{keywords}

%\medskip

%\nopagebreak

\setcounter{myfigure}{0}

%{\small\baselineskip10pt{\it Abstract -- }..............}

%\medskip

%{\small
%\baselineskip10pt
%{\bf Key words:} Shear/pressure waves, global seismicity, damage, rupture, 
%tectonic earthquakes, solid-fluid interaction, 
%staggered finite-element schemes, computational modelling.}

%\baselineskip17pt

%\medskip

\section{Introduction}
% ~~~~~~~~~~~~~~~~~~~~

%TESTOVACI VETA PRO SIRKU TEXTU V Pure & Appl Geophysics:
%The self-potential (SP) method is one of the oldest methods and has wide applications in sulphides and graphits exploration and in geophysical groundwater investigation.

Dynamic fracture mechanics is an area of continuum mechanics of solids
with wide applications in engineering and particularly also in geophysics. 
Global geophysical models typically deal with several very 
different phenomena and couple various models due to the layered character 
of terrestrial planets (including our planet Earth as well as our Moon). 
This paper demonstrates the philosophy that a single 
model can be used instead of several specialized models 
that otherwise would need to be mutually coupled in a rather complicated way. 
Such a single universal (we say ``monolithic'') model can also be 
straightforwardly implemented on computers \COL{omitting any interfaces which 
usually complicate implementations. In particular, the transient conditions
on fluid/solid interfaces are automatically involved and need not be 
specified. This can be a sound advantage of the presented model}. Of course, 
computationally, such a monolithic-type model may not always make it easier to 
produce really relevant simulations on the computers we have at our disposal 
nowadays. 
Routine calculations are performed separately either for
earthquake sources locally (see, e.g.,\ \cite{CTLC13EUFE})  or for purely seismic global 3D models, cf.,\ e.g.,\ 
\cite{KomTro02SESG,KomTro02SESG2,LayWal95MGS,ToCaMa09SSLV}.
Their mutual coupling is already treatable well in the literature
%%with earthquake source at least locally, cf.\ e.g.\ 
%%\cite{KomTro02SESG,KomTro02SESG2,LayWal95MGS,ToCaMa09SSLV}
%%as well as 
\cite{HBAD09DERC,HuAmHel14ERMW,KaLaAm08SEMS,PPAB12TDDR}.
%respectively, 
Therefore, there is promising potential 
%hope
that the \COL{presented} coupling monolithic 
approach may become \COL{even} more amenable in the future with ever 
increasing computer efficiency.

The phenomena we have in mind in this paper involve {\it global seismicity} and 
{\it tectonics}. In particular the latter involves, e.g.,\ {\it ruptures of 
lithospheric faults} or a birth of new faults, generating seismic waves which 
then propagate through the solid-like silicate mantle and iron-nickel inner 
core both in the shear (S) or the pressure (P) modes. In contrast to the 
P-waves (also called primary or compressional waves), the S-waves (also called 
secondary waves) are suppressed in the fluidic iron-nickel outer core and 
also in the water oceans.
% (where P-waves emitted from the earthquakes in the 
% crust may manifest as Tsunami at the end on the surface). 

A very low attenuation of seismic waves is the ultimate phenomenon.
We, therefore, take the {\it Jeffrey's rheology} 
model (i.e., a serial combination of the 
Maxwell and Kelvin-Voigt rheology as in 
Figure~\ref{fig-mixed-response}/bottom-left,
cf., e.g., \cite{LyHaBZ11NLVE} or also 
\cite{Roub17GMHF}) as a basic global ``monolithic'' 
ansatz. In various limits in the deviatoric and the volumetric parts, 
we model different parts of planet Earth. 
Jeffrey's rheology also seems more realistic in particular because it 
covers (in the limit) also the Kelvin-Voigt model applied to the volumetric 
strain (actually considered as a starting model in 
Figure~\ref{fig-mixed-response}/left)
whereas the pure Maxwell rheology, allowing for big creep
during long geological periods, is not a relevant effect in the volumetric 
part.

Respecting the solid parts of the model, we use the {\it Lagrangian 
description} even in the fluid regions, i.e.\ here all equations are 
formulated in terms of displacements rather than velocities. The reference
and the actual space configurations automatically coincide with each 
other in our small strain (and small displacement) ansatz, which is 
well relevant in geophysical short-time scales of seismic events. 

In the solid-like part, various inelastic processes are considered 
to model tectonic earthquakes on lithospheric faults together with
long-lasting healing periods in between them, as well as aseismic slips, and
various other phenomena. To this goal, many internal variables
may be involved such as aging/damage, inelastic strain, porosity, water content, 
breakage \bbb(i.e.,\ essentially like another damage-like internal variable in modeling of granular materials)\eee, and temperature,
cf.\ \cite{LyHaBZ11NLVE,LyaBZ14DBRM,LyaBZ14CDBF}.
On the other hand, those sophisticated models are focused 
rather on local events around the tectonic faults without ambitions 
to be directly coupled with the global seismicity. Here, rather for 
the lucidity of the exposition, we reduce the set of internal variables 
to only one scalar variable and one matrix-valued variable, 
namely damage/aging and an inelastic strain, respectively. 
The calculations in Sect.~\ref{sec-comput} involve the only former one.
Even this simple scenario, however, has a capacity to trigger a spontaneous 
rupture (so-called {\it dynamic triggering}) with emission of seismic waves and,
in a certain simplification, can serve as a seismic source coupled 
with the overall global model. 
%{\tiny Also, this simple model already 
%will well illustrate mathematical difficulties related to nonlinearities 
%in the solid parts coupled with linear but possibly hyperbolic fluidic regions.}
The mentioned inelastic strain can capture Maxwell-type rheologies 
relevant in the solid mantle and inner core to capture long-term creep 
(aseismic) effects up to $10^5$ yrs. 

Let us emphasize that the usual models are focused only on either the propagation of 
seismic waves along the whole globe while their source is considered given, or 
on the description of seismic sources due to tectonic events, 
but not their mutual coupling.
%\COMMENT{SOME CITATIONS??}
If a coupling is considered, then it concerns rather local models 
not considering the layered structure of the whole planet, cf.\ e.g.\  
\cite{BeZi01DRRM,BZamp09SRRS,HuAmHel14ERMW,KaLaAm08SEMS,LHAB09NDRW}. The 
reality ultimately captures very different mechanical properties of different 
layers of the Earth, in particular the mantle and the inner core which are 
solid %from
\bbb on \eee the short-time scales versus the outer core and the oceans 
which are fluidic even on the short-time scales.
Some other coupled models implement a pre-existing fault (in contrast to 
our model allowing for new faults birth, cf.\ Section~\ref{experinent2})
and use a slip-dependent friction law \cite{TCVE12ADGM}.

The goal of this contribution is threefold:
\begin{itemize}
\vspace*{-.0em}\item[$\upalpha$)\!]to present a model that might simultaneously
capture the sources  of seismic waves that necessarily 
behave nonlinearly (like ruptures of tectonic faults) and 
the propagation of seismic waves possibly even 
over the whole planet \bbb(when further special algorighmic and
computational methods would be used, not handled in this article, however)\eee, 
%here modelled only in a very simplified way for a relatively lucid 
%illustration of the model procedure), 
both \bbb phenomena being \eee mutually coupled. 
\vspace*{-.0em}\item[$\upbeta$)\!]
by proper scaling to approximate viscoelastic (so\COL{-}called Boger's 
\cite{Bog77HECV}) or merely elastic fluids that are relevant in \COL{the} 
outer core and in the oceans (with a very low or just zero viscosity) where 
S-waves cannot propagate 
%can then only slightly penetrate the outer core or
%the oceans but are fast attenuated, 
while P-waves are only refracted on the solid/fluid interfaces (in particular
\bbb on \eee Guttenberg's core-mantle discontinuity).
\vspace*{-.0em}\item[$\upgamma$)\!]
%by limiting further the viscosity in the outer core or the oceans to zero, 
%further to approximate this viscoelastic fluid towards  elastic (completely
%inviscid) fluid, 
%%sometimes called also a Cowling approximation \cite{KomTro02SESG2}
%respecting the phenomenon that 
%S-waves cannot penetrate into these fluidic regions and are fully reflected 
%on the interfaces between the outer core and the mantle 
%(=Gutenberg's discontinuity) and the inner core
%or on the ocean beds, i.e.\ between $\OS$ and $\OF$, while P-waves 
%propagate through these interfaces, being both refracted and reflected on them.% 
%\vspace*{-.0em}\item[$\updelta$)\!]
%perform the rigorous analysis as far as the existence of solutions,
%a-priori estimates in specific norms, and convergence towards other models
%that justifies the particular models, their energetics and asymptotics,
%and can support numerical stability and convergence when discretised
%and implemented on computers.  
to document computational efficiency of the monolithic model at least on
2-dimensional \COL{rather local} simulations \COL{of various, quite distinct 
geophysical events}.
\end{itemize}
We refer to \cite{Roub??SWEG} for the rigorous analysis as far as 
asymptotics of the scaling of viscosities in even a more complex model
involving also self-induced gravity, tidal, Coriolis, and centrifugal 
forces; cf.\ Sect.~\ref{sec-more-geo}.
%..................................Sometimes, 
%more attenuation of Kelvin-Voigt type is also involved, which leads 
%to the {\it Jeffreys rheology}, cf.\ e.g.\ \cite{LyHaBZ11NLVE} or also 
%\cite{Roub17GMHF}.
%This seems more realistic in particular because it covers (in the limit)
%also the Kelvin-Voigt model applied to the volumetric 
%strain whereas the pure Maxwell rheology allowing for big creep
%during long geological periods is not a relevant effect in the volumetric 
%part.

%Madariaga, R., 1976. Dynamics of an expanding circular fault, Bull. seism.
%Soc. Am., 65, 163–182.

%Virieux, J., 1984. SH-wave propagation in heterogeneous media: velocity-
%stress finite-diffe4ence method, Geophysics, 49, 1933–1942.
%Virieux, J., 1986. P-SV wave propagation in heterogeneous media: Velocity-
%stress finite-difference method, Geophysics, 51, 889–901.

\section{Energy-based modelling approach}\label{sec-energetic}
%        ~~~~~~~~~~~~~~~~~~~~~~~~~~~~~~

Mechanical models in general (and those used in geophysics in particular)
typically are (or should be) believed to be governed by energies and, most 
often, in a way that the conservative and the dissipative parts are separated.
In the isothermal variant, the systems have a simple general structure of 
\COL{an abstract dissipative dynamic equation}
%\\\vspace*{-1.5em} 
\begin{align}\label{Biot}
{\mathscr M}'\DDT{q}+{\mathscr R}_{\DT q}'(q,\DT q)+
%\partial_{\bfq}
{\mathscr E}'(q)={\mathscr F}(t)%\COMMENT{????}{\mathscr F}(t,q) 
\end{align}
%\vspace*{-.3em} \noindent
with a kinetic energy ${\mathscr M}$, a (pseudo) potential 
of dissipative forces ${\mathscr R}(q,\cdot)$, 
a stored energy ${\mathscr E}$, and external forcing 
${\mathscr F}$ as a time-dependent functions of
the state $q$ \COL{in the reference domain $\Omega$}. In \eqref{Biot}, ${\mathscr R}_{\DT q}'$ denotes the
partial differential of ${\mathscr R}={\mathscr R}(q,\DT q)$ with respect to 
$\DT q$. In fact, ${\mathscr R}(q,\cdot)$ may be non-differentiable at 
$\DT q=0$ when some activated processes (like damage or plasticity)
occur so that 
${\mathscr R}_{\DT q}'(q,\cdot)$ may be set-valued and \eqref{Biot}
is to be an inclusion rather than equality; yet we will ignore these
technicalities in this presentation.

This state $q$ typically involves, beside of the displacement,
also some internal variables like the inelastic strains describing 
creep and a ``permanent'' deformation resulting from re-occurring 
shifts during earthquakes, damage/aging, etc., complying with the concept 
of {\it generalized standard materials 
with internal variables} \cite{HalNgu75MSG}. In \eqref{Biot}, we use the 
notation $\DT q=\frac{\d}{\d t}q$ and $\DDT q=\frac{\d^2}{\d t^2}q$, and 
$(\cdot)'$ denotes the differential. 
%while ``$\partial$'' denotes
%a generalized differential (typically a convex subdifferential)
%of functionals which can be nondifferentiable typically because
%they describe some unilateral, unidirectional, or activated phenomena.
Also, \eqref{Biot} takes the structure of {\it dissipative Hamiltonian system},
and the \emph{Hamilton variational principle}~\COL{\cite{Ham34GMD}}
extended to dissipative systems as in \cite{Bedf85HPCM}
says that the solution $q$ to \eqref{Biot} on a fixed time interval $[0,T]$
is a critical point of the integral functional 
\begin{align}\label{Hamilton}
q\mapsto\int_0^T\!\!\!\mathscr{M}(\DT q)-{\mathscr E}(q)-\big\langle
\mathfrak{f},q\big\rangle\,\d t
\end{align}
with a nonconservative force $\mathfrak{f}={\mathscr R}_{\DT q}'(q,\DT q)
-{\mathscr F}(t)$
considered fixed on the affine manifold 
respecting some initial or terminal conditions.
\COL{The notation $\langle\cdot,\cdot\rangle$ in~\eqref{Hamilton}  means the value of a functional (in our case $\mathfrak f$) on a test function (in our case $q$), which for sufficiently smooth functions can be understood as an integral, see e.g.\ \eqref{E-R-M-F} below.}
%$\{q\!\in\!L^\infty(I;V);\
%\DT q\!\in\!L^\infty(I;H),\ \frac{\d^2u}{\d
%  t^2}\!\in\!L^2(I;V^*),\ u(0)=u_0,\ \frac{\d u}{\d t}=v_0\}$.

This energy-governed structure \eqref{Biot} allows to control the 
energetics: indeed, testing  \eqref{Biot} by $\DT q$ and using the calculus 
$\langle{\mathscr M}'\DDT q,\DT q\rangle=\frac{\d}{\d t}{\mathscr M}(\DT q)$ 
and $\langle{\mathscr E}'(q),\DT q\rangle=\frac{\d}{\d t}{\mathscr E}(q)$, we 
arrive (at least formally) to the energy balance on a time interval $[0,t]$:
\begin{subequations}\label{engr-balance}\begin{align}%\nonumber
&\hspace{-0.0cm}\ddd{{\mathscr M}(\DT
q(t)){+}{\mathscr E}(q(t))_{_{_{_{_{_{_{}}}}}}}\!\!\!}
{kinetic\,+\,stored energy}{at time $t$}
\hspace{-0.em}{+}\hspace{-0.em}\ddd{\int_0^{\,t}\!\!\Xi(q(s),\DT q(s))\,\d
s}{dissipated energy over}{the time interval
$[0,t]$}\hspace{-0.8em}
%\\&\ \ \ \ 
=\hspace{-.2em}
\ddd{{\mathscr M}(\DT q(0)){+}{\mathscr E}(q(0))_{_{_{_{_{_{_{_{}}}}}}}}\!\!\!}
{\ kinetic+stored energy}{at time $0$}\hspace{-0.2em}
+\hspace{-0.2em}\ddd{\int_0^{\,t}\!\!\!
%{\mathscr E}_t'(t,q){+}
\big\langle{\mathscr F}(s),\DT q(s)\big\rangle\,\d s}{work done by
loading}{over time interval $[0,t]$}\!\!\!\!\!
\end{align}
%\COMMENT{preco pri $q$ nepises $q(t)$?}
with the dissipation rate
\begin{align}\label{def-of-Xi}
%$\bfq_0=(u_0,\bfzeta_0,\bfpi_0,\bfeps_0)$
\Xi\big(q,\DT q\big)
=\big\langle{\mathscr R}_{\DT q}'(q,\DT q),\DT q\big\rangle. 
\end{align}\end{subequations}
%More precisely, \eqref{engr-balance} is usually obtained from the 
%sub-differential
%formulation rather as an inequality only, while the equality in
%\eqref{engr-balance} needs some data qualification.
% (e.g.\ to ensure ${\mathscr M}'\DDT u$ in duality with $\DT u$ etc.).

Moreover, this structure allows for various numerically stable time 
discretisations, and for rigorous analysis as far as convergence and existence 
of solutions to \eqref{Biot} concerns.

%\section*{\large\it 3. Rupture in viscoelastic continua}
%                       ~~~~~~~~~~~~~~~~~~~~~~~~~~~~~~~~~

\section{Coupling of rupturing solids with elastic fluids}\label{sec-coupling}
%        ~~~~~~~~~~~~~~~~~~~~~~~~~~~~~~~~~~~~~~~~~~~~~

Denoting the displacement by $u$, we use the usual concept of small strains, 
with the total strain $e=e(u)=\frac12\nabla u+\frac12(\nabla u)^\top$
and the inelastic, symmetric, trace-free strain $\pi$.
As standard in small-strain theories, we consider the additive 
split, \COL{sometimes called Green-Naghdi split~\cite{GreNag65GTEP}}, of the total strain $e=e(u)$ as
\begin{align}\label{eta-split}
e(u)=e_{\rm el}+\pi
\end{align}
with $e_{\rm el}$ denoting the elastic strain and $\pi$ the inelastic
strain, the latter being considered as trace free. 
To distinguish the response on the shear or compression
load, we further introduce its orthogonal decomposition to the 
spherical and the deviatoric part: 
\begin{subequations}\begin{align}\label{sph-dev}
&e(u)={\rm sph}\,e(u)+{\rm dev}\,e(u)
\ \ \text{ with }\ {\rm dev}\,e=e-\frac{{\rm tr}\,e}3\mathbb I,
\\[-.2em]&\label{sph-dev+}
{\rm dev}\,e={\rm dev}(e_{\rm el}{+}\pi)=e_{\rm dev}{+}\pi
\ \text{ with }\ e_{\rm dev}\!={\rm dev}e_{\rm el},
\end{align}\end{subequations}
where $\mathbb I$ is the identity matrix while 
${\rm tr}\,e=\sum_{i=1}^3e_{ii}$ denotes the trace of $e$.
%Note that the deviatoric strain ${\rm dev}\,e(u)$ is trace-free. 
%In \eqref{sph-dev+}, we used the usual additive decomposition \eqref{eta-split},
%so that $e_{\rm dev}={\rm dev}\,e(u)-\pi$ is in the position of
%the deviatoric part of the elastic strain.

\subsection{Solid model for mantle and inner core}\label{sec-solid}
\begin{figure*}
\centering
\includegraphics[width=47em]{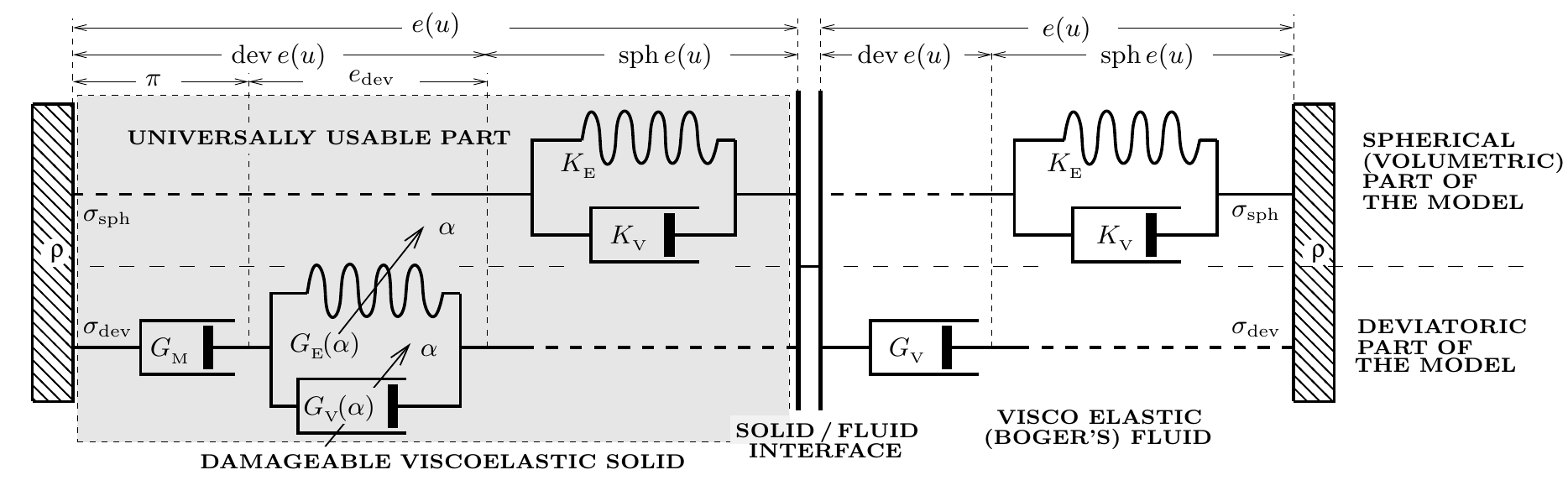}
\caption{\sl Schematic diagram for solid-fluid interaction. 
The spherical response is the Kelvin-Voigt rheology both in 
the solid and in the fluid parts.
The deviatoric response of the solid uses a viscoelastic Jeffreys rheology 
subjected to damage $\alpha$, while the fluid uses the mere Stokes model.
%has a viscoelastic spherical Kelvin-Voigt or Jeffreys deviatroric rheology,
%the later one subjected to damage $\alpha$, 
%in the deviatoric part while the spherical (volumetric) part is not 
%subjected to damage, 
%while the fluid is mere Stokes model in the deviatoric part while
%while viscoelastic in spherical part.
}
\label{fig-mixed-response}
\end{figure*}

%           ~~~~~~~~~~~~~~~~~~~~~~~~~~~~~~~~~~~~~
A rather minimal scenario for a simplified seismic source 
model in the solid part (the mantle) is an isothermal damage 
(also called aging) in the deviatoric part. 
To model also the aseismic slip (creep) or a permanent displacement during 
cumulation of subsequent shift (which is important in particular if healing
is considered in between re-occurring earthquakes), as well as a low 
attenuation of seismic waves, the Maxwell viscoelastic 
rheology is considered in the deviatoric response. For well-posedness
of the problem, it is suitable to combine it with the Kelvin-Voigt
rheology, which results to the Jeffreys rheological model, as used in
\cite{LyHaBZ11NLVE}. The spherical response in the solid part 
(the mantle and inner core) exhibits just the undamageable Kelvin-Voigt
rheology, reflecting the phenomenon that neither any big permanent deformation 
nor damage by compression is possible. Cf.\ Figure~\ref{fig-mixed-response}/left
for the overall rheological model.

The state $q=(u,\pi,\alpha)$ then involves the displacement $u$,
the (trace-free) inelastic strain $\pi$, and the scalar damage variable 
$\alpha$; alternatively, $\alpha$ is called also {\it aging} and 
we follow the usual convention that $\alpha=0$ means no damage while 
$\alpha=1$ is the maximal degradation. Consistently with 
Section~\ref{sec-energetic}, we must specify 
the energies governing the model through the (generalized) Hamiltonian 
principle \eqref{Hamilton}. 
%
%Additive decomposition ${\rm sym}(\nabla u)=:e(u)=e_{\rm el}+\pi$.
The {\it specific stored energy} 
%$\mathscr{E}=\mathscr{E}(q)$ 
is postulated as 
\begin{align}%\nonumber
&\varphi(e,\pi,\alpha)=\frac 32K_{_{\rm E}}|{\rm sph}\,e|^2
+G_{_{\rm E}}(\alpha)|e_{\rm dev}|^2+\gamma_{_{\rm DAM}}(\alpha)
%\\&\qquad
\label{def-of-phi}
=\frac 32K_{_{\rm E}}|{\rm sph}\,e|^2
+G_{_{\rm E}}(\alpha)|{\rm dev}\,e{-}\pi|^2+\gamma_{_{\rm DAM}}(\alpha)
\end{align}
with $K_{_{\rm E}}$ the elastic bulk modulus 
%($K=\lambda+\frac2d G$),
and $G_{_{\rm E}}=G_{_{\rm E}}(\alpha)$ the elastic shear modulus \COL{introduced for a damageable Lam\'{e} material},
%(=second Lam\'e constant $\mu$),
and $\gamma_{_{\rm DAM}}=\gamma_{_{\rm DAM}}(\alpha)$ damage stored energy
(also understood as the damage toughness).
The {\it specific dissipation potential} is
%(Maxwell + Kelvin-Voigt = Jeffreys rheology):
\begin{align}\label{zeta-split}
&\zeta(\alpha;\DT{e},\DT\pi,\DT\alpha)=
\zeta_{_{\rm VSC}}(\alpha;\DT{e},\DT\pi)+\zeta_{_{\rm DAM}}(\DT\alpha)\ \ \ 
\text{ with } 
\\&\nonumber
\zeta_{_{\rm VSC}}(\alpha;\DT{e},\DT\pi)=%\frac d2
\frac 32K_{_{\rm V}}|{\rm sph}\,\DT e|^2\!
+G_{_{\rm V}}(\alpha)|{\rm dev}\,\DT e{-}\DT\pi|^2\!
%\\
+\,G_{_{\rm M}}|\DT\pi|^2
%+\eta(\DT\alpha)
\end{align}
and with $\zeta_{_{\rm DAM}}=\zeta_{_{\rm DAM}}(\DT\alpha)\ge0$ a convex function of 
the damage rate \COL{ and with the convex quadratic functional $\zeta_{_{\rm VSC}}$.
  The former one is the potential of non-conservative forces related to damage
  evolution while the latter one together with elastic parts in the stored
  energy introduces Kelvin-Voigt rheology for the spherical part and Jeffrey
  rheology for the deviatoric part}. 
We consider still the specific {\it kinetic energy} in the 
usual form $\displaystyle{\frac\varrho2|\DT u|^2}$ with $\varrho=\varrho(x)$ 
the mass density.

To introduce a length-scale into the model that allows for 
``controlling'' a typical width of the damaged zone around 
fault core \bbb(usually varying in between 1--100 m) \eee
and of the narrow cataclasite fault core (i.e.\ the principal slip zone\bbb, usually
varying in between 0.1--10 m\eee) 
where the inelastic strain is accommodated, the gradient theories are used. 
This means that $\varphi$ from \eqref{def-of-phi} is augmented both by 
$|\nabla\alpha|^2$ and $|\nabla\pi|^2$ terms. In a special choice 
$\gamma_{_{\rm DAM}}(\alpha)=G_{\rm c}\frac{1}{2\eps}
%(1{-}\alpha)
\alpha^2$
together with $G_{_{\rm E}}(\alpha)=({\eps^2}/{\eps_0^2}{+}(1{-}\alpha)^2)\COL{G_{_{\rm 0}}}$ \COL{  where $G_{_{\rm 0}}$ is the shear modulus of the undamaged material (up to a small value of ${\eps^2}/{\eps_0^2}$) }
and $\eps>0$ small (with respect to $\eps_0$),
the ``augmented'' specific stored energy $\varphi_{_{\rm A}}$ takes the form
\begin{align}%\nonumber
&\!\!
%\calE(u,z):=\int_\Omega
\varphi_{_{\rm A}}(e,\pi,\alpha,\nabla\pi,\nabla\alpha)
:=\frac32K_{_{\rm E}}|{\rm sph}\,e(u)|^2
%\\[-.3em]&\nonumber\qquad\qquad
+
\Big(\frac{\eps^2}{\eps_0^2}{+}(1{-}\alpha)^2\Big)G_{_{\rm 0}}|{\rm dev}\,e(u){-}\pi|^2
%\bbC e{:}e
%\\[-.3em]
\label{AT-ansatz}
%&\qquad\qquad
+\frac\kappa2|\nabla\pi|^2
+\!\lineunder{G_{\rm c}\Big(\frac{1}{2\eps}
%(1{-}\alpha)
\alpha^2\!
+\frac\eps2|\nabla\alpha|^2\Big)}{crack surface density}\!
\end{align}
with $\eps_0>0$ (dimension [m]), \COL{$\kappa$ (dimension [J/m]) allows to control the width of the mentioned cataclasite fault core while $\eps$ (dimension $[m]$) allows to control the width of the mentioned damaged zone around the fault core}, and a so-called fracture toughness 
$G_{\rm c}>0$ (dimension [J/m$^2$]) fixed.
This is, in fact, the usual {\it phase-field crack} model, here considered
only in the deviatoric component reflecting the phenomenon that pure tension
(and Mode I cracks) is avoided in geophysically relevant situations while 
compression does not lead directly to cracks, so that only shear may lead to 
rupture (in Mode II). In the static situation, for $\eps$ converging to zero, 
this so-called Ambrosio-Tortorelli functional 
%\shortcite{AmbTor92AFDP} 
approximates the {\it Griffith crack model}~\COL{\cite{Gri21PRFS},
originally shown in the scalar case \cite{AmbTor92AFDP} 
and later for the vectorial case by \cite{Foca01VAFD}.
In the dynamical case, this approximation property is not rigorously justified, 
but nevertheless this phase-field model is routinely used, cf.\ e.g.\ 
\cite{BoLaRi11TDMD,LaOrSu10ESRM,MieRou15RIST,Roub??MDDP}}. Such a 
crack model is relevant particularly if a new fault is nucleated in the 
compact rock, although it is a rare event geophysically. 

%rarely, new faults may arise in compact rocks: 

%\hrule

%Bourdin at al.\ \cite{BMMS14MPCC}
%devised $\gamma_{_{\rm DAM}}(\alpha)=3G_{\rm c}\alpha/(8\eps)$ and $\kappa=3G_{\rm c}\eps/4$

%....................................

\subsection{Fluid model for outer core and oceans}\label{sec-fluid}
%           ~~~~~~~~~~~~~~~~~~~~~~~~~~~~~~~~~~~~~

Besides solid regions, there are layers in our planet that are naturally
fluidic, namely the outer core and, of course, the oceans. From the 
seismic wave propagation, they must exhibit 
some spherical elastic response (otherwise, ideally, incompressible fluids
would lead to unphysically infinitely large P-wave speed)  
but no deviatoric elastic response to prevent S-wave propagation in 
such regions. In any case, damage becomes irrelevant in these regions.

One scenario is to keep still a (presumably small) viscous response
both in the spherical and the deviatoric part, leading respectively
to a combination of the Kelvin-Voigt solid rheology and the Stokes fluid
rheology. In the notation from Section~\ref{sec-solid}, this means
that we take $G_{_{\rm E}}=0$, $\gamma_{_{\rm DAM}}=0$, $G_{_{\rm V}}>0$ independent of 
$\alpha$, $\kappa=0$, $G_{\rm c}=0$, and we formally put $G_{_{\rm M}}=\infty$.
The last action results to $\pi=0$ provided the initial conditions 
are $\pi|_{t=0}=0$. In fact, an equivalent option would be 
to put $G_{_{\rm V}}=\infty$ and keep $G_{_{\rm M}}>0$ finite (small).
This leads to the so-called {\it viscoelastic} ({\it Boger}'s \cite{Bog77HECV}) 
{\it fluid}, cf.\ Figure~\ref{fig-mixed-response}/right-part. The S-waves actually 
can slightly penetrate into such fluids but are soon attenuated 
and thus cannot propagate practically.

Another scenario is to suppress any viscosity, obtaining, thus, the merely
{\it elastic fluid}. This can be achieved, in addition to 
taking  $G_{_{\rm E}}=0$, $\gamma_{_{\rm DAM}}\!=0$, $\kappa=0$, and $G_{\rm c}=0$, 
%and $G_{_{\rm M}}=\infty$, 
by putting both $K_{_{\rm V}}=0$ and $G_{_{\rm V}}=0$.
Such fluids are fully conservative, leading to a linear hyperbolic
problem. The S-waves cannot propagate through such fluidic regions at all.

\subsection{The coupled equations}\label{sec-coupled}
%           ~~~~~~~~~~~~~~~~~~~~~~~~~~~~~~~~~~~~~

Let us \COL{consider} the reference domain $\Omega$ occupied by a viscoelastic body in question (e.g., the whole planet Earth). Of course, the local 
potentials $\varphi_{_{\rm A}}$ and $\zeta$, and 
also the mass density $\varrho$ are allowed (and supposed) to depend
on $x\in\Omega$, and in particular, they can vary on the 
solid and the fluid domains; let us denote them by $\OS$ and $\OF$,
respectively. We will not indicate this dependence explicitly for notational 
simplicity.

Taking into account the additive split \eqref{eta-split},
the overall functionals used in Section~\ref{sec-energetic} are now:
\begin{subequations}\label{E-R-M}\begin{align}
&\mathscr{E}(q)=\mathscr{E}(u,\pi,\alpha)=\int_\Omega
\varphi_{_{\rm A}}(e(u),\pi,\alpha,\nabla\pi,\nabla\alpha)\,\d x,
\\&
%\nonumber
\mathscr{R}(q,\DT q)=\mathscr{R}(\alpha;\DT u,\DT\pi,\DT\alpha)
%=\mathscr{R}_{_{\rm VISCO}}(\alpha;\DT u,\DT\pi)+\mathscr{R}_{_{\rm DAM}}(\DT\alpha)
%\\&\hspace{3em}
=\int_\Omega\zeta_{_{\rm VSC}}(\alpha;e(\DT{u}),\DT\pi)
+\zeta_{_{\rm DAM}}(\DT\alpha)\,\d x,
\label{E-R-M-split}
\\&\mathscr{M}(\DT q)=\mathscr{M}(\DT u)=\int_\Omega\frac\varrho2|\DT u|^2\,\d x,
\\&\label{E-R-M-F}
%\REPLACE{{\mathscr F}(t,u)=\int_\Omega g(t)\cdot u\,\d x}{%
\langle{\mathscr F}(t),u\rangle=\int_\Omega g(t)\cdot u\,\d x
\end{align}\end{subequations}
where $g$ is a bulk force, typically the gravitational force independent
of time or some more general force as, e.g., in \eqref{grav-force} or 
\eqref{Coriolis-force} below.
Note the \eqref{E-R-M-F} does not involve the internal variables 
$(\pi,\alpha)$, in accord with the conventional concept that the 
internal variables are not directly subjected to outer forcing.

More specifically, the system that results from \eqref{Biot}
by considering \eqref{E-R-M} together with \eqref{AT-ansatz} looks as:
\begin{subequations}\label{system++}\begin{align}\label{system-u++}
&\!\varrho\DDT u-{\rm div}(\sigma_{\rm sph}{+}\sigma_{\rm dev})
%(%\bbD_{_{\rm V}}\DT e_{\rm el}
%\zeta_{\DT e}'(e(\DT u),\DT \ein)+\varphi_e'(e(u),\ein,\alpha))
=g&&\!\!\text{in }\ \Omega
\\&\hspace{.5em}\text{ with }\ 
\sigma_{\rm sph}=
%\begin{cases}3K_{_{\rm V}}{\rm sph}\,e(\DT u)
%%(e(\DT u){-}\DT\pi)
%+3K_{_{\rm E}}{\rm sph}
%%(e(u){-}\pi)
%\,e(u)\!\!\!
%&\text{in }\ \OS,\\
3K_{_{\rm V}}{\rm sph}\,e(\DT u)
+3K_{_{\rm E}}{\rm sph}\,e(u)\!\!\!\!\!\!
&&\!\!\text{in }\ \Omega,
%&\text{in }\ \OF,
%\end{cases}\hspace{-9em}
\label{system++sph}
\\&\hspace{.5em}\text{ and }\,\ \sigma_{\rm dev}=
\begin{cases}2G_{_{\rm V}}\!(\alpha)\big({\rm dev}\,e(\DT u){-}\DT\pi\big)&
%(e(\DT u){-}\DT\pi)
\\[-.2em]\hspace*{.7em} +
2G_{_{\rm E}}\!(\alpha)
%,e_{\rm el}){\rm dev}\,e_{\rm el}
%\partial_{e}\calG_{_{\rm E}}(\alpha,{\rm dev}\,e(u))
({\rm dev}\,e(u){-}\pi)
&\text{in }\ \OS,\\
2G_{_{\rm V}}{\rm dev}\,e(\DT u)&\text{in }\ \OF,\end{cases}\hspace{-10em}
%\displaybreak
\\[.2em]&\label{system-pi}
\!\COL{G_{_{\rm M}}\DT\pi
=
 \sigma_{\rm dev}+{\rm div}(\kappa\nabla\pi)} &&\!\!\text{in }\ \OS,
\\[.2em]&%\nonumber
%\!\!\!\zeta_{_{\rm DAM}}'(\DT\alpha)
%%-\kappa_{\rm v}\Delta\DT\alpha
%+
%%\frac12\bbC'(\alpha)e_{\rm el}\Colon e_{\rm el}
%%p_{\rm dam}\partial_\alpha G_{_{\rm E}}(\alpha,{\rm dev}\,e(u))
%G_{_{\rm E}}'(\alpha)|({\rm dev}\,e(u){-}\pi)|^2\!
%={\rm div}(\kappa\nabla\alpha)\!\!\!\!
%%\kappa\Delta\DT\alpha
\!\zeta_{_{\rm DAM}}'(\DT\alpha)+
G_{_{\rm E}}'(\alpha)|({\rm dev}\,e(u){-}\pi)|^2
+\frac{G_{\rm c}}\eps\alpha
%&&{}\\
\label{system-alpha++}
%&\hspace{10.5em}
={\rm div}(\eps G_{\rm c}\nabla\alpha)\!\!\!\!
&&\!\!\text{in }\ \OS.
%\\[-.5em]&\hspace{8.8em}\text{with }\ p_{\rm dam}=\begin{cases}
%G'(\alpha)|{\rm dev}(e(u){-}\pi)|^2\ \ \ \ \ \ \,
%&\text{in }\ \OS,\\[-.2em]
%\quad 0&\text{in }\ \OF,\end{cases}\hspace{-9em}
\end{align}\end{subequations}
Note that the spherical and the deviatoric stresses are orthogonal, i.e.
$\sigma_{\rm sph}:\sigma_{\rm dev}=0$. Also note that $\pi$ and $\alpha$
are relevant only on the solid domain $\OS$, cf.\ also 
Figure~\ref{fig-mixed-response}. 

There are no transient conditions for 
displacement/stress on the solid-fluid interior interfaces (i.e.\ between 
outer core and mantle and inner core and possibly also between mantle and 
oceans) because \eqref{system-u++} is considered on the whole domain $\Omega$. 
%so that the tractions on these interfaces are equilibrated automatically.
In particular, the stress-vector equilibrium and 
continuity of the displacement across these interfaces are automatically
involved and does not need to be written explicitly.
On the other hand, there are boundary conditions to be read from the 
abstract equation \eqref{Biot}, namely 
%The interface conditions for displacement/stress on the interior boundaries 
%between the mantle and the 
%outer core and the inner and the outer core as well between the 
%mantle (crust) and the oceans are automatically involved as \eqref{system-u++}
%holds on the whole $\Omega$. On the other hand,
%note in particular that both $\pi$ and $\alpha$ in \eqref{system++}
%are now needed and defined only in the solid part $\OS$.
%, while damage $\alpha$ is also (rather formally) defined also in the
%fluid part $\OF$ but is irelevant there and, as the driving ``pressure'' 
%$p_{\rm dam}$ is zero there and (pressumably) $\kappa>0$ is small, 
%it stays nearly constant (equal to $\alpha_0$).
%Those formally defined $\alpha$ in $\OF$ may possibly only slightly 
%influence $\OS$ (where it really occurs in $\sigma_{\rm dev}$) near 
%the boundaries between $\OS$ and $\OF$.
%as \eq{system-alpha++} is not considered only on $\OS$ instead of $\Omega$,  
%Therefore, the condition for the ``flux'' of $\pi$ and $\alpha$ is now to be prescribed 
%on all interior boundaries (i.e.\ on the mantle/core and mantle/oceans and
%inner/outer-core interfaces). More specifically,
%\eqref{BC-IC-alpha} is to be replaced by
\begin{subequations}\label{BC}\begin{align}\label{BC1}
&(\sigma_{\rm sph}{+}\sigma_{\rm dev})\vec{n}=0&&\text{on }\ \partial\Omega,
\\&\label{BC2}
\COL{\kappa\frac{\partial\pi}{\partial\vec{n}}=0}\ \ \ \text{ and }\ \ \ 
\COL{\eps G_{\rm c}\frac{\partial\alpha}{\partial\vec{n}}=0}\!\!\!&&\text{on }\ \partial\OS,
\end{align}\end{subequations}
where $\vec{n}$ denotes the normal to the boundary $\partial\Omega$
of $\Omega$ (e.g.\ the surface of the Earth) 
the interior boundaries $\partial\OS$ (i.e.\ on the mantle/core 
and mantle/oceans and inner/outer-core interfaces) \COL{and $\frac\partial{\partial\vec{n}}=\Vec{n}\cdot\nabla$ denotes normal derivative}.

We have in mind an initial-value problem, so that we have to complete
the system still by initial conditions:
\begin{subequations}\label{IC}\begin{align}
&u|_{t=0}^{}=u_0\ \ \text{ and }\ \ \DT u|_{t=0}^{}=v_0
&&\text{ in }\ \Omega,
\\&\label{IC2}
\pi|_{t=0}^{}=\pi_0\ \ \text{ and }\ \ \alpha|_{t=0}^{}=\alpha_0
&&\text{ in }\ \OS.
\end{align}\end{subequations}

Although it is usually not an aspect under attention in geophysical
modelling, let us mention that it is possible to prove rigorously 
that, under a suitable data qualification, the initial-boundary-value 
problem \eqref{system++}--\eqref{IC} has a solution which also satisfies
the energy conservation \eqref{engr-balance}, cf.\ \cite{Roub??SWEG}
for the case $G_{_{\rm V}}$ independent of $\alpha$ while 
\cite[Sect.\,7.5]{KruRou18MMCM}
outlines modifications if $G_{_{\rm V}}(\cdot)=\tau_{_{\rm V}}G_{_{\rm E}}(\cdot)$
for some relaxation time $\tau_{_{\rm V}}=\tau_{_{\rm V}}(x)>0$.

\subsection{Towards fluids from solids: a monolithic model}
%           ~~~~~~~~~~~~~~~~~~~~~~~

One can approximate the fluid models from Sect.\,\ref{sec-fluid} 
asymptotically when sending corresponding parameters of the
solid model in Sect.\,\ref{sec-solid} to their limits. More
specifically, Boger's viscoelastic fluid can be approached 
from the Jeffreys solid by sending 
\begin{align}
G_{_{\rm M}}\!\to\infty,\ \ 
G_{_{\rm E}}\!\to0,\ \ \kappa\to0,\,\text{ and }\,\zeta_{_{\rm DAM}}\!\to0
\ \,\text{ in }\,\OF.\!
\end{align}
When $\pi_0=0$ in \eqref{IC2}, the inelastic strain $\pi$ 
converges to 0 on the fluidic domain $\OF$. Under suitable data
qualification and scaling, this convergence can rigorously 
be justified together with convergence of the energy balance, 
cf.\ \cite{Roub??SWEG}.

Sending further 
\begin{align}
G_{_{\rm V}}\to0\ \ \text{ and }\ \ K_{_{\rm V}}\to0\ \ \text{ in }\ \OF,
\end{align}
we approach the merely elastic fluid in $\OF$. Again, this convergence
can rigorously be proved together with convergence of the energy balance
(i.e.\ the energy dissipated via viscous attenuation in the Bogger fluid
actually converges to zero) but for a slightly modified model with 
the bounded elastic stress and constant $G_{_{\rm V}}$ in the solid part $\OS$, 
cf.\ \cite{Roub??SWEG}.

The idea behind the ``monolithic'' approach is to take,
instead of the limit fluid in Sect.\,\ref{sec-fluid}, 
its approximation and to implement computationally only the solid model
in the whole domain $\Omega$. Again, the interface conditions between $\OS$ 
and $\OF$ expressing here continuity of displacements and the traction stresses 
will thus be covered automatically \bbb without any extra effort in
coding\eee.

\section{Staggered\,/\,FEM  discretisation}\label{sec-staggered}
%    ~~~~~~~~~~~~~~~~~~~~~~~~~~~~~~~~~~~~~

We write the abstract equation \eqref{Biot} as a first-order 
system, which is more suitable for time discretisation than
the original 2nd-order system, in particular because it allows 
for varying time steps during simulations. 
In view of \eqref{E-R-M},
%and the aditive split \eqref{E-R-M-split}
it has the structure
\begin{subequations}\label{Biot+}\begin{align}
&\ \DT u=v\,,\\ 
\label{Biot+u}
&\COL{
{\mathscr M}'\DT{v}+{\mathscr R}_{v}'(\alpha;v,\DT\pi)
+{\mathscr E}_{u}'(u,\pi,\alpha)={\mathscr F}(t)\,,}\\ 
\label{Biot+p}
&\COL{
{\mathscr R}_{\DT\pi}'(\alpha;v,\DT\pi)
+{\mathscr E}_{\pi}'(u,\pi,\alpha)=0\,,}\\ 
\label{Biot+a} &
{\mathscr R}_{\DT\alpha}'(\DT\alpha)+{\mathscr E}_\alpha'(u,\pi,\alpha)=0\,.
\end{align}\end{subequations}
It is important that ${\mathscr R}$ is additively split, cf.\ 
\eqref{E-R-M-split}, which suggests the splitting with the time
discretisation of the state $q$ to the components $(u,\pi)$ and $\alpha$ \COL{and in the same manner also the calculation is performed}. 
Although ${\mathscr E}$ is necessarily nonconvex to facilitate modelling 
of sudden rupture events, it is also advantageous that both 
${\mathscr E}(\cdot,\cdot,\alpha)$ and ${\mathscr E}(u,\pi,\cdot)$
are convex or, here in the ansatz \eqref{AT-ansatz}, even quadratic.
%\INSERT{ It should be noted that the differentials in~\eqref{Biot+up} have two components related to the couple $(u,\pi)$ meant as
%${\mathscr R}_{(v,\DT\pi)}'(\alpha;v,\DT\pi)=\left({\mathscr R}_{v}'(\alpha;v,\DT\pi),{\mathscr R}_{\DT\pi}'(\alpha;v,\DT\pi)\right)^\top$, ${\mathscr E}_{(u,\pi)}'(u,\pi,\alpha)=\left({\mathscr E}_{u}'(u,\pi,\alpha),{\mathscr E}_{\pi}'(u,\pi,\alpha)\right)^\top$, while the other terms in the same equation relation have in this sense only the first one non-vanishing. 
%The equation could have been written as a system of two equations with separated  $u$ and $\pi$ differentials, but it was advantageous to keep them coupled due to proposed numerical scheme which follows.}

Considering a time step $\tau=(\tau_k)_{k\in\N}$ with $k=1,2,....$ indexing  
time levels $t^k$ in the time-discrete system so that $\tau^k=t^k-t^{k-1}$, and 
using also the Crank-Nicholson mid-point strategy, we devise the staggered 
(also called fractional-step splitting) system
\begin{subequations}\label{Biot++}\begin{align}
&\frac{u_\tau^k-u_\tau^{k-1}}{\tau_k}=v_\tau^{k-1/2}\ \ \text{ with }\ \ 
v_\tau^{k-1/2}=\frac{v_\tau^k+v_\tau^{k-1}}2,
\label{Biot++1}
\\
%\nonumber
&\COL{
{\mathscr M}'\frac{v_\tau^k-v_\tau^{k-1}}{\tau_k}+
%[{\mathscr R}_{_{\rm VISCO}}]
{\mathscr R}_{v}'\Big(\alpha_\tau^{k-1};v_\tau^{k-1/2},
\frac{\pi_\tau^k-\pi_\tau^{k-1}}{\tau_k}\Big)}
%\\&\hspace{4em}
\COL{\ +\,
%\partial_{\bfq}
{\mathscr E}_{u}'(u_\tau^{k-1/2},\pi_\tau^{k-1/2},\alpha_\tau^{k-1})
={\mathscr F}(t^k),}
\label{Biot++2u}
\\%\nonumber
&\COL{
{\mathscr R}_{\DT\pi}'\Big(\alpha_\tau^{k-1};v_\tau^{k-1/2},
\frac{\pi_\tau^k-\pi_\tau^{k-1}}{\tau_k}\Big)}
%\\&\hspace{4em}
\COL{\ +\,
%\partial_{\bfq}
{\mathscr E}_{\pi}'(u_\tau^{k-1/2},\pi_\tau^{k-1/2},\alpha_\tau^{k-1})
=0\,,
\label{Biot++2p}}
\\&
%{\mathscr R}_{_{\rm DAM}}'
{\mathscr R}_{\DT\alpha}'
\Big(\frac{\alpha_\tau^k-\alpha_\tau^{k-1}}{\tau_k}\Big)
+{\mathscr E}_\alpha'(u_\tau^k,\pi_\tau^k,\alpha_\tau^{k-1/2})=0\,.
\label{Biot++3}
\end{align}\end{subequations}
The system \eqref{Biot++} is to be solved, recursively,
for $k=1,2,...$, starting from the initial conditions 
$u_\tau^0=u_0$, $v_\tau^0=v_0$, $\pi_\tau^0=\pi_0$, and $\alpha_\tau^0=\alpha_0$,
cf.\ \eqref{IC}. \COL{The system is decoupled in the sense that the 
calculation is performed separately for 
$(u_\tau^k,v_\tau^k,\pi_\tau^k)$ from (\ref{Biot++}a-c)
and $\alpha_\tau^k$ from \eqref{Biot++3}, as mentioned already above.}

This scheme exhibits energy conservation \cite{RouPan17ECTD}, which can be 
seen by multiplying \eqref{Biot++2u} by 
\COL{the mid-point velocity 
$v_\tau^{k-1/2}$, \eqref{Biot++2p} by the inelastic-strain rate}
$(\pi_\tau^k{-}\pi_\tau^{k-1})/\tau_k$, and \eqref{Biot++3} by 
$(\alpha_\tau^k{-}\alpha_\tau^{k-1})/\tau_k$.
For the ${\mathscr M}$-term in the former test, we use \eqref{Biot++1} 
and the binomial formula to obtain an analog of the calculus
$\langle{\mathscr M}'\DT v,v\rangle=\frac{\d}{\d t}{\mathscr M}(v)$
as the equality 
\begin{align}%\nonumber
\Big\langle{\mathscr M}'\frac{v_\tau^k-v_\tau^{k-1}}{\tau_k},
v_\tau^{k-1/2}\Big\rangle
%\\&\nonumber
&=\Big\langle{\mathscr M}'\frac{v_\tau^k-v_\tau^{k-1}}{\tau_k},
\frac{v_\tau^k+v_\tau^{k-1}}2\Big\rangle
%\\&
=\frac{{\mathscr M}\big(v_\tau^k\big)-{\mathscr M}\big(v_\tau^{k-1}\big)}{\tau_k},
\end{align}
while another binomial formula for the quadratic 
functional ${\mathscr E}(\cdot,\cdot,\alpha_\tau^{k-1})$ 
gives 
\begin{align}%\nonumber
\Big\langle{\mathscr E}_{(u,\pi)}'(u_\tau^{k-1/2},\pi_\tau^{k-1/2},\alpha_\tau^{k-1}),
\Big(v_\tau^{k-1/2},\frac{\pi_\tau^k-\pi_\tau^{k-1}}{\tau_k}\Big)\Big\rangle\!\!\!\!\!\!
%\\
={\mathscr E}(u_\tau^{k},\pi_\tau^{k},\alpha_\tau^{k-1})
-{\mathscr E}(u_\tau^{k-1},\pi_\tau^{k-1},\alpha_\tau^{k-1}).
\label{binom1}\end{align}
Similarly, the latter test gives 
\begin{align}%\nonumber
&\Big\langle
{\mathscr E}_\alpha'(u_\tau^k,\pi_\tau^k,\alpha_\tau^{k-1/2}),
\frac{\alpha_\tau^k-\alpha_\tau^{k-1}}{\tau_k}\Big\rangle
%\\[-.2em]&\qquad\qquad
={\mathscr E}(u_\tau^{k},\pi_\tau^{k},\alpha_\tau^{k})
-{\mathscr E}(u_\tau^{k},\pi_\tau^{k},\alpha_\tau^{k-1})\,.
\label{binom2}\end{align}
Summing \eqref{binom1} with \eqref{binom2}, we 
can enjoy cancellation of $\pm{\mathscr E}(u_\tau^{k},\pi_\tau^{k},\alpha_\tau^{k-1})$.
Then, summing it over $k=1,2,...$, we obtain a discrete
analog of the energy balance \eqref{engr-balance} as an (exact!) equality, 
not only as an inequality. This eliminates a spurious numerical
attenuation usually exhibiting by implicit discretization schemes and
is helpful during coding because accidental mistakes can thus be immediately
detected.   

After applying still a space discretisation, the recursive scheme 
\eqref{Biot++} can be implemented on computers, leading to 
one linear and one linear-quadratic-programming problem at 
each time level. Therefore, this scheme can be solved by finite 
(non-iterative) algorithms, is robust (i.e.\ numerically stable), 
convergent, energy conserving. It can be interpreted as a combination of
the (generalized) Newmark time discretisation with staggered
(fractional-step split) for the damage flow-rule, cf.\ also 
\cite{HofMie12CPFM} for a similar scheme. Here, for the space disretisation
in the examples in the following Section~\ref{sec-comput},  we used 
just the simplest P1 finite elements. \bbb On the other hand, a more
detailed numerical analysis as far as rate of convergence or error
estimates is another challenge not addressed in this article, and it is
standardly considered as very difficult in such highly nonlinear
problems, however. \eee

\section{Illustrative 2D computational simulations}\label{sec-comput}
%    ~~~~~~~~~~~~~~~~~~~~~~~~~~~~~~~~~~~~~~~~~

We present rather academic (not entirely in real 
geophysical scale and only 2-dimensional) computational
examples documenting efficiency and applicability range
of the above presented monolithic model even in 
its simplified variant when $G_{_{M}}\to\infty$ so that 
$\pi\to0$. In other words, we neglect the inelastic strain,
which is relevant for short-time/range events like 
ongoing ruptures and earthquakes around hypocentres if 
the (low) Maxwellian attenuation or aseismic creep are neglected.
Also, for the illustrative calculations, we have neglected the 
viscosity, which is justified in the fluidic part (where it leads 
to the linear hyperbolic problem for a merely elastic fluid
\cite{Roub??SWEG}) but not in the nonlinear solid part. 
Regardless, there is a belief that this shortcut does not influence
substantially the presented simulations. 
%{\tiny Also, this simple model already 
%will well illustrate mathematical difficulties related to nonlinearities 
%in the solid parts coupled with linear but possibly hyperbolic fluidic regions.}

%\INSERT{
We consider a solid rock with the elastic bulk modulus $K_{_{\rm ES}}=600$\,GPa, 
the shear modulus $G_{_{\rm ES}}=250$\,GPa, and density 
$\varrho_{_{\rm S}}=5000$\,kgm$^{-3}$. Thus, in this two-dimensional situation,
the speed of the P-waves (=\,the sound speed) is $v_{_{\rm PS}}=\sqrt{
%\left(K_{_{\rm ES}}+2\frac{d{-}1}dG_{_{\rm ES}}\right)
(K_{_{\rm ES}}\!+G_{_{\rm ES}})/\varrho_{_{\rm S}}}=13.04$\,kms$^{-1}$, and the speed of the 
S-waves is $v_{_{\rm SS}}=\sqrt{G_{_{\rm ES}}/\varrho_{_{\rm S}}}=7.07$\,kms$^{-1}$.
As for the fluids, we consider the elastic bulk modulus $K_{_{\rm EF}}=1100$\,GPa, 
and density $\varrho_{_{\rm F}}=11000$\,kgm$^{-3}$ which provides the speed of the 
P-waves $v_{_{\rm PF}}=\sqrt{K_{_{\rm EF}}/\varrho_{_{\rm F}}}=10$\,kms$^{-1}$. 
\COL{The other parameters of the model which influence the damage propagation are $G_{\rm c}=2$\,Jm\textsuperscript{-2}, $\eps=10^{-5}$m, $\eps_0=1$m.}

%}

%
%......$v_{_{\rm P}}$ the speed of the P-waves (=sound speed) is $\sqrt{M/\varrho}$ 
%where $M=\lambda+2G_{_{\rm E}}=K_{_{\rm E}}+2\frac{d{-}1}dG_{_{\rm E}}$ 
%is the so-called {\it P-wave modulus}
%
%......$v_{_{\rm S}}$ the speed of the S-waves is $\sqrt{G_{_{\rm E}}/\varrho}$ 
%%(always $v_{_{\rm P}}\ge v_{_{\rm S}}$)
%

As already emphasized, the model is truly academical. A real 
3-dimensional global seismic model involving larger domains (or 
possibly the whole multilayered globe) should use sophisticated 
discretisation techniques and powerful supercomputing facilities.

Here, we present three 2-dimensional computational experiments on a rather
local domain spanning ``only'' hundred(s) of kilometers.
%performed on a PC with Intel(R) Core(TM) i5 CPU@2.3GHz and 8GB RAM.
%The calculation took three or four hours.
Therefore, instead of zero-traction boundary condition \eqref{BC1}, we
consider boundary conditions allowing for a loading evolving in time. 
In particular,  the load is controlled by Dirichlet boundary conditions
(one can check it in~Figs.\ref{fig-EQ-geom},\ref{fig-EQ-geom+} and
\ref{fig-reverse-geom+} below): the prescribed nonzero 
displacements increase at the velocity $v_{\rm g}=1$\,mms$^{-1}$.
Moreover, the coefficient $3$ in \eqref{def-of-phi}--\eqref{AT-ansatz}
and \eqref{system++sph} is to be 2 in the two-dimensional setting.
Also in these three calculations, the interface between solid and fluid is
considered distinctly,
with the intent to provide various possibilities for reflected/refracted waves. Nevertheless,
it should be emphasized that in all three calculations the same computational
model is used and the differences appear only in initial/boundary conditions
and geometry of the solid and the fluid phases.

\begin{figure}
\centering
\includegraphics[width=27em,clip=true]{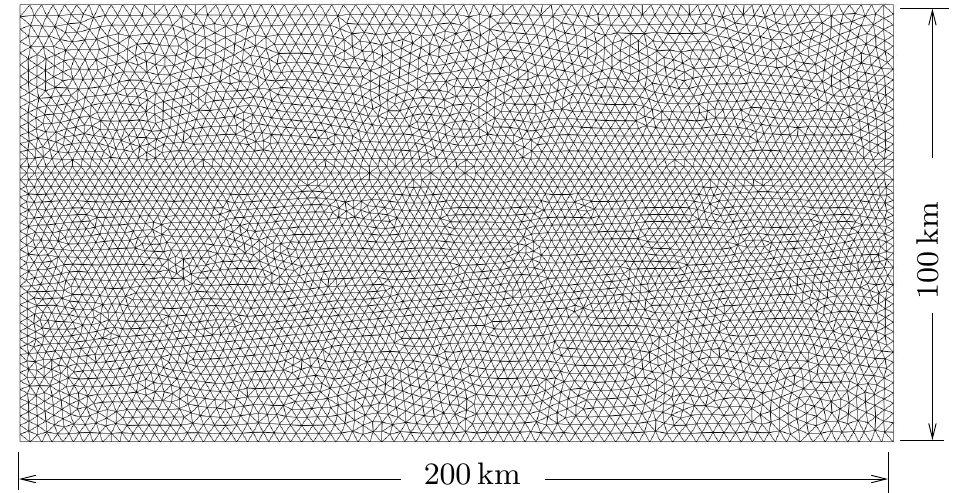}
\caption{\sl An example of triangular meshing of the rectangular
%$200{\times}100\,{\rm km}$ 
domain used for Figure~\ref{fig-EQ-geom} below. The element size is about 
$2\,{\rm km}$.
}
\label{fig-EQ-mesh}
\end{figure}

%\INSERT{
The space discretisation of the 2-dimensional body made by the P1 finite
elements includes an irregular but more or less uniformly sized mesh, whose
smallest element size $h_{\rm min}=2$\,km is a hundredth of the domain dimension,
cf.\ Figure~\ref{fig-EQ-mesh}.
The size of the elements $h_{\rm min}$ is required at the interface between 
solid and fluid and also on pre-existing faults. The largest elements close 
to the top face are only a half larger.
The size of the mesh (about 2\,km) was chosen in accordance with 
%aforementioned
computational hardware \bbb(namely Intel(R) Core(TM) i5 CPU 2.3GHz and 8GB RAM) \eee
used to compromise the visualization of
the expected geophysical events with CPU time and memory needed. Although, 
the shown mesh pertains \bbb to \eee the first calculations.
%, to one of the calculations,
the 
%others
second and the third ones are very similar with the same smallest element 
size $h_{\rm min}$ so that they are not shown explicitly.

%}

The time discretisation
\eqref{Biot++} is implemented with adaptively varying time step $\tau_k$ 
shrinking when fast rupture and wave propagation start. To capture
the latter phenomenon properly, the so-called Courant-Friedrich-Lewy (CFL) 
condition \cite{CoFrLe28PDGM} has been respected, i.e.\ here 
\COL{$\tau_k<h_{\rm min}/\max(v_{_{\rm PS}},v_{_{\rm PF}})$};
%\DELETE{ where $h_{\rm min}$ is the size of the smallest 
%element in the space discretisation}; 
in fact, the CFL-condition is ultimately 
needed for explicit time discretisation to ensure stability and convergence
but also desired for our implicit one \bbb when waves are emitted and
propagating\eee.
%\INSERT{
In particular, we took $\tau_k=100$\,s for the initial phase of the prescribed 
loading and then 
%it is 
decreased to $\tau_k=0.146$\,s when the rupture and wave propagation were 
triggered. \bbb
In absolute values, the time instant $t$ of the switch to the shorter time step
was chosen as follows:   the first calculation $t=32\,$ks, the second calculation
$t=49\,$ks, the third calculation $t=27.2$\,ks according to the rupture triggering.
The scale of temporal events in the experiments is then more then $2\times10^5$
in the performed academic calculations. In real structures it can be  several
orders higher. The prescribed horizontal shift of the upper plate at the moment
of rupture initiation regarding the velocity $v_{\text{g}}$ is in any case more
then 27\,m which is almost $10^4$ smaller then the size of the considered domain. \eee
%}

\subsection{Rupture on pre-existing \COL{horizontal} fault}\label{sec-old-fault}
%           ~~~~~~~~~~~~~~~~~~~~~~~~~~~~~

\begin{figure}
\centering
\includegraphics[width=27em]{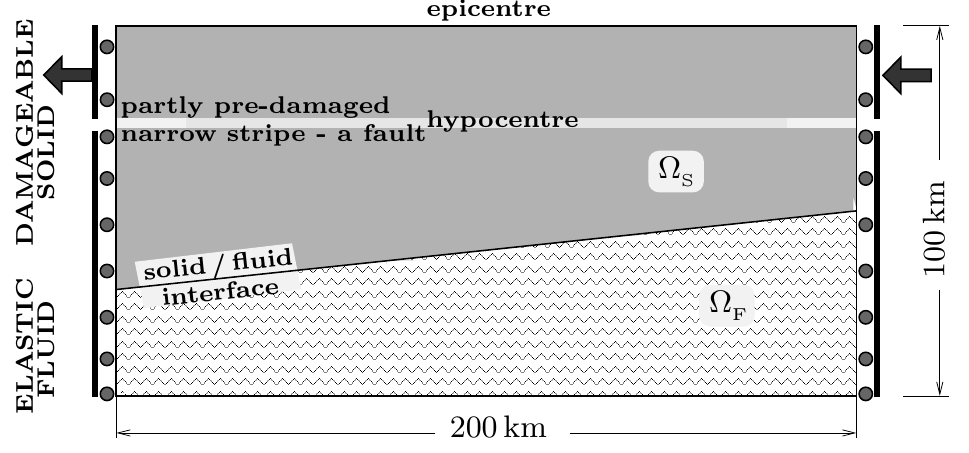}
\caption{\sl A computational 2-dimensional domain  and boundary conditions 
for the experiment in 
%Figure~\ref{fig-EQ}
Sect.\,\ref{sec-old-fault}. Upper trapezoidal part is the solid from 
Figure~\ref{fig-mixed-response}/left-part while the lower part is the fluid
from Figure~\ref{fig-mixed-response}/right-part. The upper solid part is
shifted by the 
%Dirichlet
 boundary conditions to the left while the lower solid/fluid 
part %\REPLACE{to the right}{
is horizontally constrained.
%}\COMMENT{prosim oprav podla toho aj ten obrazok}. 
At time $t=0$, the fault
is ``compact'' only at the middle part (where the hypocentre of an earthquake 
will be) while the rest of the fault (i.e.\ both sides) is (partly)
damaged.
}
\label{fig-EQ-geom}
\end{figure}

The most typical scenario leading to tectonic earthquakes is that a 
fault (as a flat usually straight stripe with weakened damage threshold) 
is gradually stretched until the shear stress reaches the critical
value to trigger the rupture. Then a fast shift occurs, within which 
a seismic (mainly S-) wave can be emitted. The geometry of the 2-dimensional 
computation region \COL{ with the (a bit hypothetically) horizontally
positioned fault } together with boundary conditions is depicted in 
Figure~\ref{fig-EQ-geom}; the boundary conditions imposes
a prescribed displacement applied from both sides of the top solid layer.

\begin{figure*}
\centering
{\footnotesize\bf \hspace{2em}KINETIC\ \,ENERGY\hspace{9.5em}SHEAR\ \,STORED\ \,ENERGY \hspace{5em} SPHERICAL\ \,STORED\ \,ENERGY}
\\
\rotatebox[origin=lt]{90}{\parbox{2cm}{\centering\COL{$t_1=32.5$\,ks}}}\hspace*{-.2em}
\includegraphics[width=0.30\textwidth,bb=100 170 640 450,clip=true]{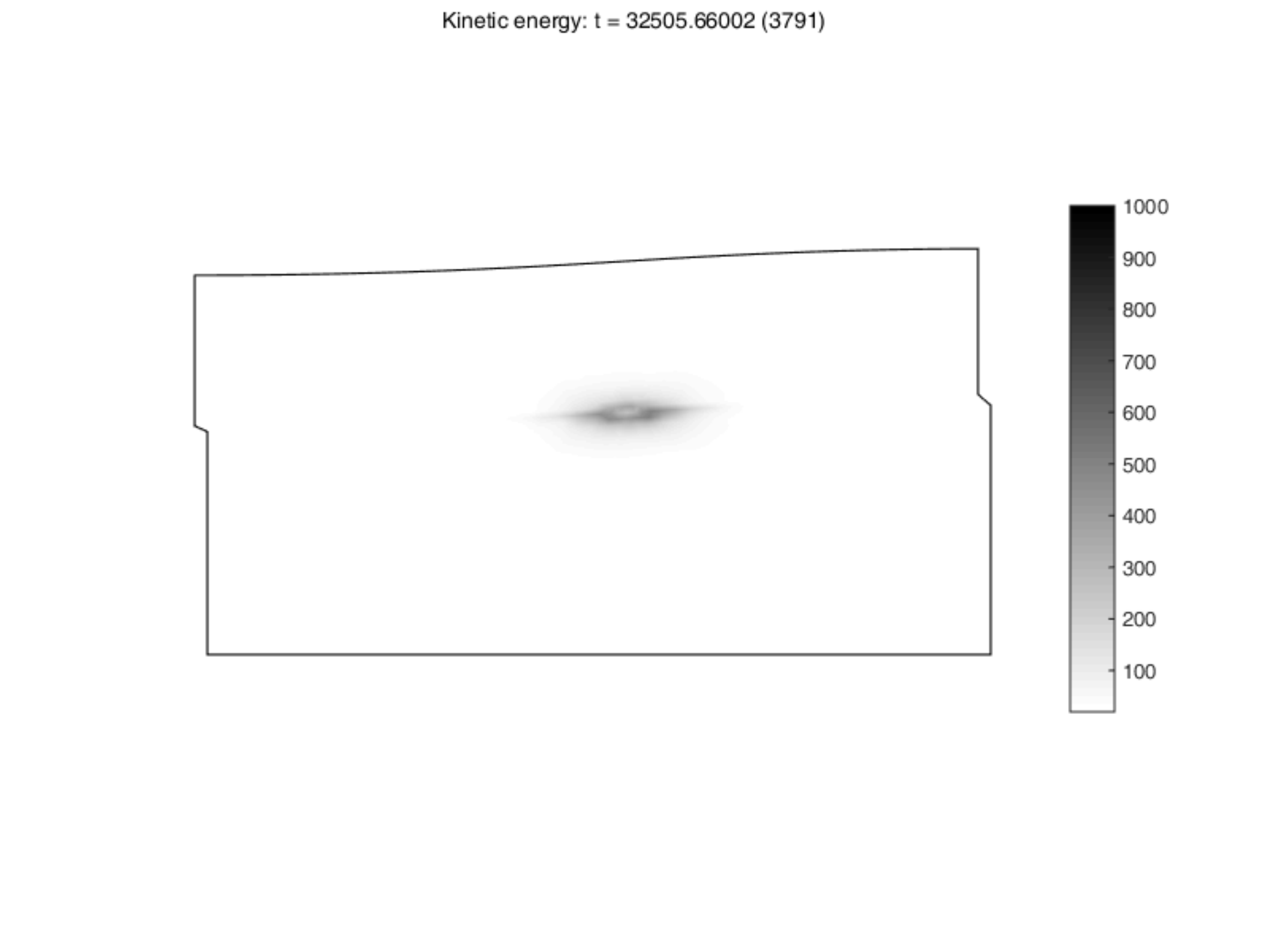}%\hspace*{.5em}
\includegraphics[width=0.30\textwidth,bb=100 170 640 450,clip=true]{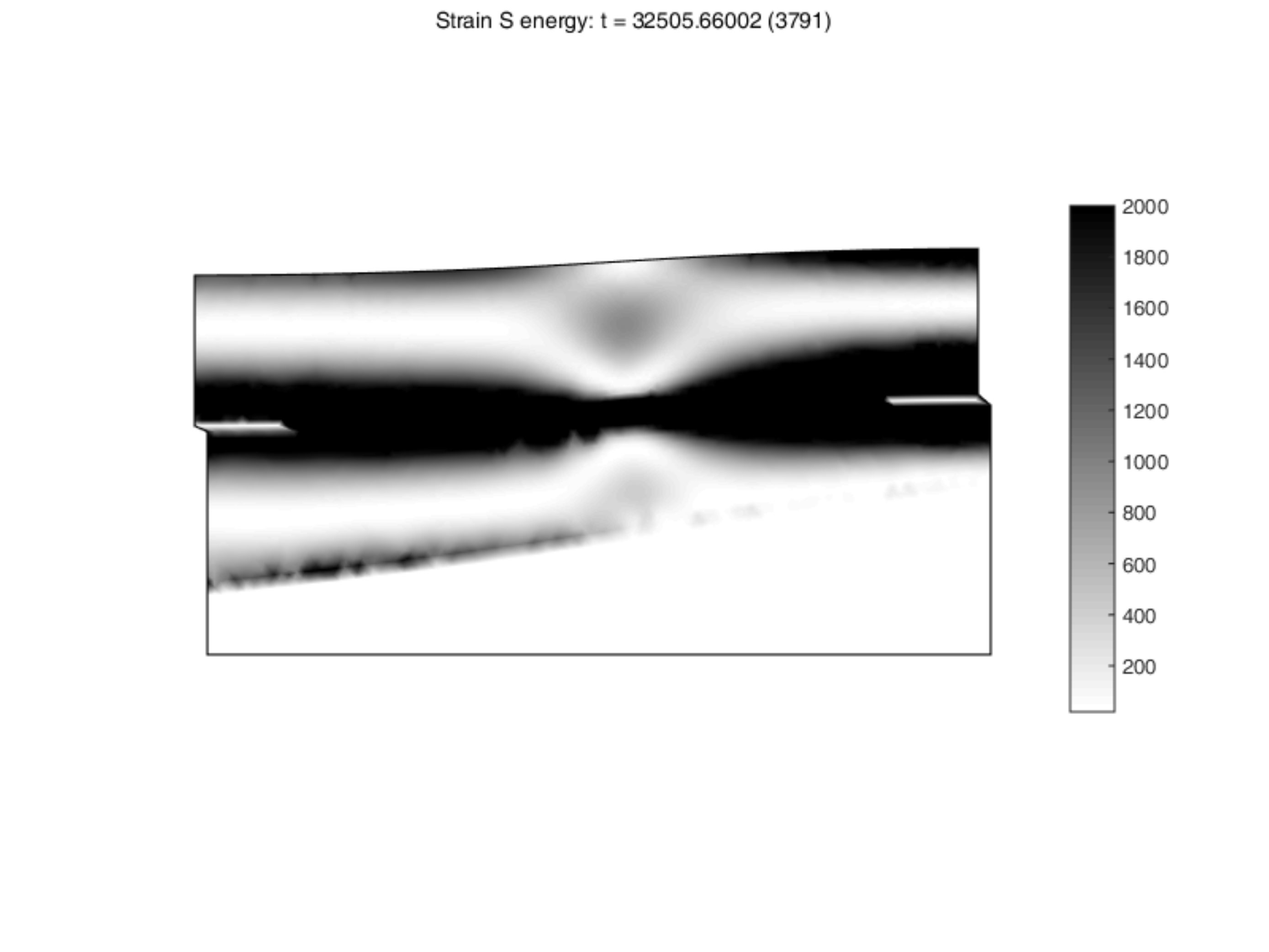}%\hspace*{.5em}
\includegraphics[width=0.30\textwidth,bb=100 170 640 450,clip=true]{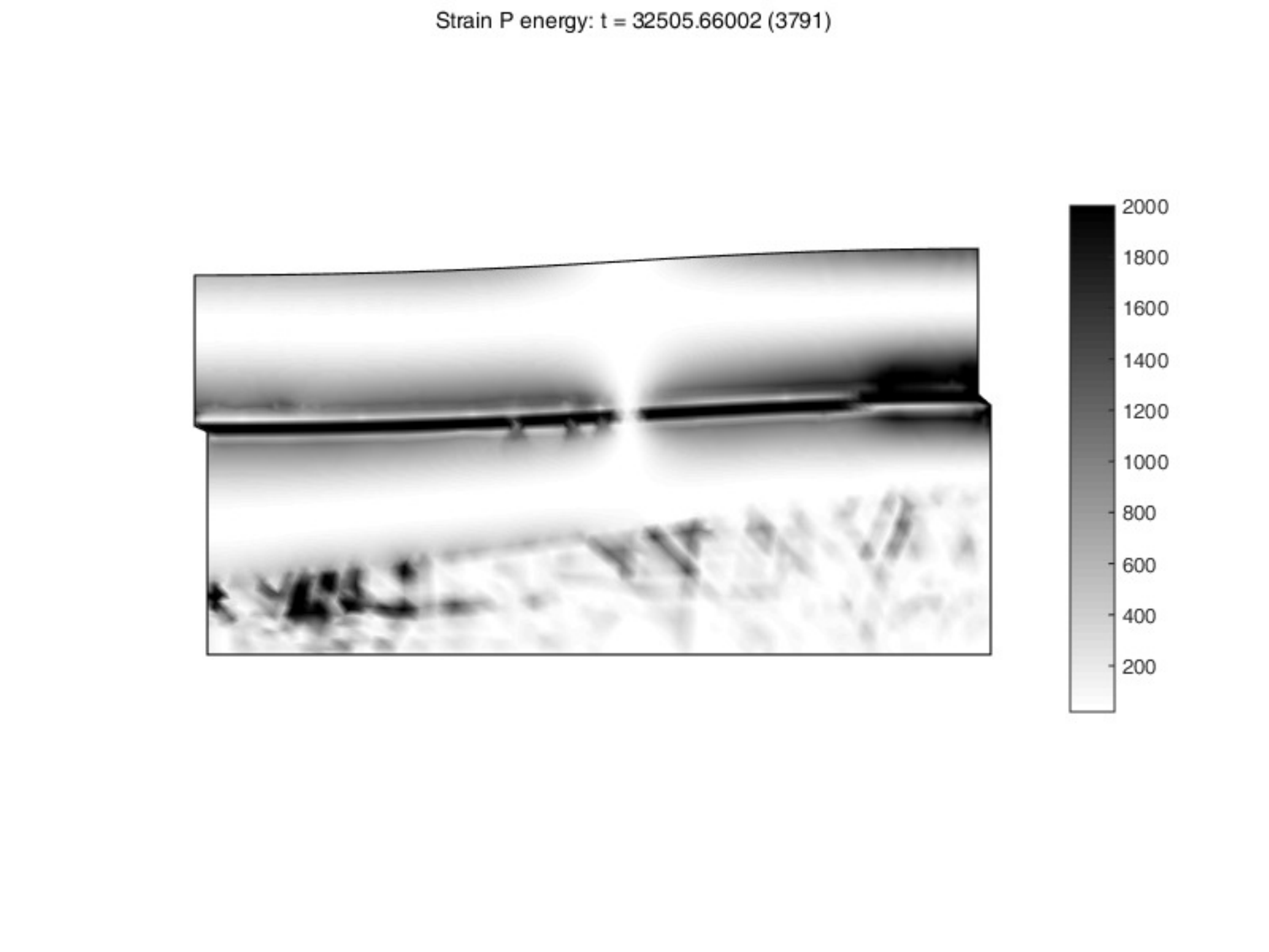}
\\
\rotatebox[origin=lt]{90}{\parbox{2cm}{\centering\COL{$t_2=t_1+1.46$\,s}}}\hspace*{-.2em}
\includegraphics[width=0.30\textwidth,bb=100 170 640 450,clip=true]{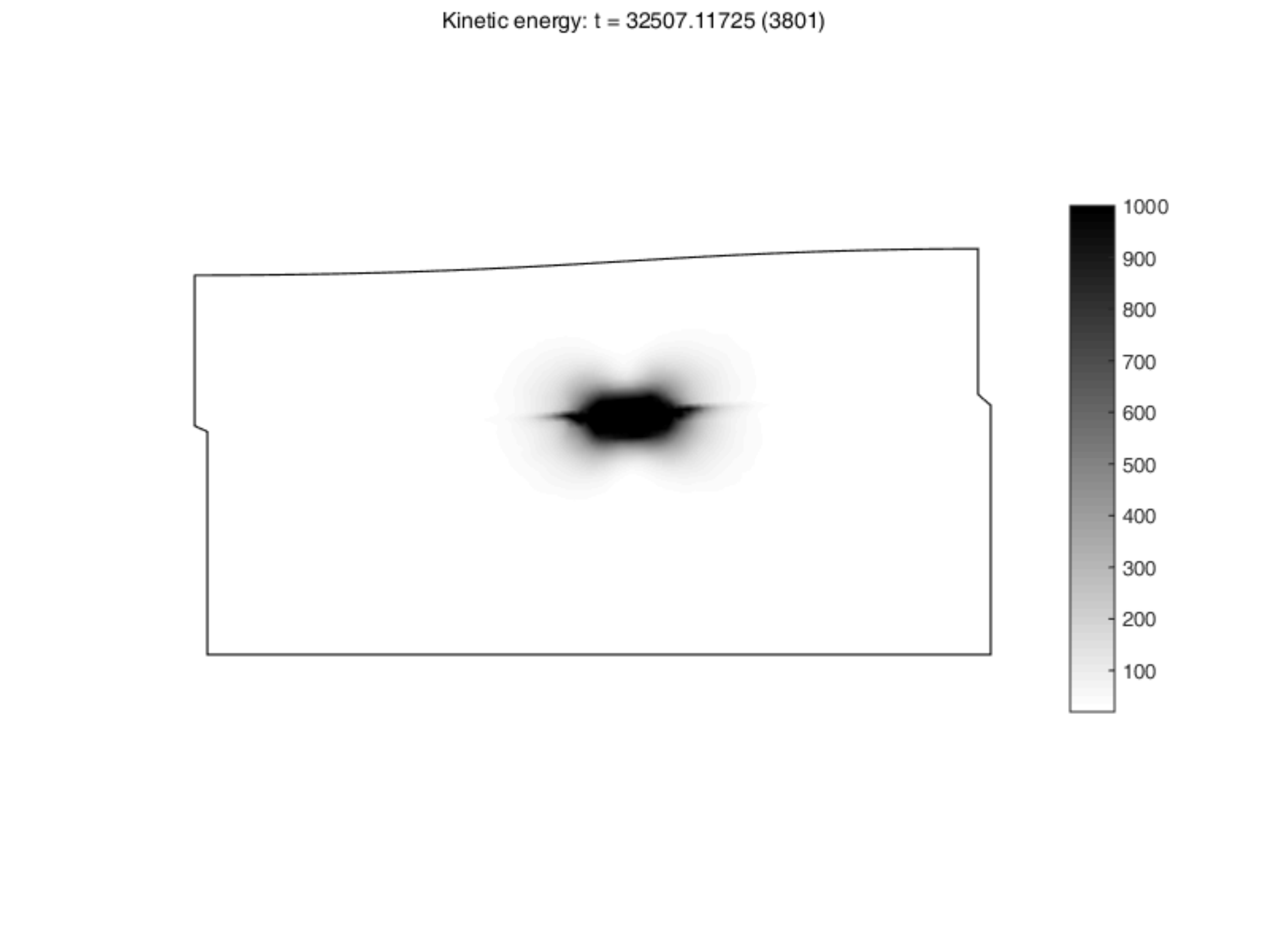}%\hspace*{.5em}
\includegraphics[width=0.30\textwidth,bb=100 170 640 450,clip=true]{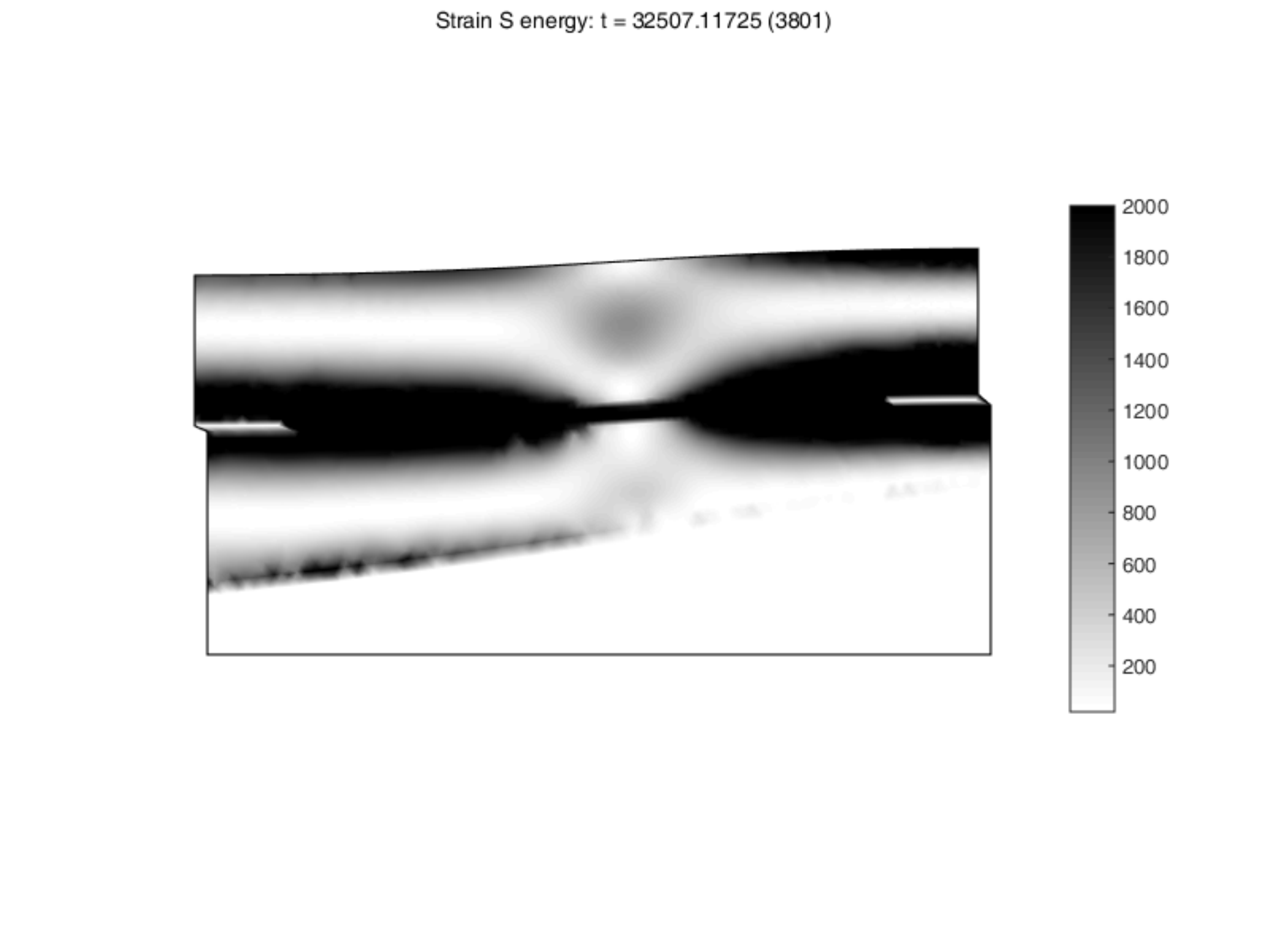}%\hspace*{.5em}
\includegraphics[width=0.30\textwidth,bb=100 170 640 450,clip=true]{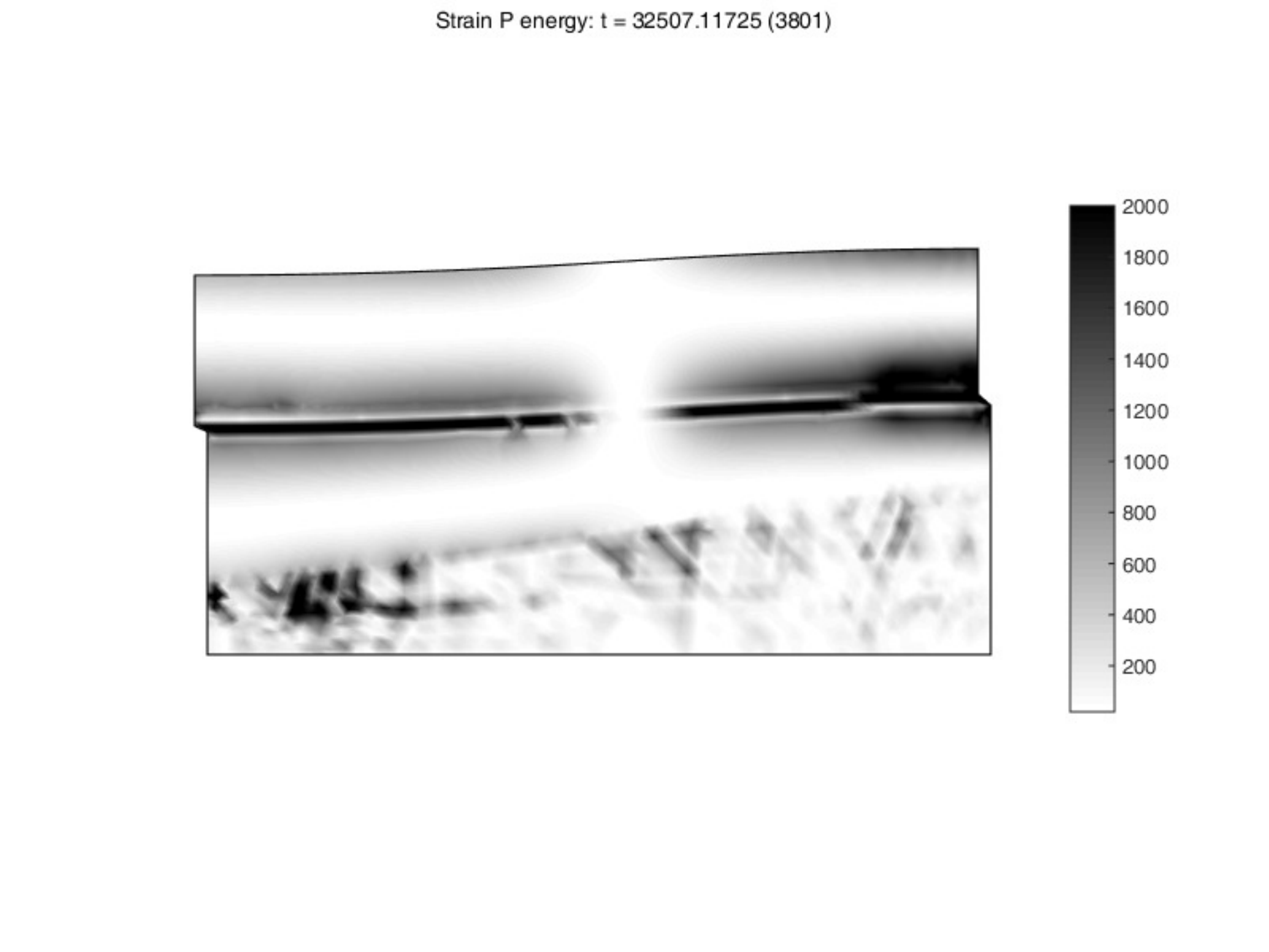}
\\
\rotatebox[origin=lt]{90}{\parbox{2cm}{\centering\COL{$t_3=t_1+2.92$\,s}}}\hspace*{-.2em}
\includegraphics[width=0.30\textwidth,bb=100 170 640 450,clip=true]{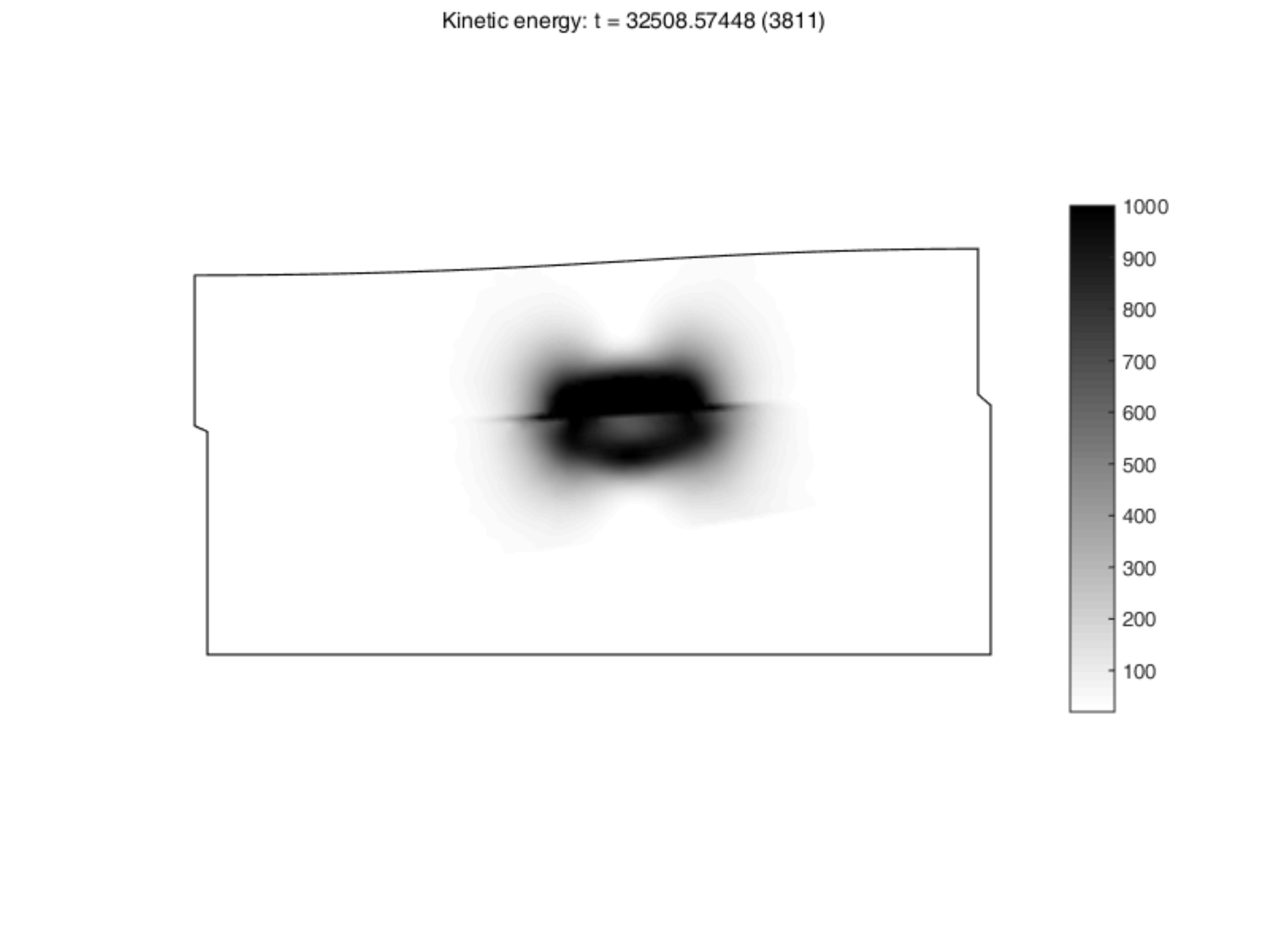}%\hspace*{.5em}
\includegraphics[width=0.30\textwidth,bb=100 170 640 450,clip=true]{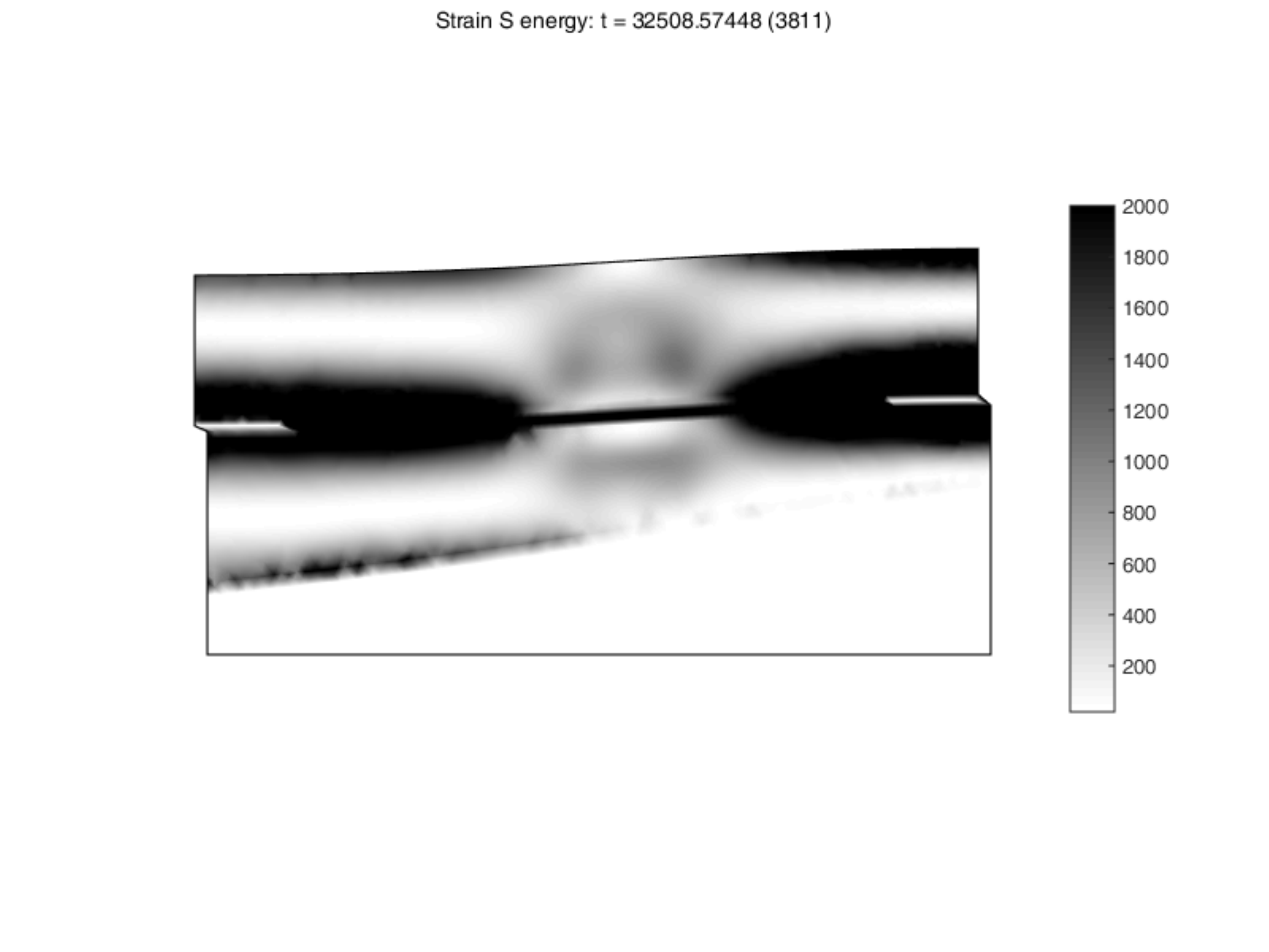}%\hspace*{.5em}
\includegraphics[width=0.30\textwidth,bb=100 170 640 450,clip=true]{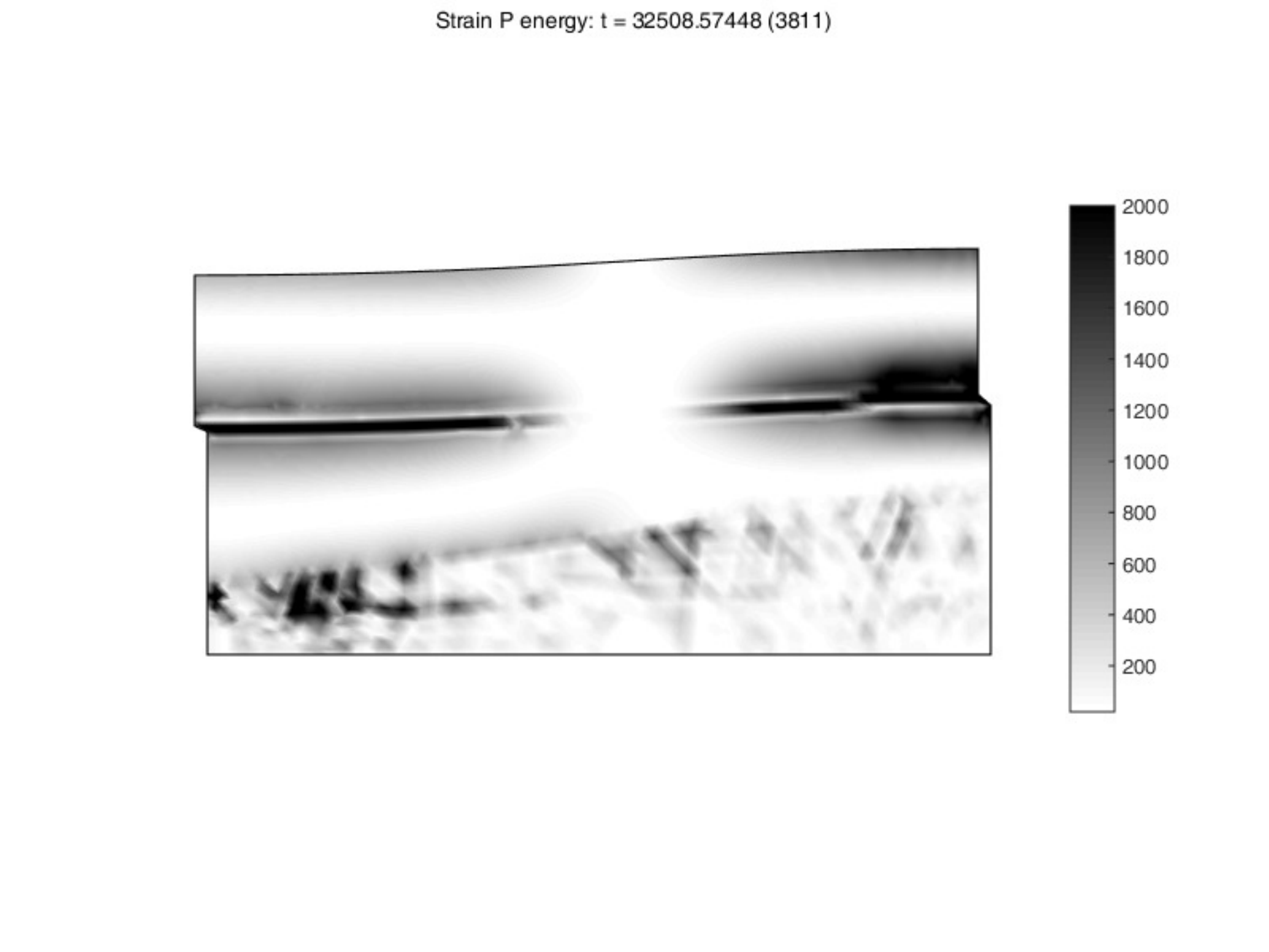}
\\
\rotatebox[origin=lt]{90}{\parbox{2cm}{\centering\COL{$t_4=t_1+4.38$\,s}}}\hspace*{-.2em}
\includegraphics[width=0.30\textwidth,bb=100 170 640 450,clip=true]{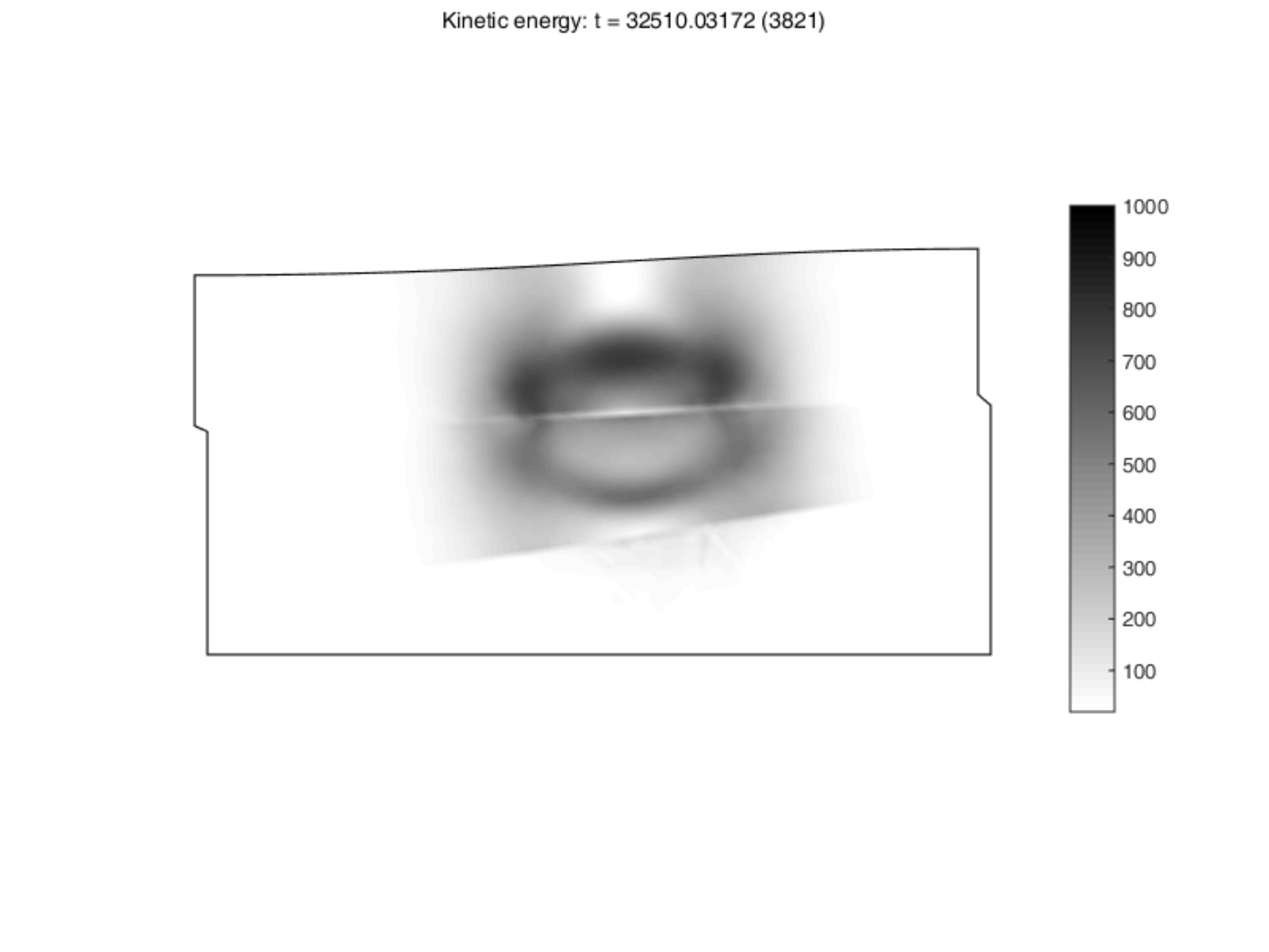}%\hspace*{.5em}
\includegraphics[width=0.30\textwidth,bb=100 170 640 450,clip=true]{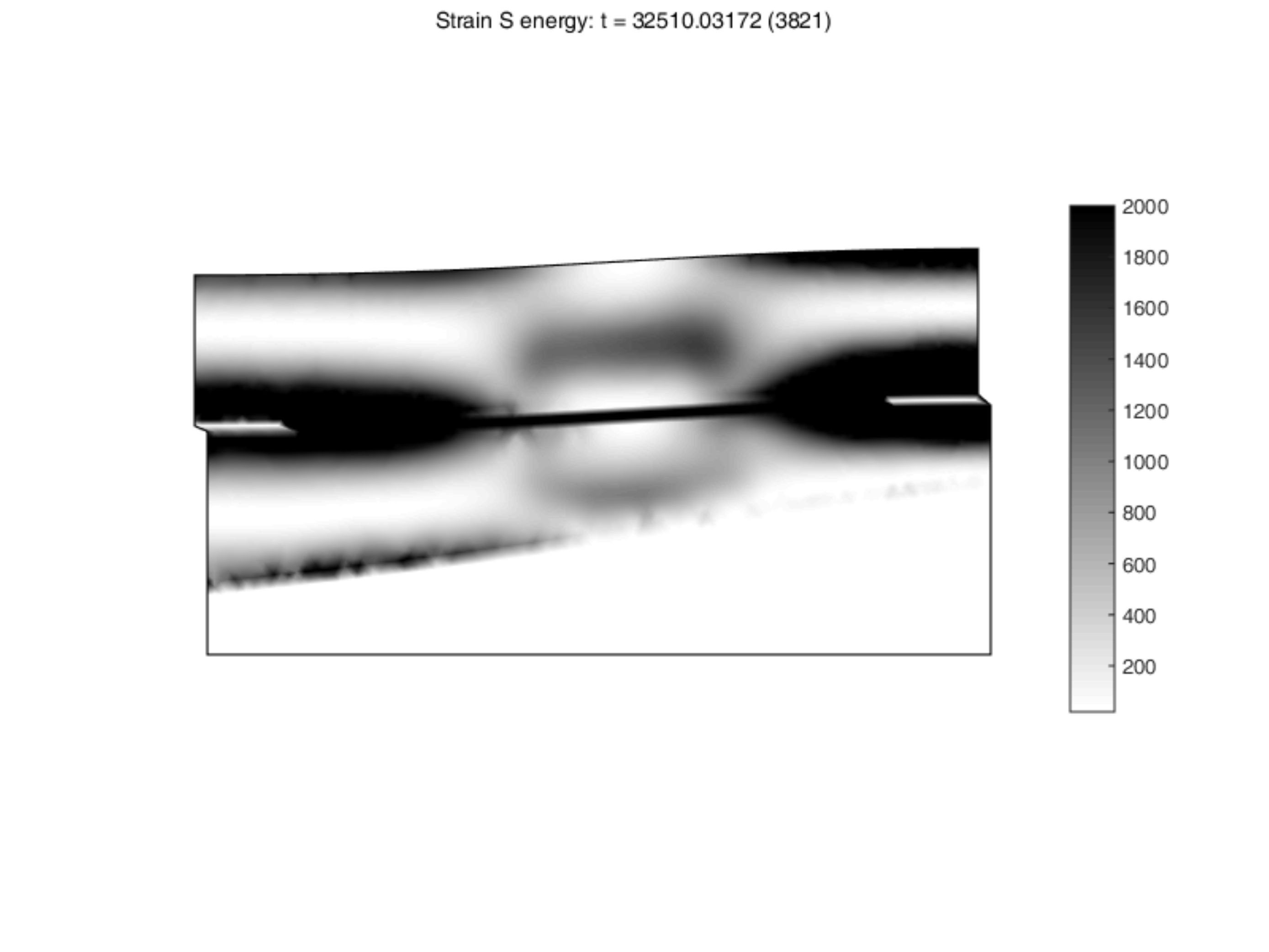}%\hspace*{.5em}
\includegraphics[width=0.30\textwidth,bb=100 170 640 450,clip=true]{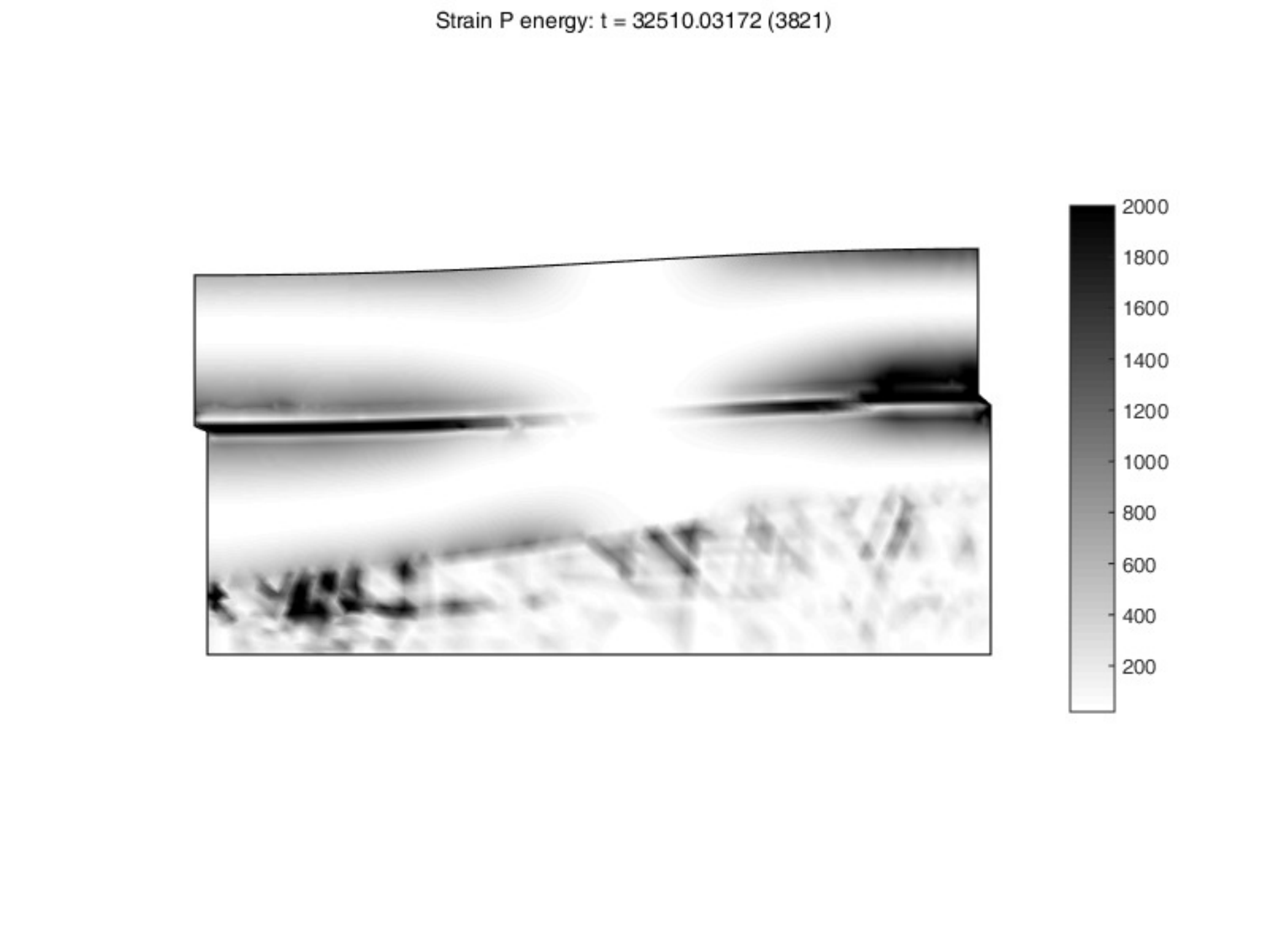}
\\
\rotatebox[origin=lt]{90}{\parbox{2cm}{\centering\COL{$t_5=t_1+5.84$\,s}}}\hspace*{-.2em}
\includegraphics[width=0.30\textwidth,bb=100 170 640 450,clip=true]{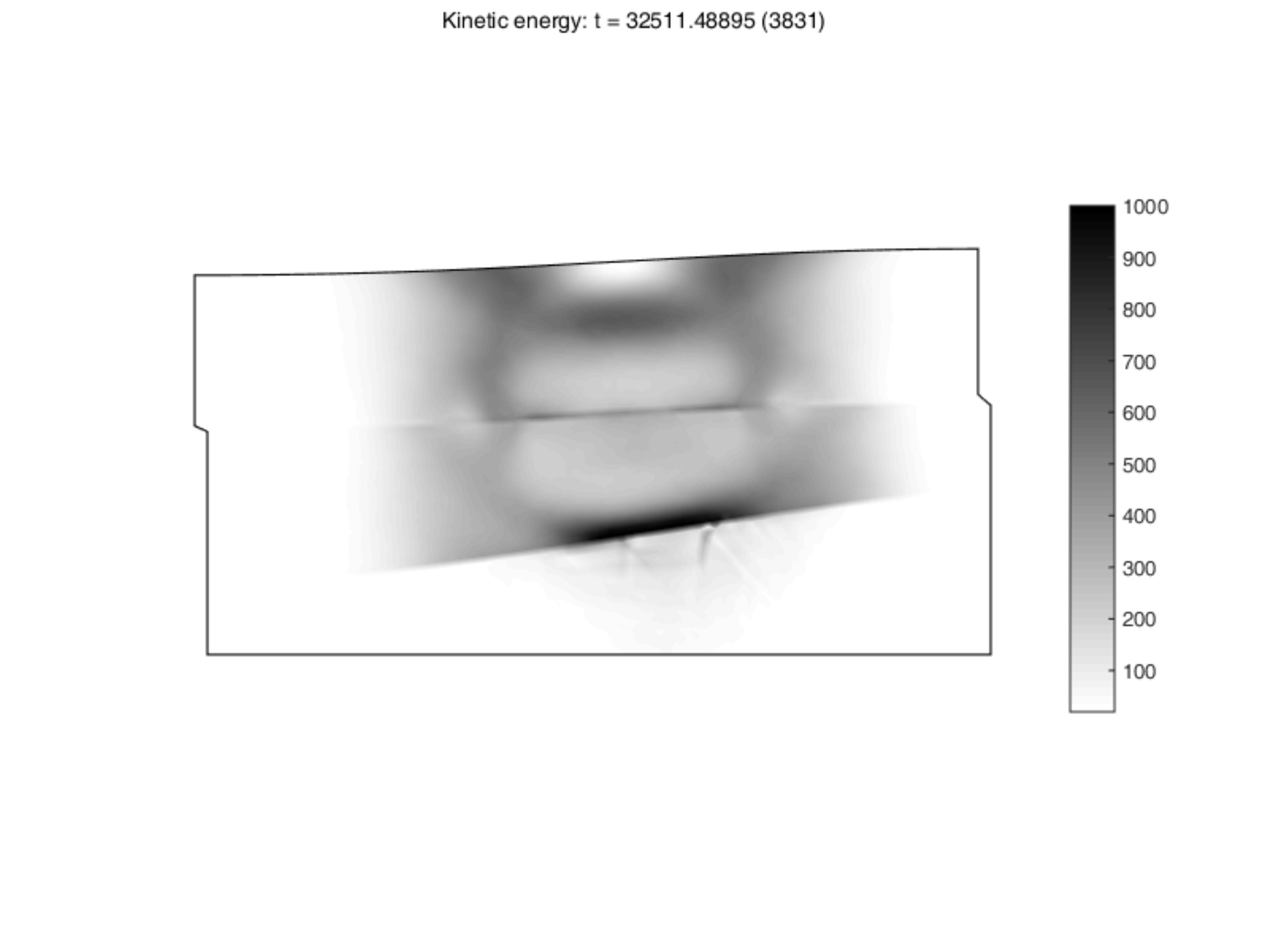}%\hspace*{.5em}
\includegraphics[width=0.30\textwidth,bb=100 170 640 450,clip=true]{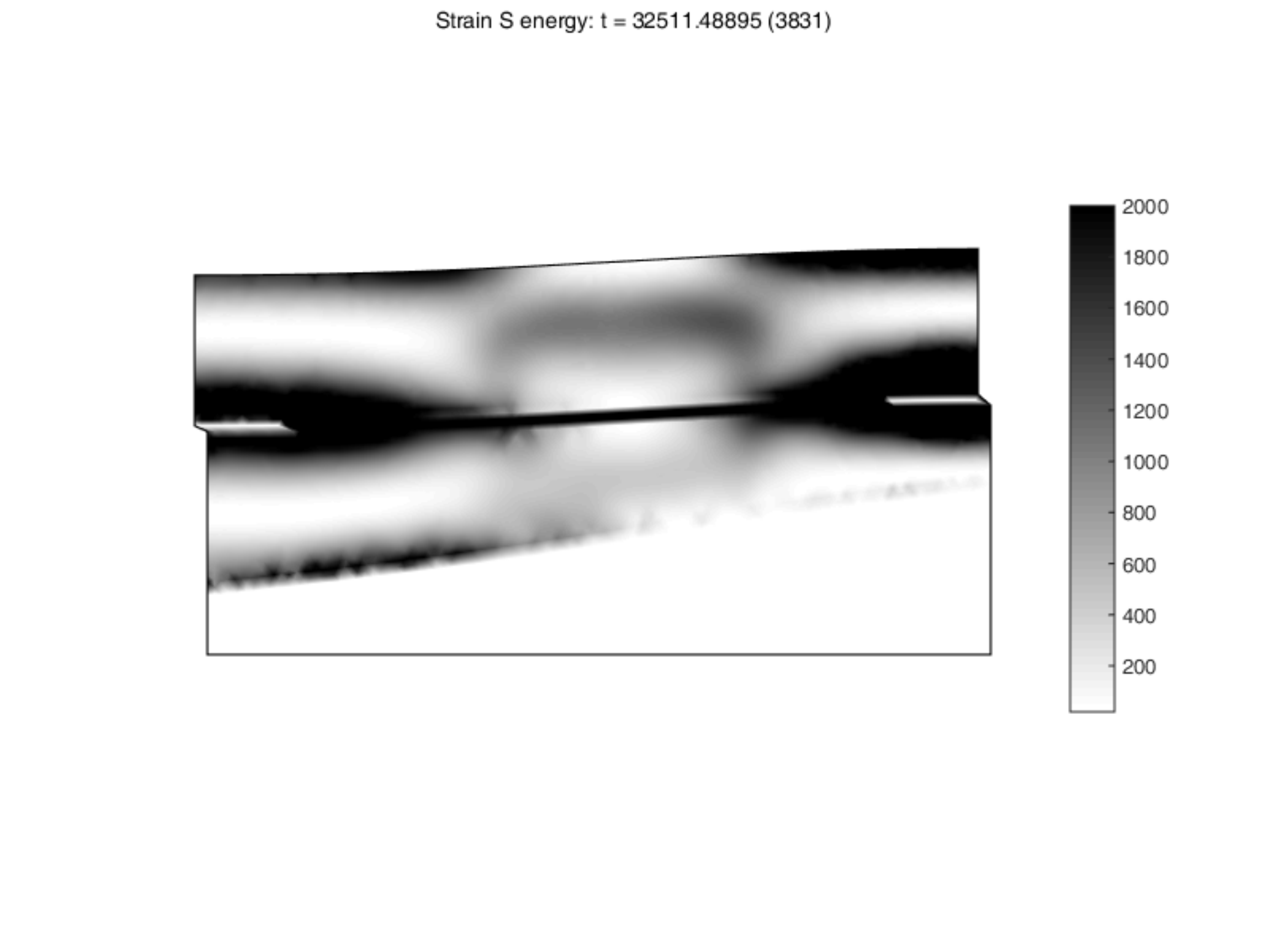}%\hspace*{.5em}
\includegraphics[width=0.30\textwidth,bb=100 170 640 450,clip=true]{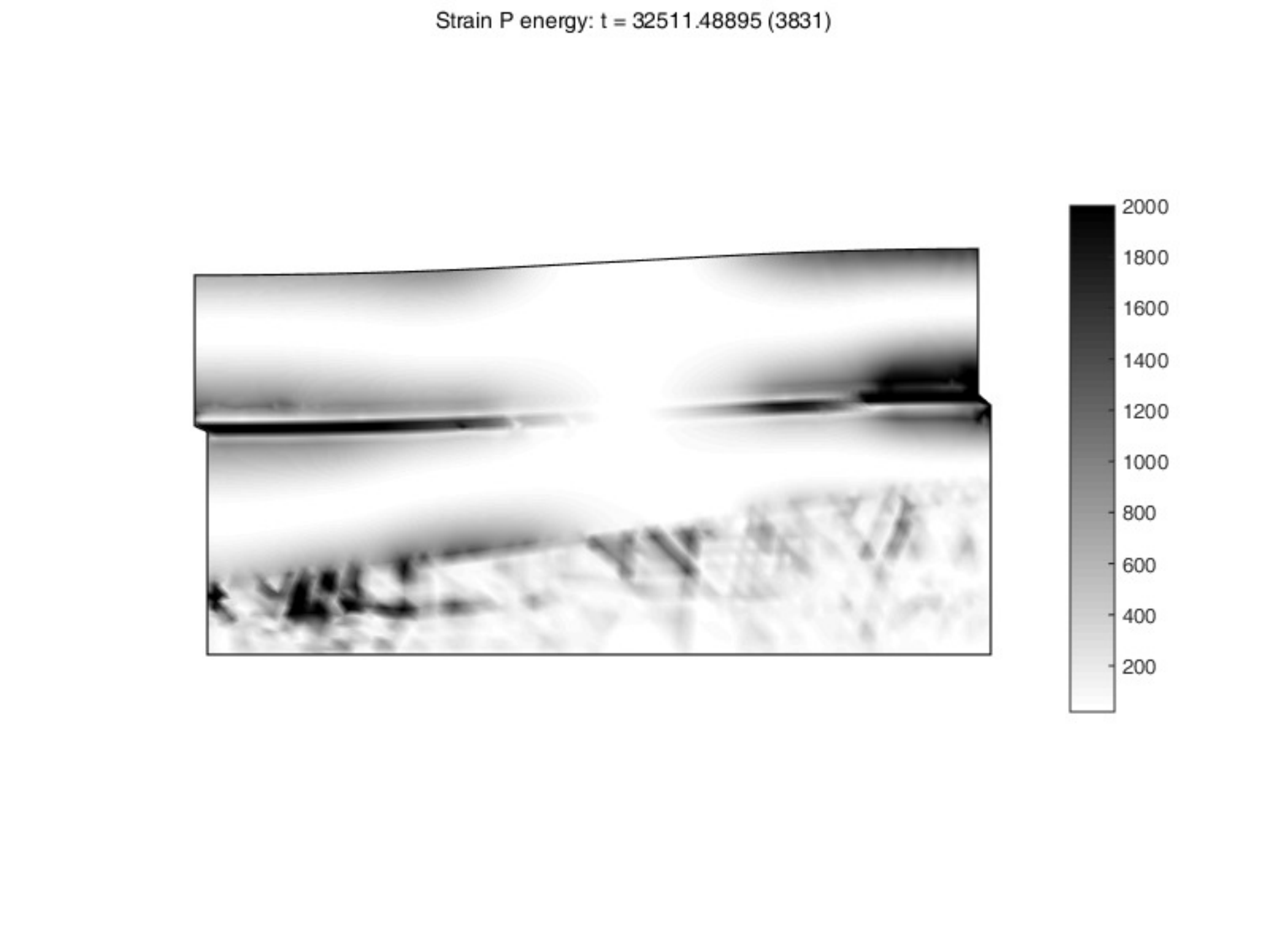}
\\
\rotatebox[origin=lt]{90}{\parbox{2cm}{\centering\COL{$t_6=t_1+7.30$\,s}}}\hspace*{-.2em}
\includegraphics[width=0.30\textwidth,bb=100 170 640 450,clip=true]{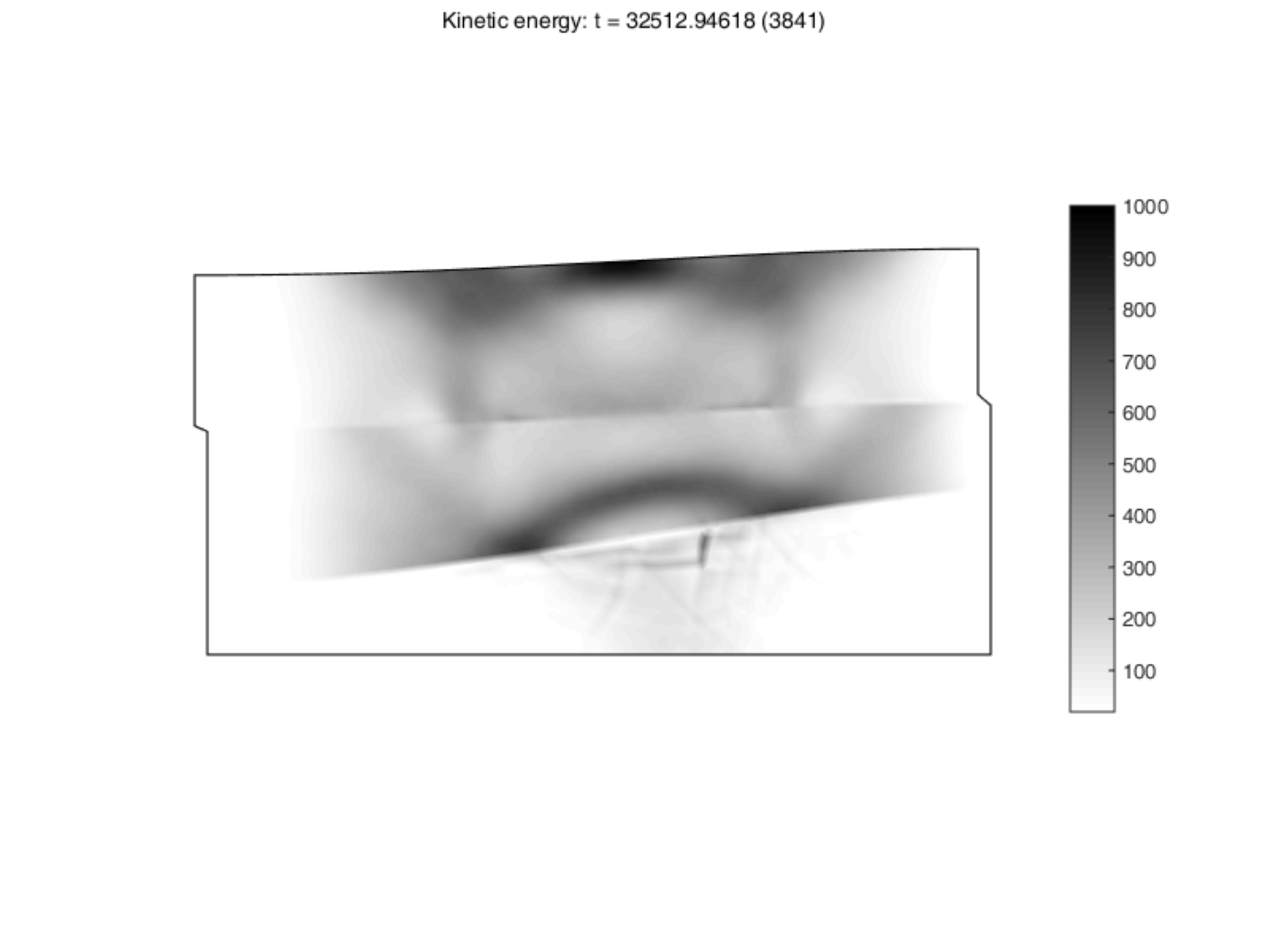}%\hspace*{.5em}
\includegraphics[width=0.30\textwidth,bb=100 170 640 450,clip=true]{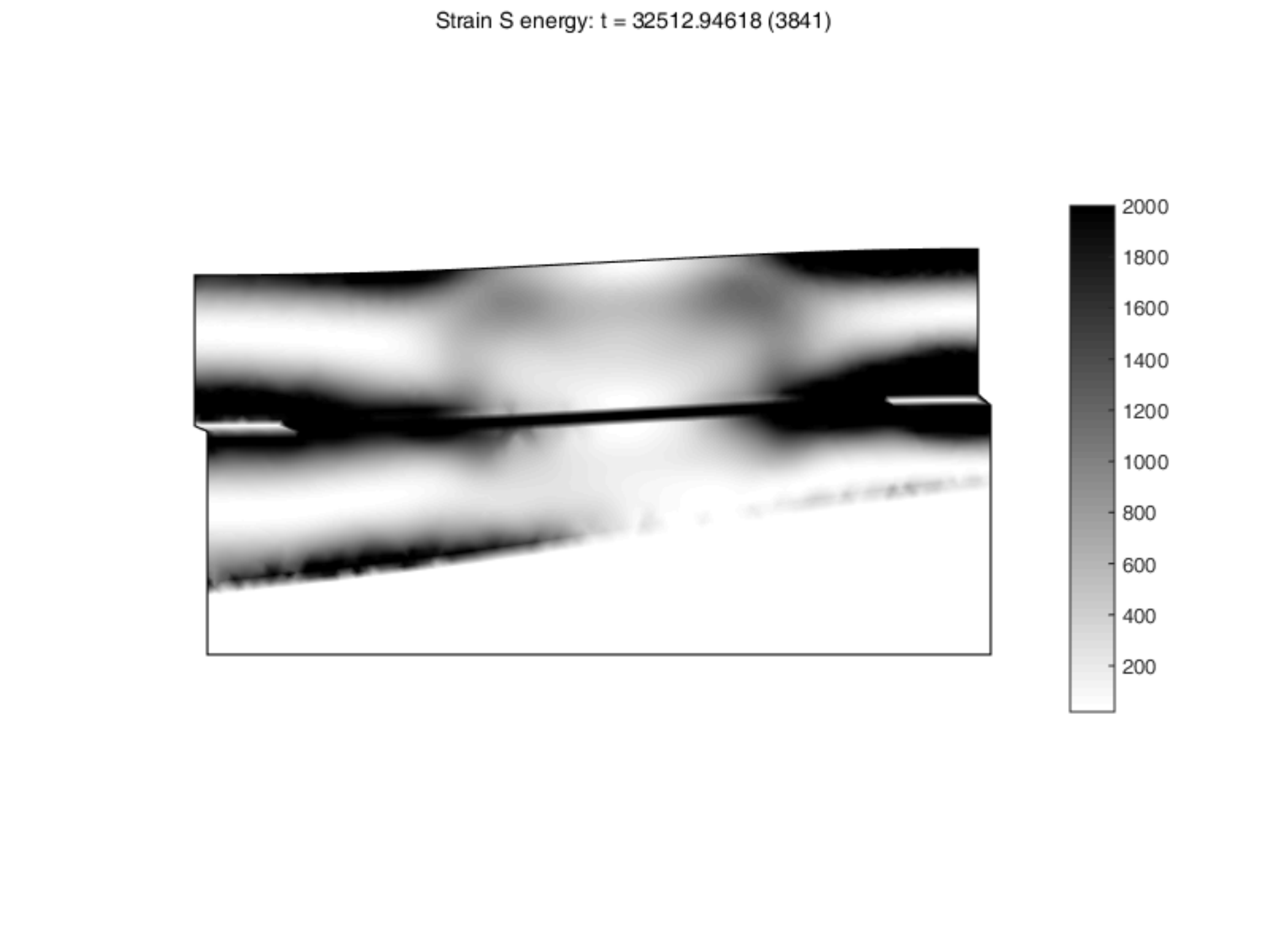}%\hspace*{.5em}
\includegraphics[width=0.30\textwidth,bb=100 170 640 450,clip=true]{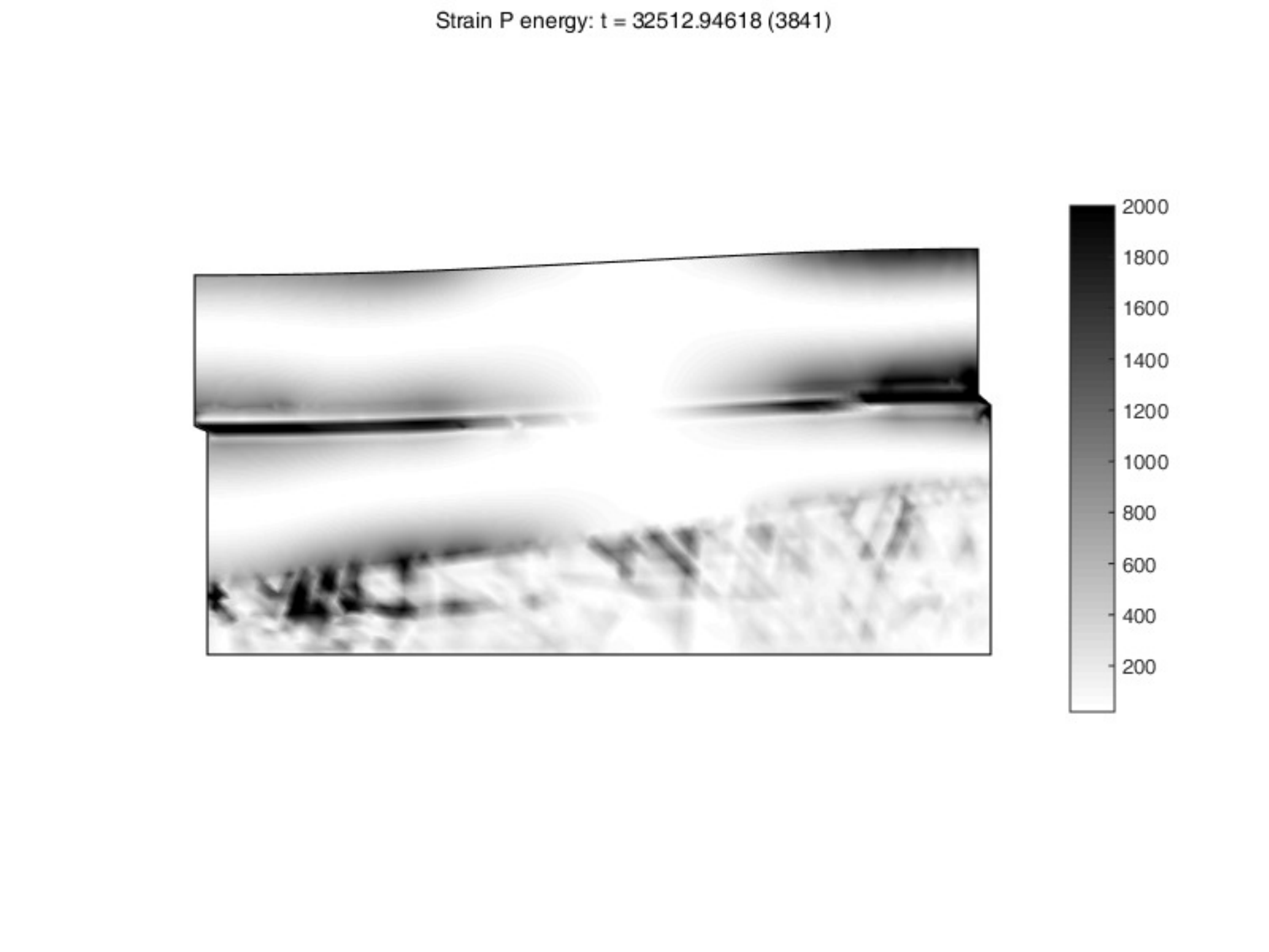}
\\
\rotatebox[origin=lt]{90}{\parbox{2cm}{\centering\COL{$t_7=t_1+8.76$\,s}}}\hspace*{-.2em}
\includegraphics[width=0.30\textwidth,bb=100 170 640 450,clip=true]{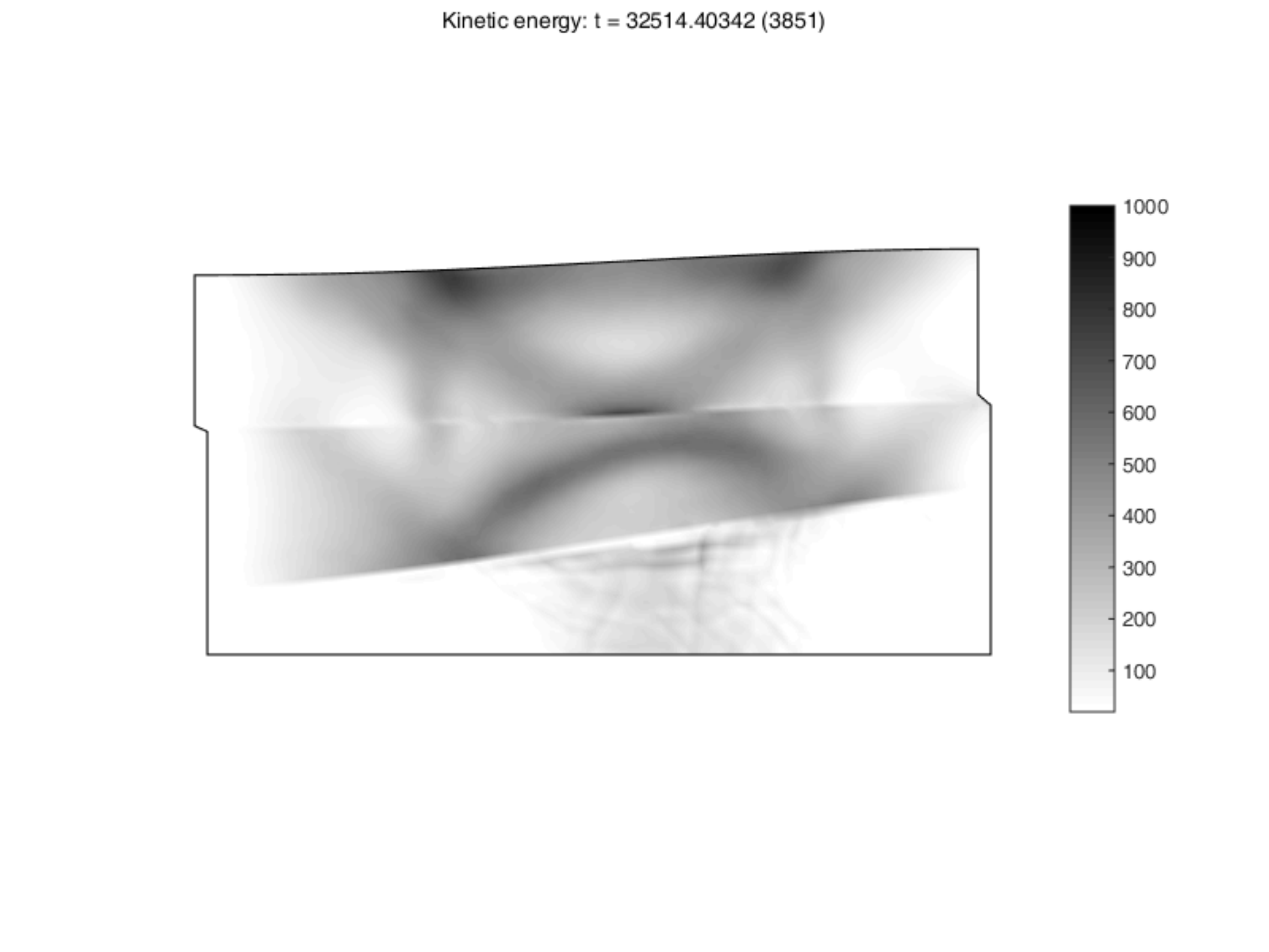}%\hspace*{.5em}
\includegraphics[width=0.30\textwidth,bb=100 170 640 450,clip=true]{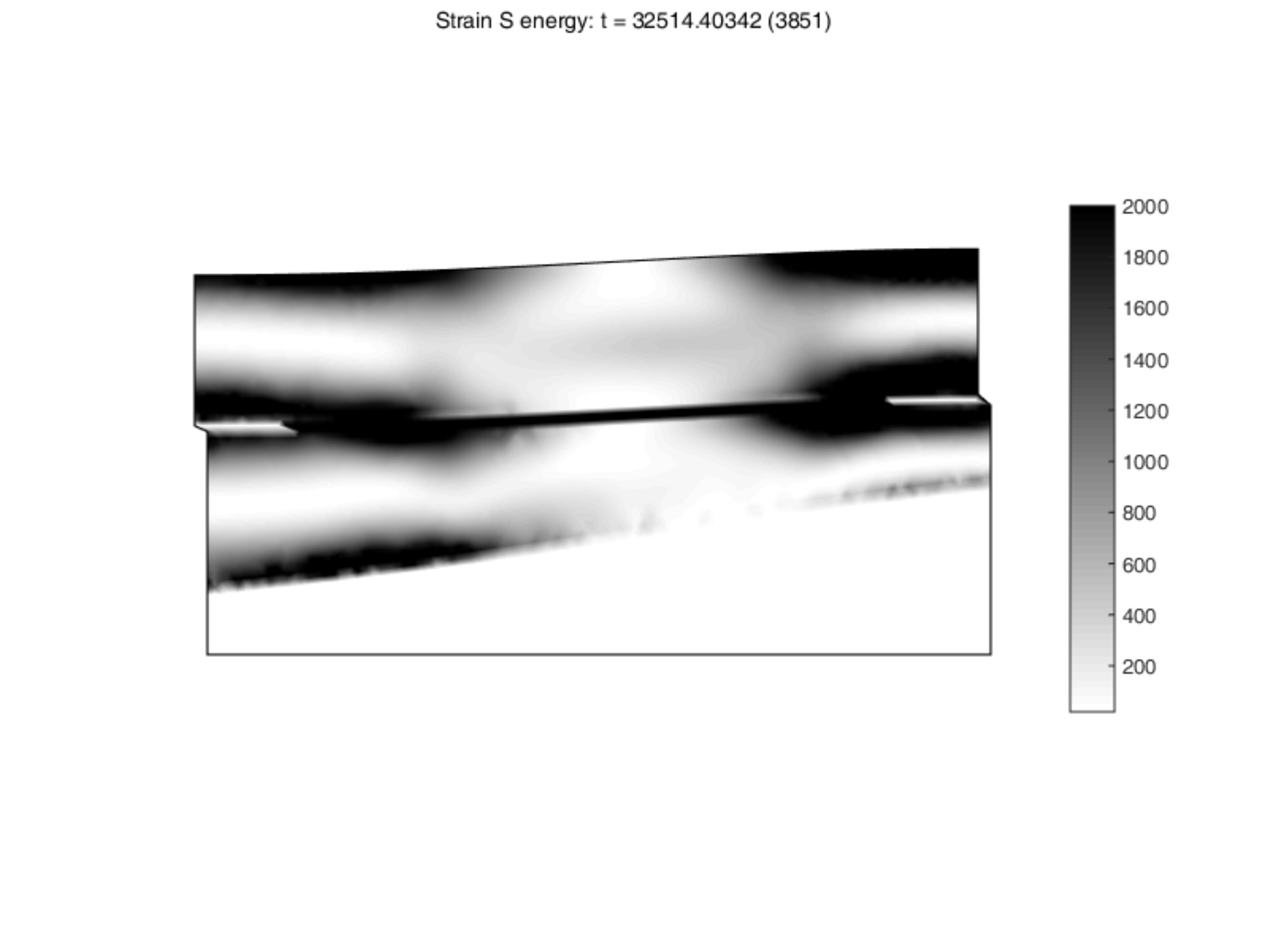}%\hspace*{.5em}
\includegraphics[width=0.30\textwidth,bb=100 170 640 450,clip=true]{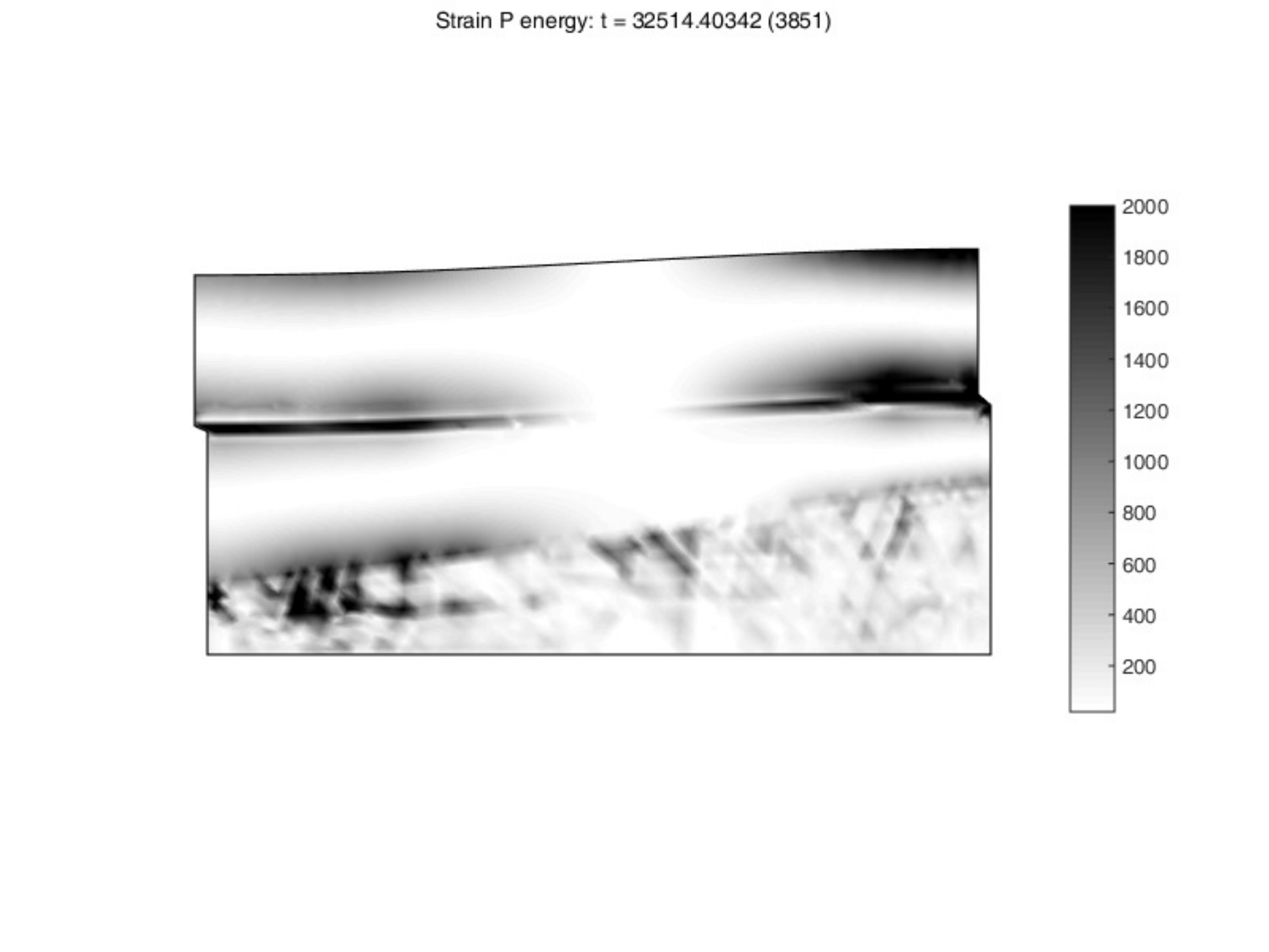}
\\
\rotatebox[origin=lt]{90}{\parbox{2cm}{\centering\COL{$t_8=t_1+10.22$\,s}}}\hspace*{-.2em}
\includegraphics[width=0.30\textwidth,bb=100 170 640 450,clip=true]{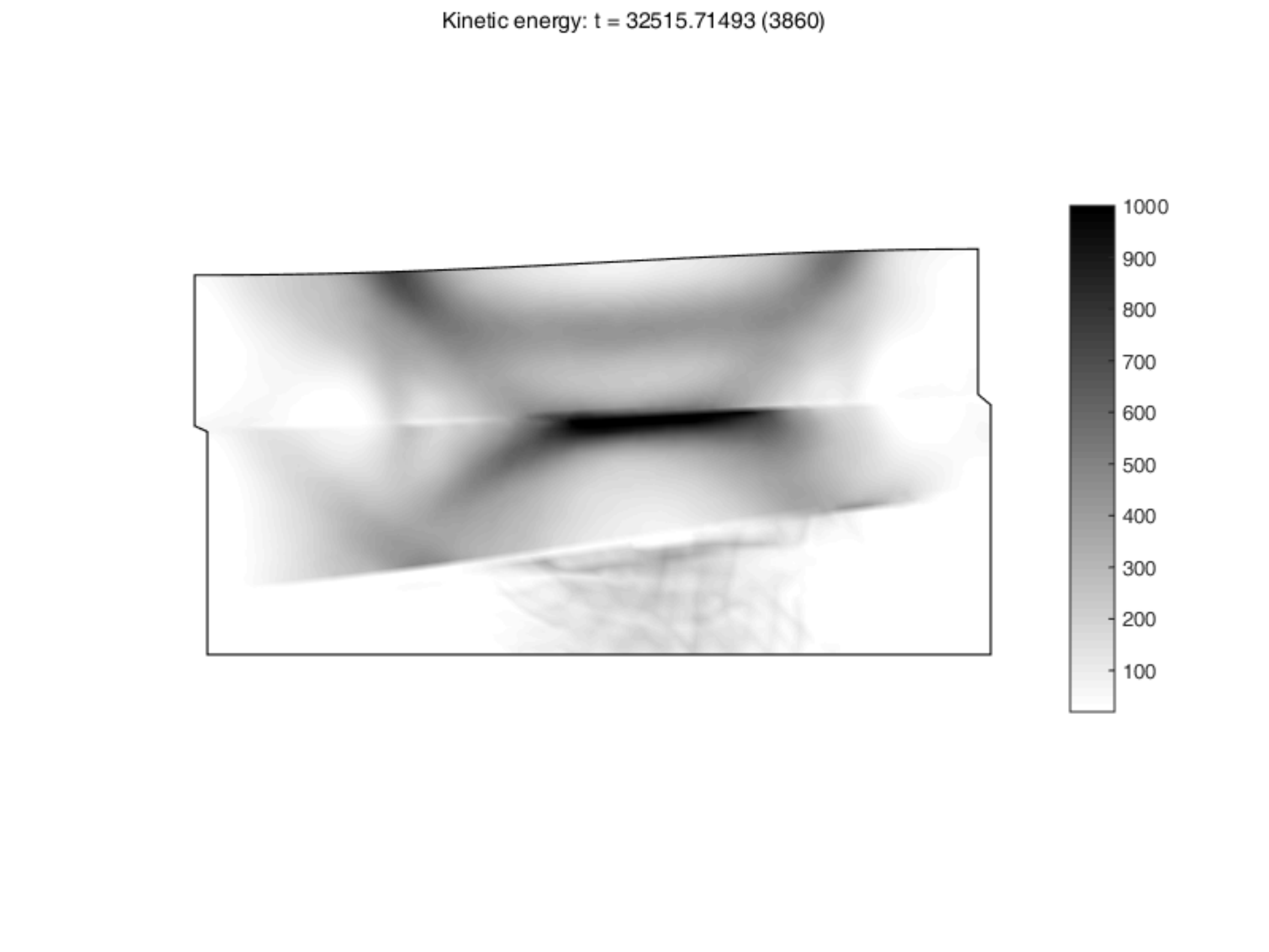}%\hspace*{.5em}
\includegraphics[width=0.30\textwidth,bb=100 170 640 450,clip=true]{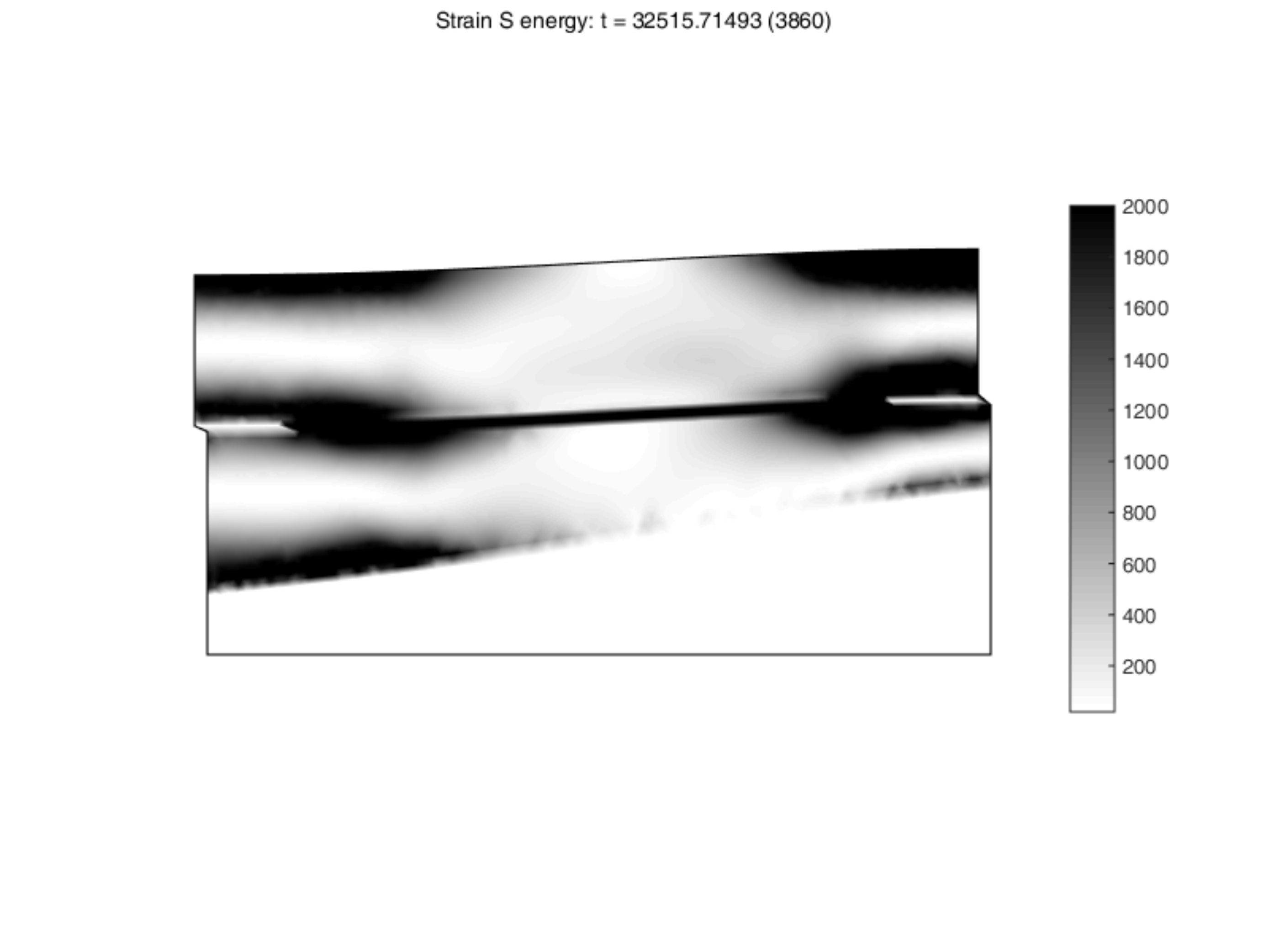}%\hspace*{.5em}
\includegraphics[width=0.30\textwidth,bb=100 170 640 450,clip=true]{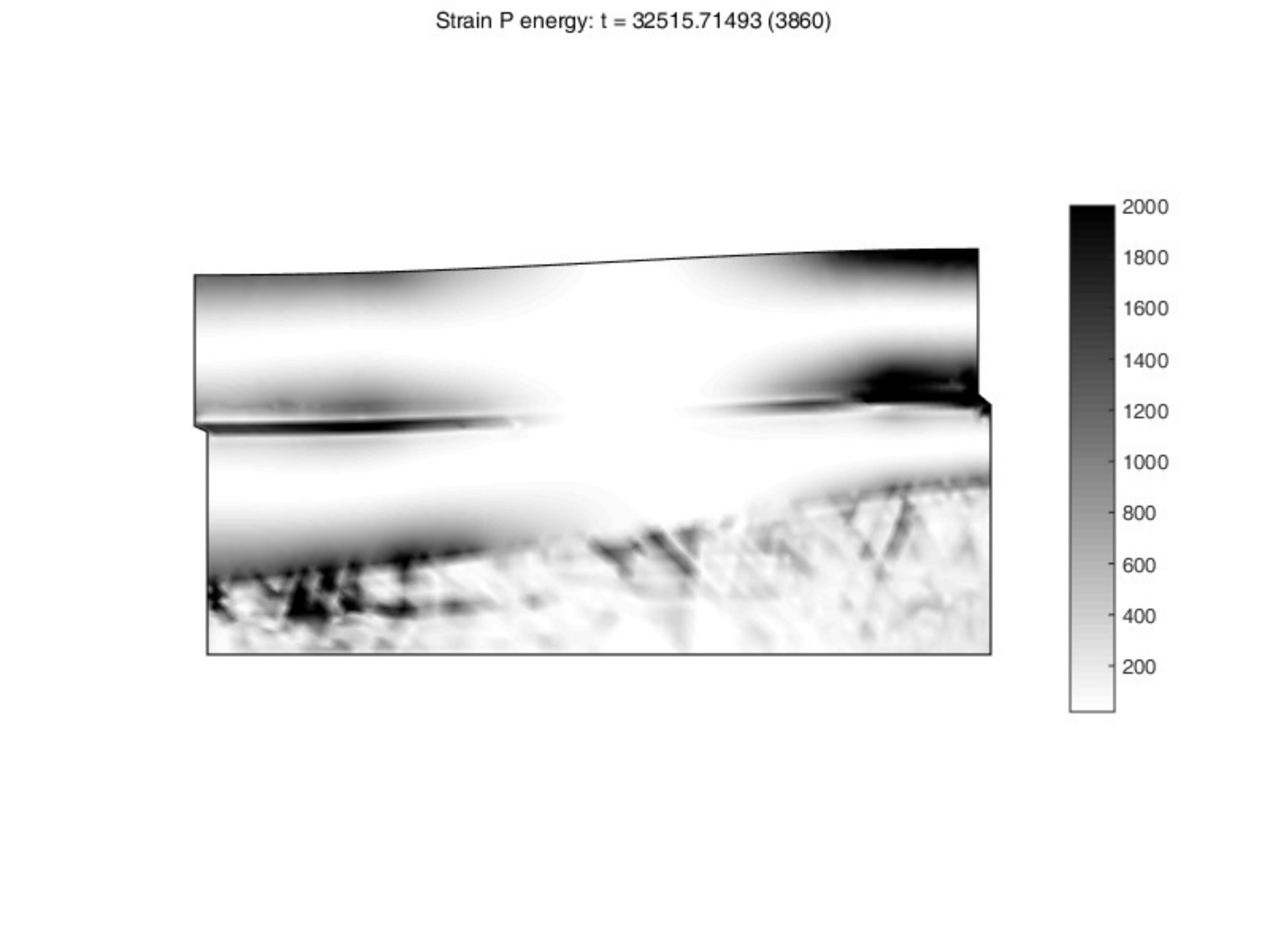}
\caption{\sl Simulations of a rupture of a fault in the middle part, emission
of a seismic (mainly S-) wave in the hypocentre, its propagation and
reflection on the solid/fluid interface where a weak P-wave can be observed
in the fluid part, and eventually S-wave reaches also the surface (in an
so-called epicentre) where the earthquake manifests. The displacement is
magnified 100$\times$ to visualize the deformation.}
\label{fig-EQ}
\end{figure*}

%.....HERE DESCRIPTION OF THE RESULTS ......  
The results of this example are depicted in Figure~\ref{fig-EQ} in eight 
selected snapshots \bbb after rupture triggering, showing the \eee
 spatial distribution of the kinetic and the stored 
energy in its decomposition to the shear and the spherical parts.
\bbb
The instants are expressed only by increments with respect to the first
one $t_1$ due to the large time scale of the whole calculation. The actual
number of numerical time steps can be computed by doing the increment
by $\tau_k=0.146$ s, which means e.g.\ 10 time steps between $t_1$ and $t_2$. \eee
The 
interesting moment is when the upper plate is moved sufficiently far towards 
left and thus the fault is stretched enough so that the rupture occurs in 
the middle; i.e.\ the earthquake starts in the hypocentre.
In fact, the fracture toughness $G_{\rm c}$ is put higher close to the boundary 
$\Gamma$ to prevent rupture starting from the boundary where there is 
necessarily a stress concentration due to the fast varying boundary
conditions. 

\begin{figure*}
\newlength{\aux}
\settoheight{\aux}{\includegraphics[width=0.32\textwidth,bb=190 070 1000 720,clip=true]{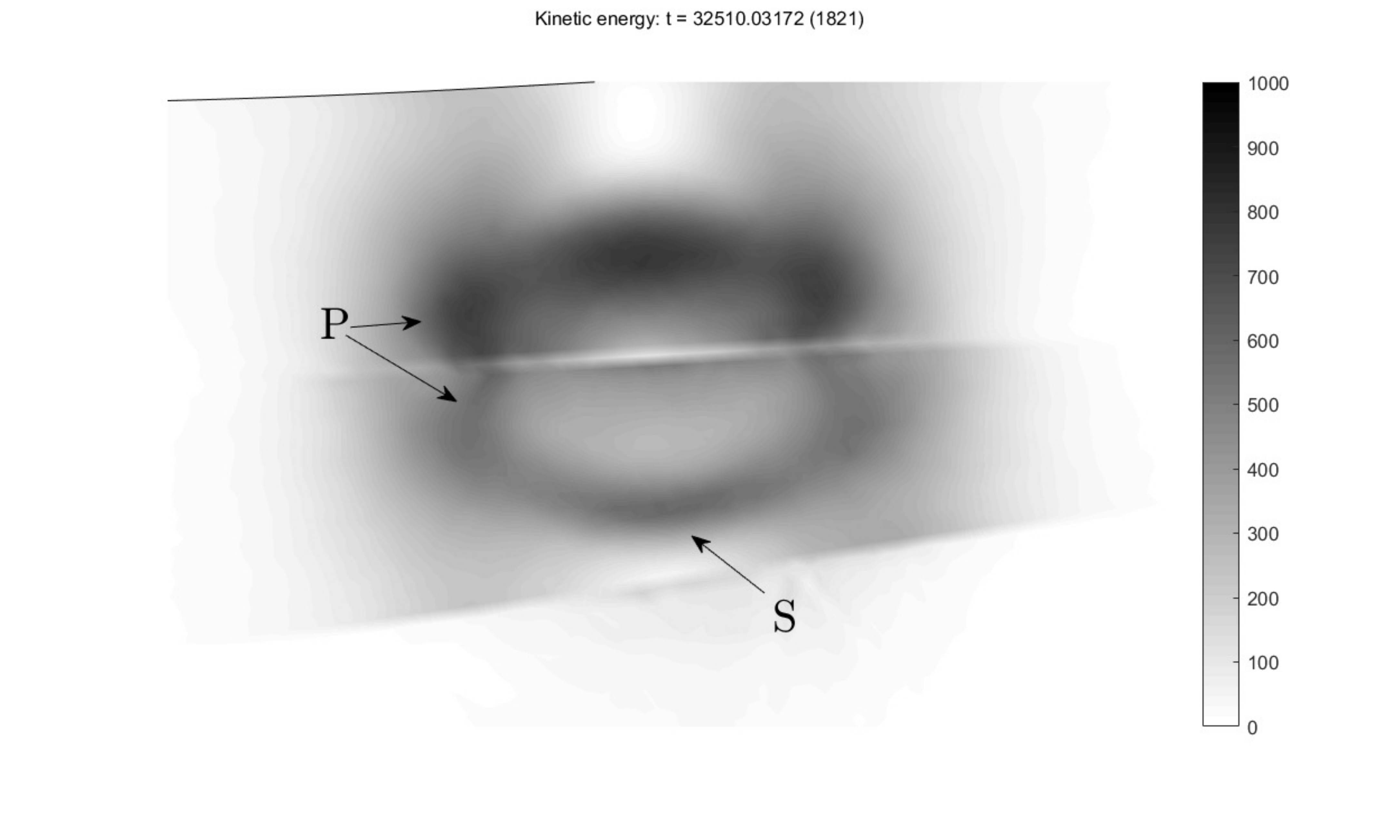}}
\centering
\raisebox{0.9\aux}{$t_4$}\hspace*{.1em}
\includegraphics[width=0.32\textwidth,bb=190 070 1000 720,clip=true]{EK_Wave1.pdf}%
\hspace*{5em}
\raisebox{0.9\aux}{$t_6$}\hspace*{.1em}
\includegraphics[width=0.32\textwidth,bb=190 070 1000 720,clip=true]{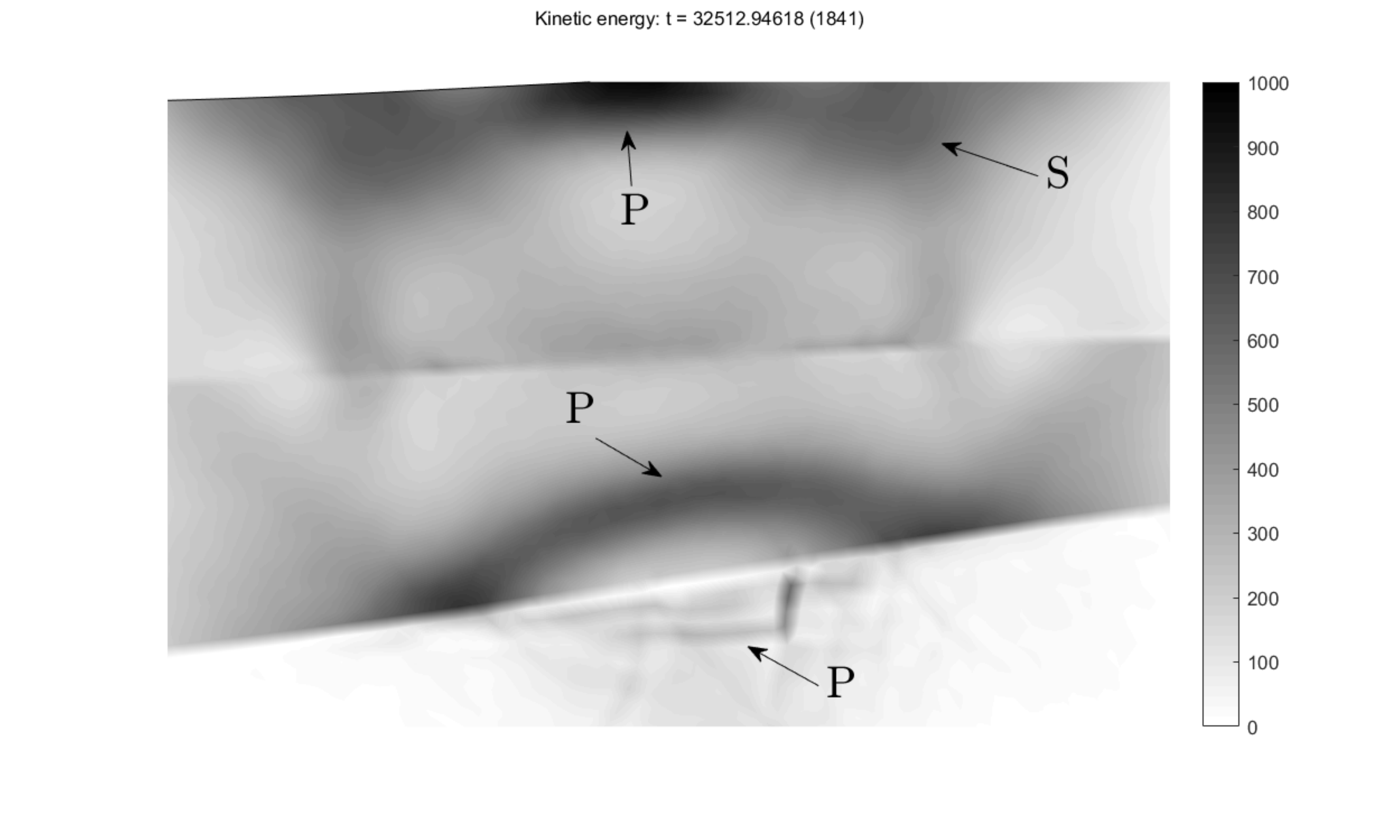}%
\caption{\COL{\sl Various types of seismic waves recognizable in the kinetic 
energy drawings for the relevant parts from two selected snapshots from 
Fig.~\ref{fig-EQ}-left. The specification is possible due to the shear energy 
(Fig.~\ref{fig-EQ}-middle) where only S-waves can be visible, while not the 
P-waves.}}
\label{fig-EQ-dtl}
\end{figure*}

%The big stretch causes big stress and eventually a rupture in the middle
%of the fault, i.e.\ the earthquake in the hypocentre. 
At that fault rupture 
moment, the strain energy around is relaxed and a seismic wave is emitted
and starts propagating through the solid part. One can see that this wave is 
not a ball-shaped partly because the seismic source (the rupturing area)
is rather a surface than a point and partly (or mainly) because the seismic 
source by a slip of the fault generates rather an S-wave towards normal 
(i.e.\ here vertical) directions (which are slower), while the horizontal-like 
fronts are rather P-waves (propagating faster). 
\COL{This is also shown in Fig.~\ref{fig-EQ-dtl}-left. The pertinent waves in 
kinetic energies around the hypocentre are shown for the snapshot corresponding 
to the time $t_4$.
%the detail reveals the S-wave which can also be seen for the same time 
%instant in the shear energy in Fig.~\ref{fig-EQ}-middle. 
The P-wave is sufficiently separated from the S-wave.
%Here we advantageously exploit that in computer we have at 
%disposal the strain energy split into the spherical and the shear parts,
%from which one can distinguish these types of waves; in particular, 
%the shear energy in Fig.~\ref{fig-EQ}-middle shows clearly the S-wave.
 } 
%\COL{.....HERE COMMENTS + REF. TO Ben-Zion }
\COL{This is also shown in Fig.~\ref{fig-EQ-dtl}-left. The pertinent waves in 
kinetic energies around the hypocentre are shown for the snapshot corresponding 
to the time $t_4$, 
the detail reveals the S-wave which can also be seen for the same time 
instant in the shear energy in Fig.~\ref{fig-EQ}-middle. The P-wave is well 
separated 
from the S-wave. Here we advantageously exploit that in computer we have at 
disposal the strain energy split into the spherical and the shear parts,
from which one can distinguish these types of waves; in particular, 
the shear energy in Fig.~\ref{fig-EQ}-middle shows clearly the S-wave.
 } 

When the seismic wave reaches the Earth surface, it is reflected and 
partly starts propagating to the sides, i.e.\ the earthquake 
starts at the epicentre. On the other side, when reaching the solid/fluid
interface below the hypocentre, the wave is mainly reflected (which documents
that it is rather an S-wave) and to a small extent it generates a slight P-wave 
in the fluidic part. Some other reflection is seen on the fault itself,
which is actually a well recognized phenomenon related to 
(or serving for identifying of) the width of the low-velocity damaged
fault zone, cf.\ e.g.\ %\cite{BeZi98PSFZ,QBRS17ISSJ}
\cite{BeZi98PSFZ}.
\COL{The detail plot in Fig.~\ref{fig-EQ-dtl}-right specifies the form of the 
reflected waves  for the snapshot corresponding to the time $t_6$ comparing it 
at least with the shear energy in Fig.~\ref{fig-EQ}-middle at the same time 
instant.} 

\subsection{A new listric normal fault birth}\label{experinent2}
%           ~~~~~~~~~~~~~~~~~~~~~~~~~~~~~~~~

\begin{figure}
\centering
\includegraphics[width=27em]{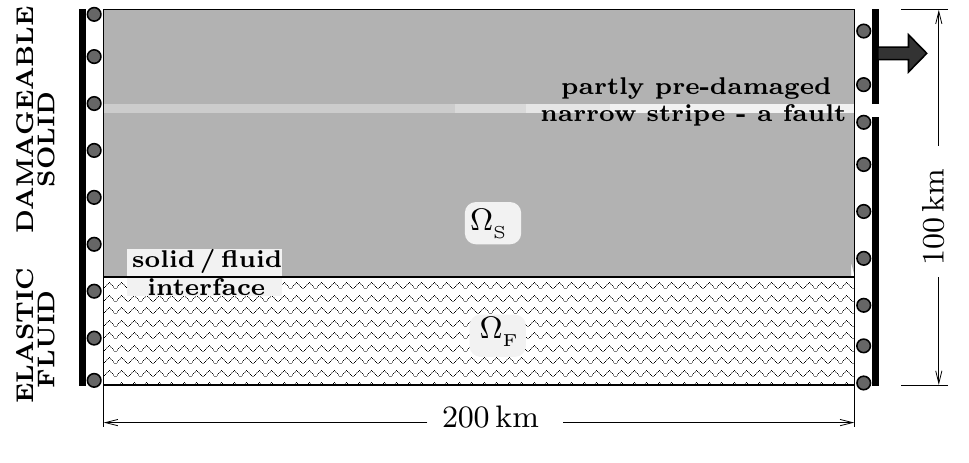}
\caption{\sl A computational 2-dimensional domain and boundary conditions 
for the experiment in 
%Figure~\ref{fig-EQ+}
Sect.\,\ref{experinent2}. 
%\COL{Notice different boundary conditions in comparison to 
%Figure~\ref{fig-EQ-geom}. Also, 
At time $t=0$, the fault is ``compact'' at the left part 
%(close to the fixed face) only 
while the right part is 
%partially
substantially damaged.
}
\label{fig-EQ-geom+}
\end{figure}

Another interesting event, although very rare from the 
mankind time scale, is a nucleation of damage in compact rocks, 
giving rise to a new damage regions, i.e.\ to a new fault.
The geometry of the 2-dimensional computation region together 
with the boundary conditions imposing (unlike the previous example) the 
increasing displacement only at the right face of the top solid layer 
is depicted in Figure~\ref{fig-EQ-geom+}.
It differs from the previous example also by the initial conditions
which makes the fault largely damaged on its right-hand part. Therefore, 
the upper plate can quite easily slide onto the lower plate while
being rather well connected with it on the left-hand part.

\begin{figure*}
\centering
%{\scriptsize\bf \hspace{2em}KINETIC\hspace{7em}SHEAR STORED} \hspace{7em}{\scriptsize\bf SPHERICAL}
%\\{\scriptsize\bf \hspace{4em}ENERGY\hspace{10.5em}ENERGY} \hspace{8em}{\scriptsize\bf STORED\ \, ENERGY}
{\footnotesize\bf \hspace{2em}KINETIC\ \,ENERGY\hspace{11.5em}DAMAGE \hspace{11em} SHEAR\ \,STORED\ \,ENERGY}
\\
\rotatebox[origin=lt]{90}{\parbox{2cm}{\centering\COL{$t_1=49.0$\,ks}}}\hspace*{-.2em}
\includegraphics[width=0.32\textwidth,bb=100 170 660 450,clip=true]{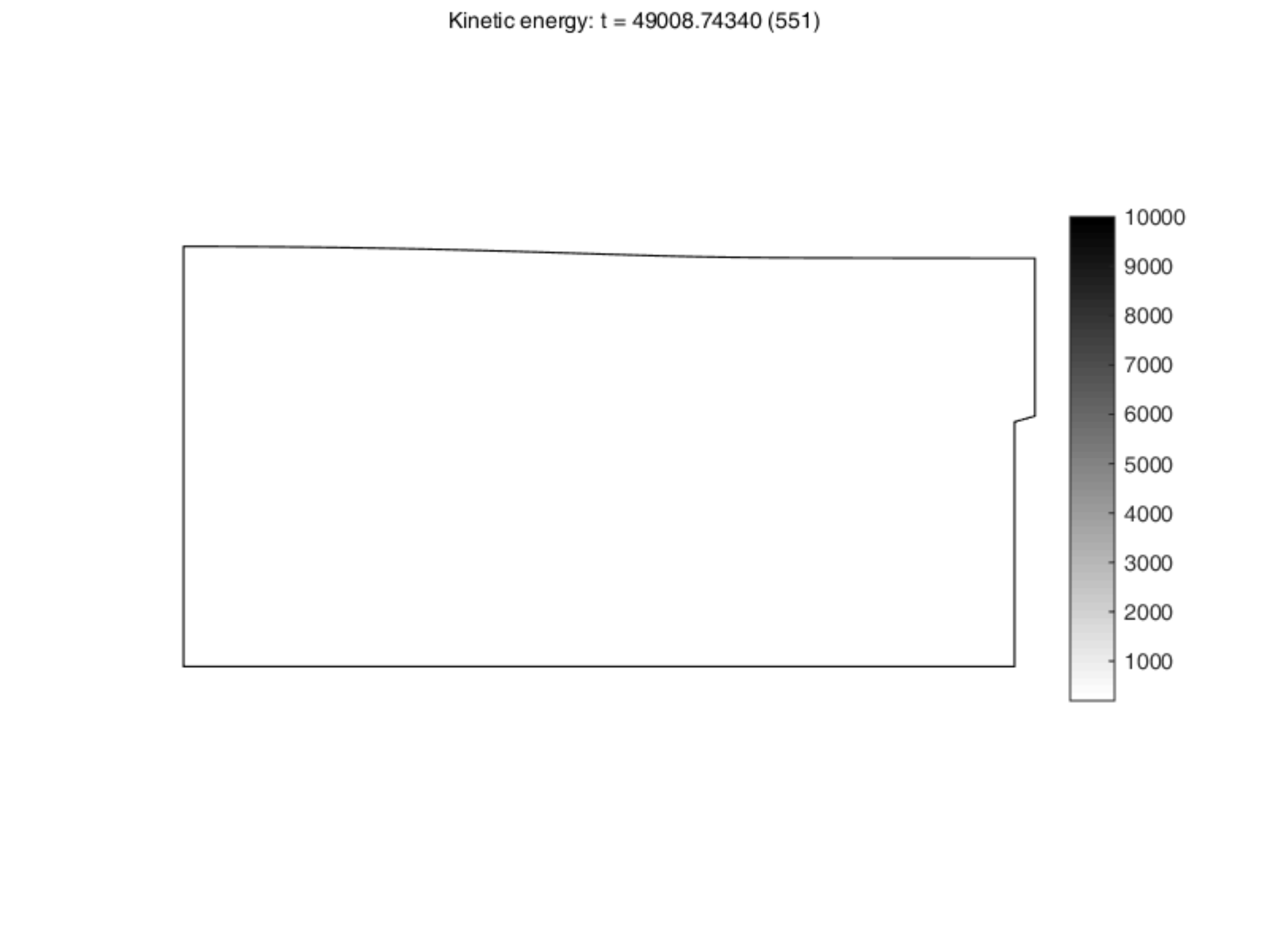}%\hspace*{.5em}
\includegraphics[width=0.32\textwidth,bb=100 170 660 450,clip=true]{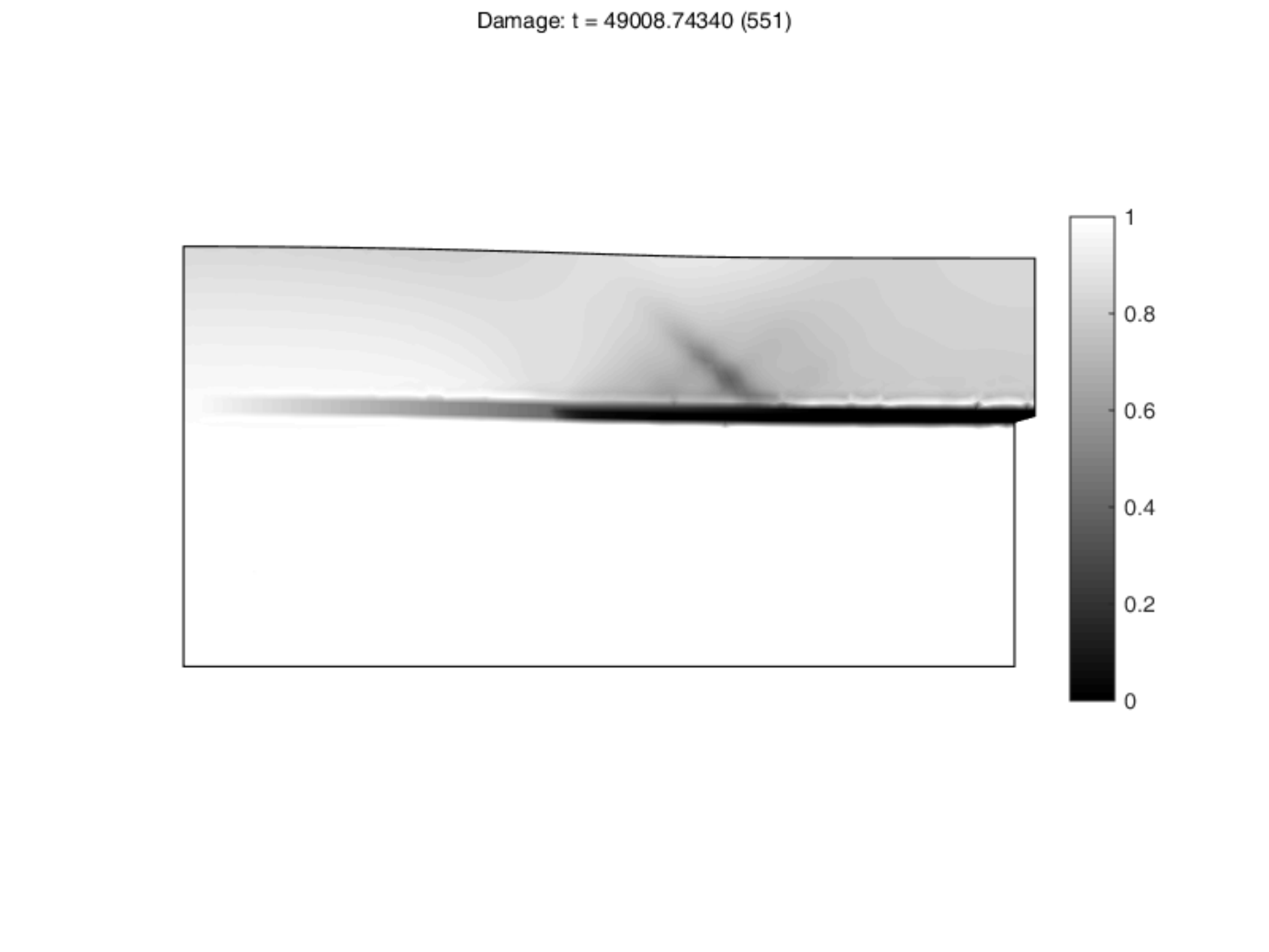}%\hspace*{.5em}
\includegraphics[width=0.32\textwidth,bb=100 170 660 450,clip=true]{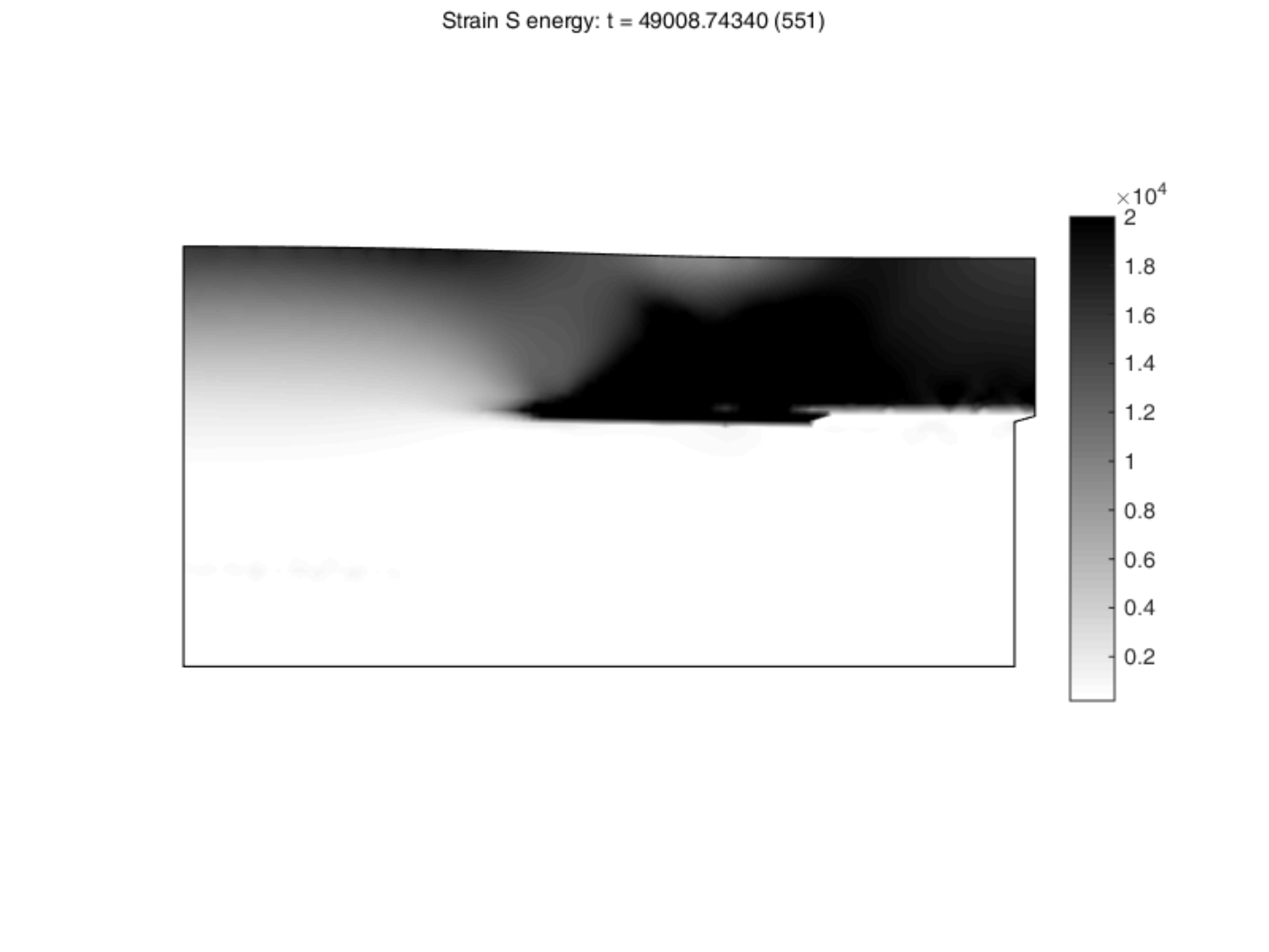}
\\
\rotatebox[origin=lt]{90}{\parbox{2cm}{\centering\COL{$t_2=t_1+7.30$\,s}}}\hspace*{-.2em}
\includegraphics[width=0.32\textwidth,bb=100 170 660 450,clip=true]{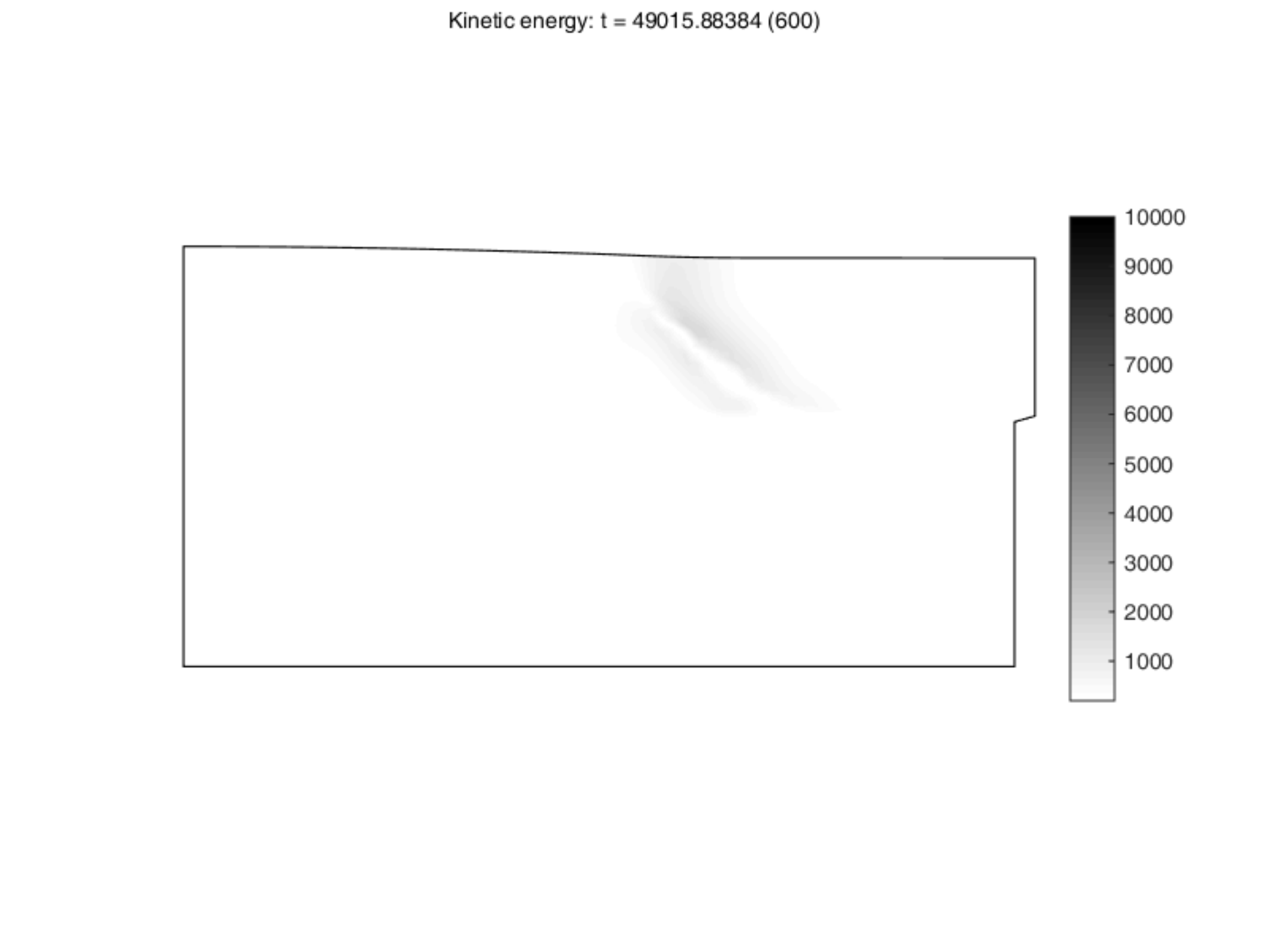}%\hspace*{.5em}
\includegraphics[width=0.32\textwidth,bb=100 170 660 450,clip=true]{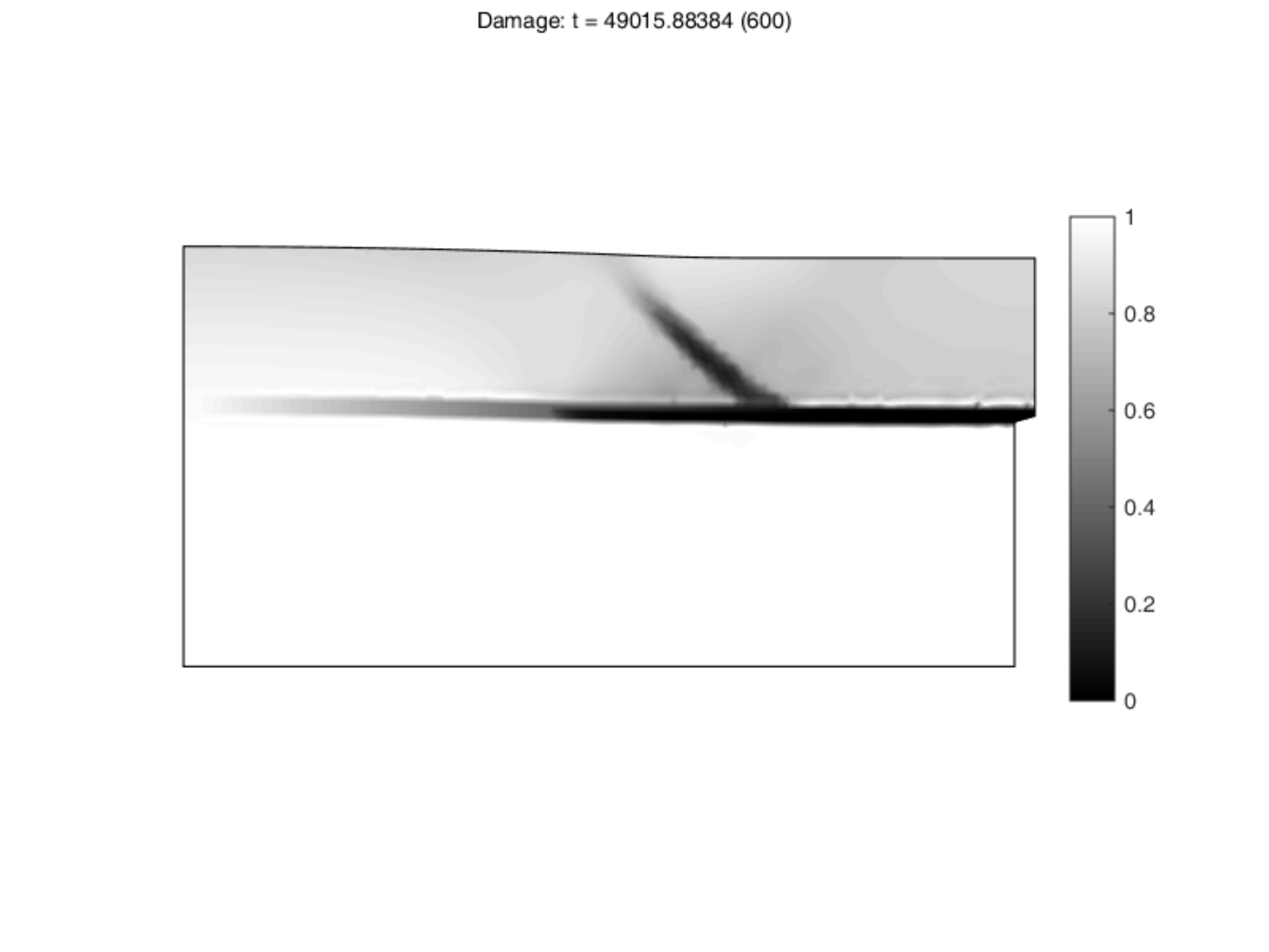}%\hspace*{.5em}
\includegraphics[width=0.32\textwidth,bb=100 170 660 450,clip=true]{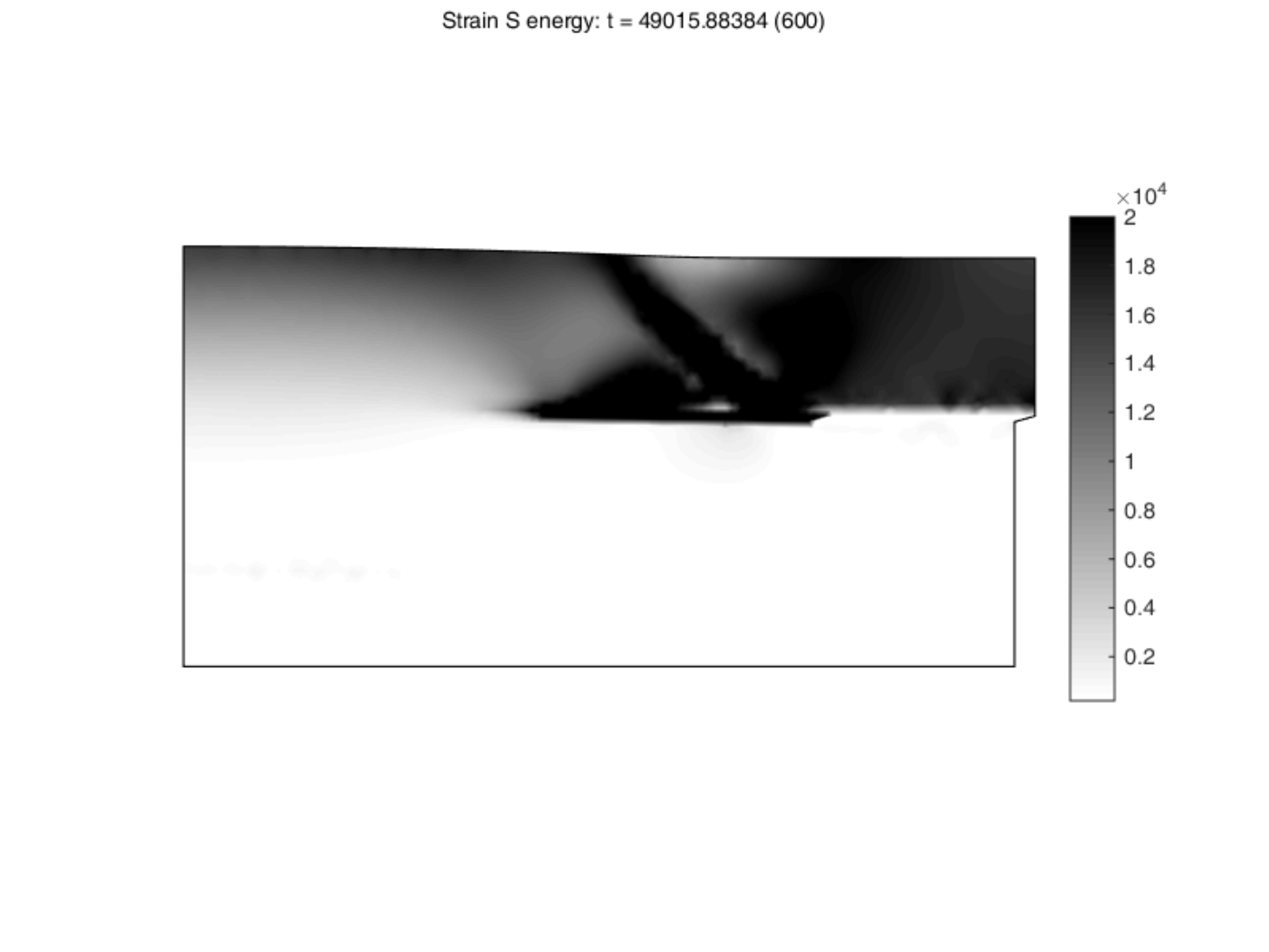}
\\
\rotatebox[origin=lt]{90}{\parbox{2cm}{\centering$t_3=t_1+11.68$\,s}}\hspace*{-.2em}
\includegraphics[width=0.32\textwidth,bb=100 170 660 450,clip=true]{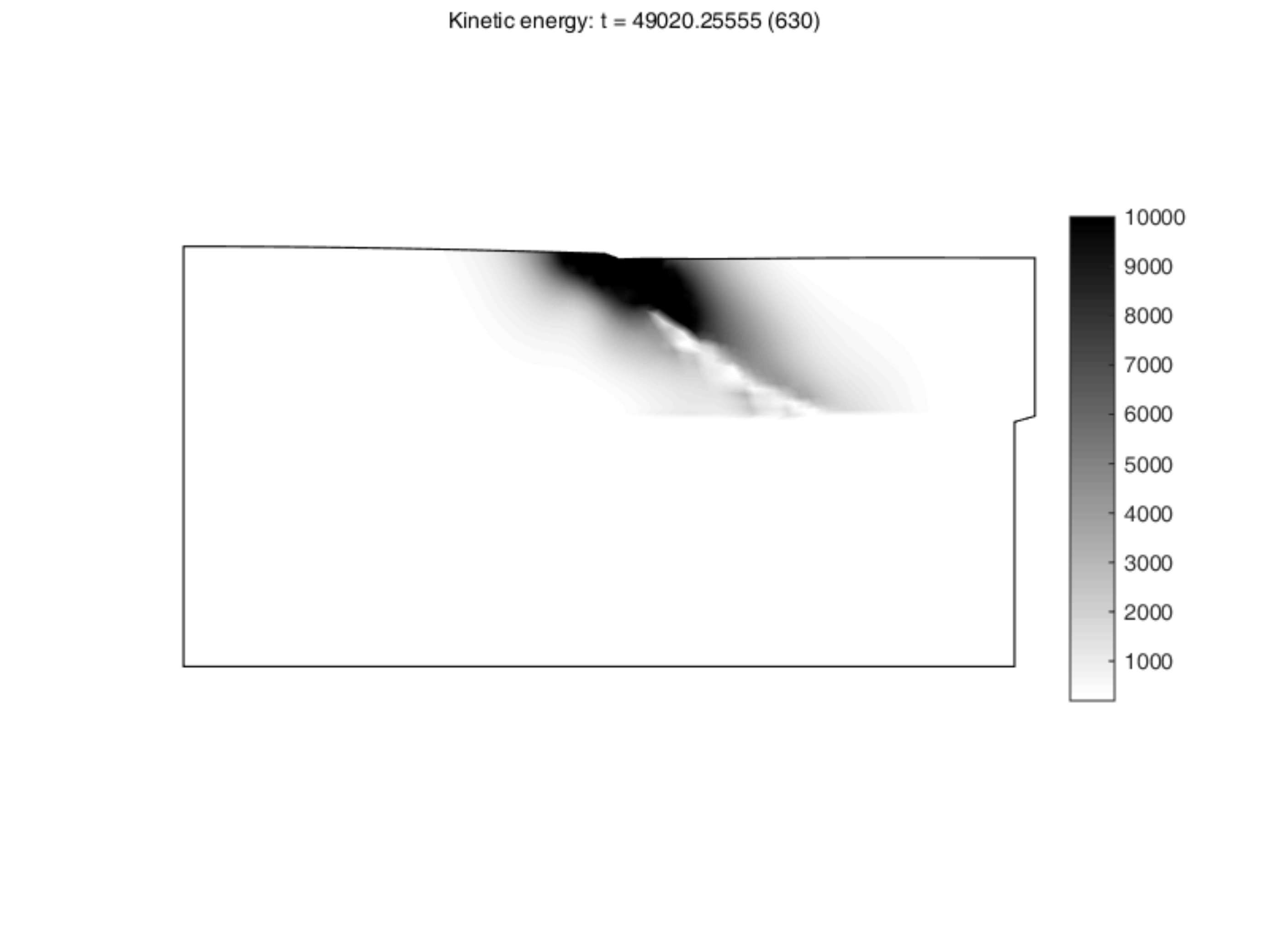}%\hspace*{.5em}
\includegraphics[width=0.32\textwidth,bb=100 170 660 450,clip=true]{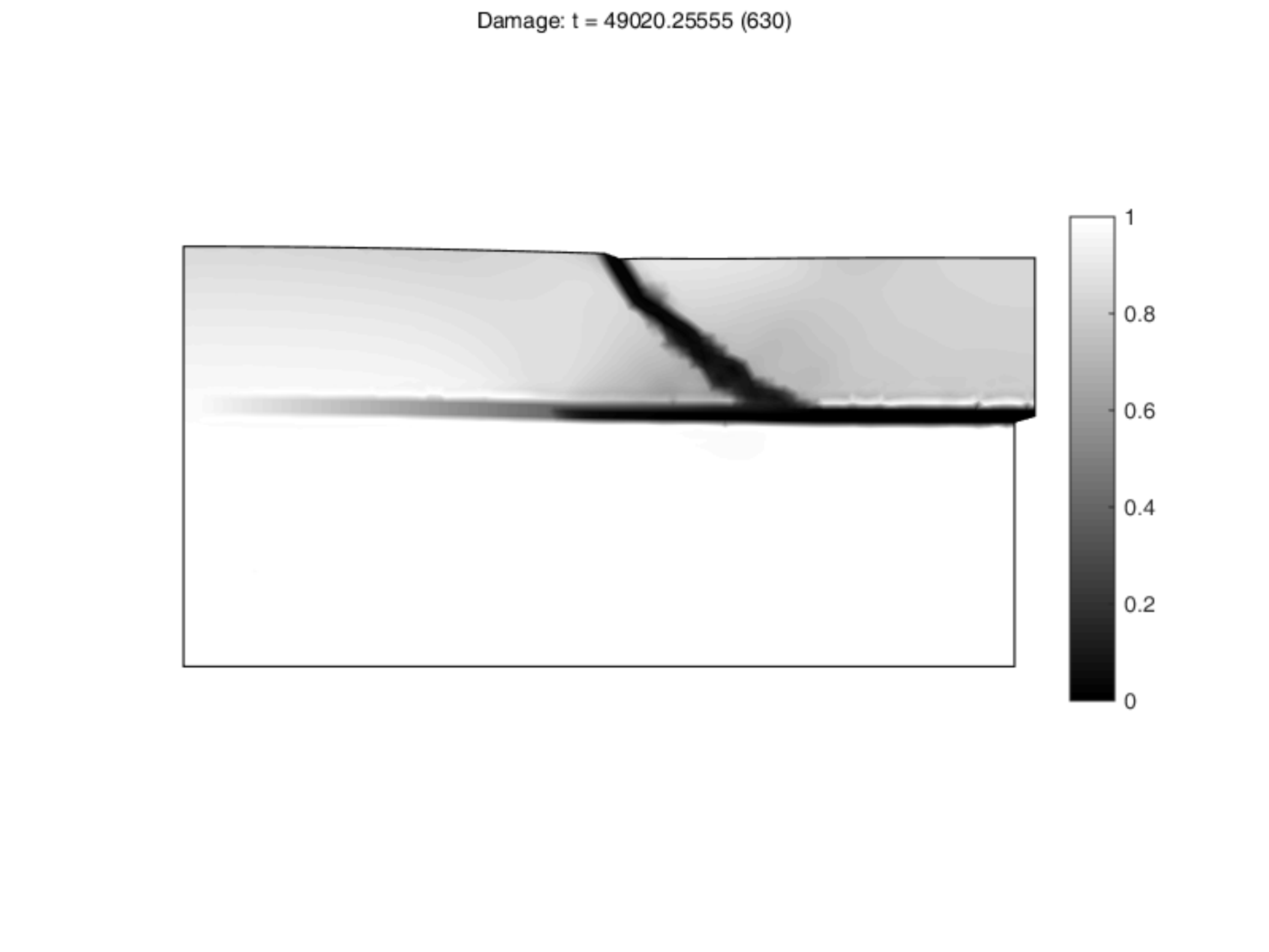}%\hspace*{.5em}
\includegraphics[width=0.32\textwidth,bb=100 170 660 450,clip=true]{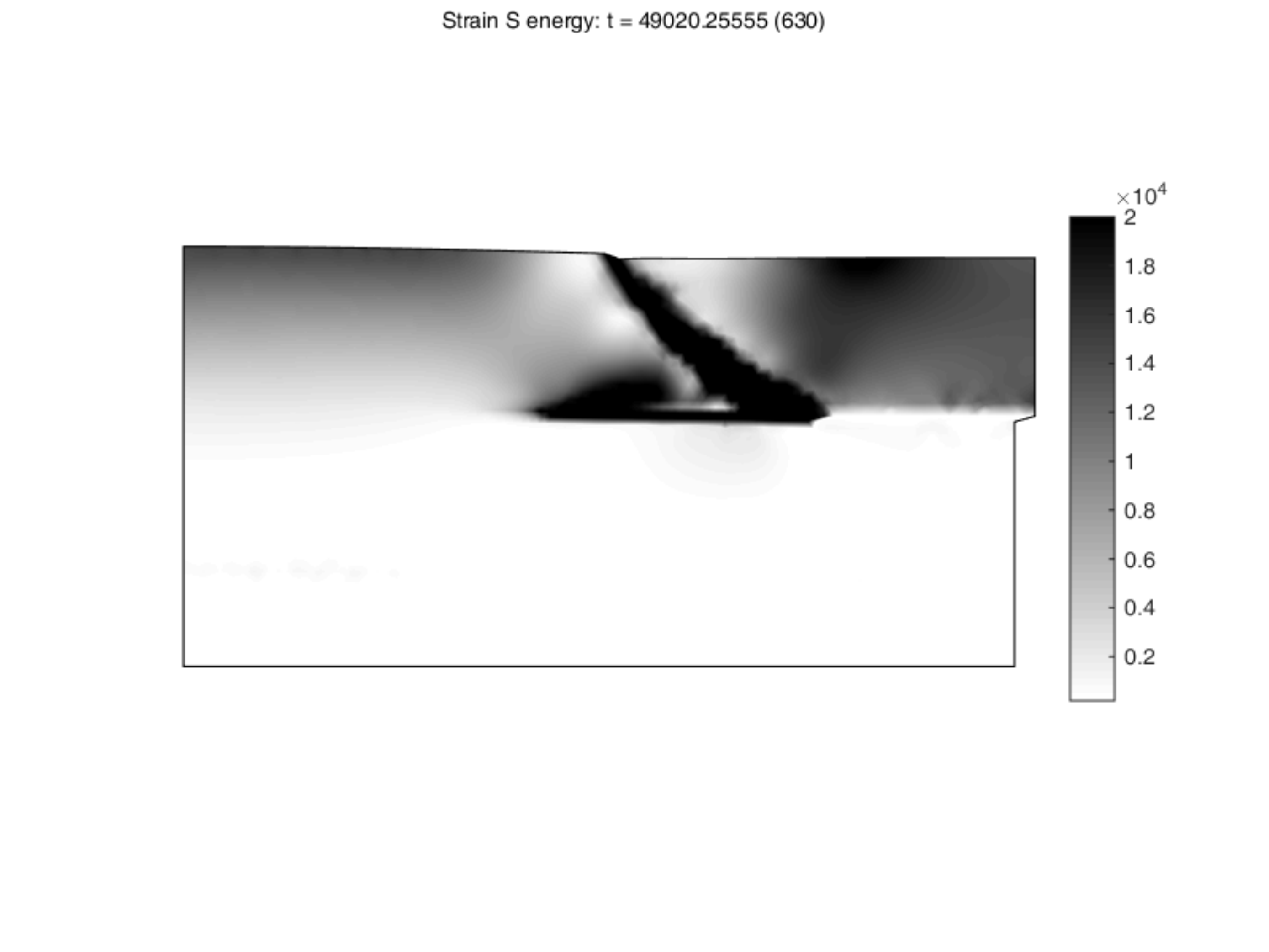}
\\
\rotatebox[origin=lt]{90}{\parbox{2cm}{\centering\COL{$t_4=t_1+16.06$\,s}}}\hspace*{-.2em}
\includegraphics[width=0.32\textwidth,bb=100 170 660 450,clip=true]{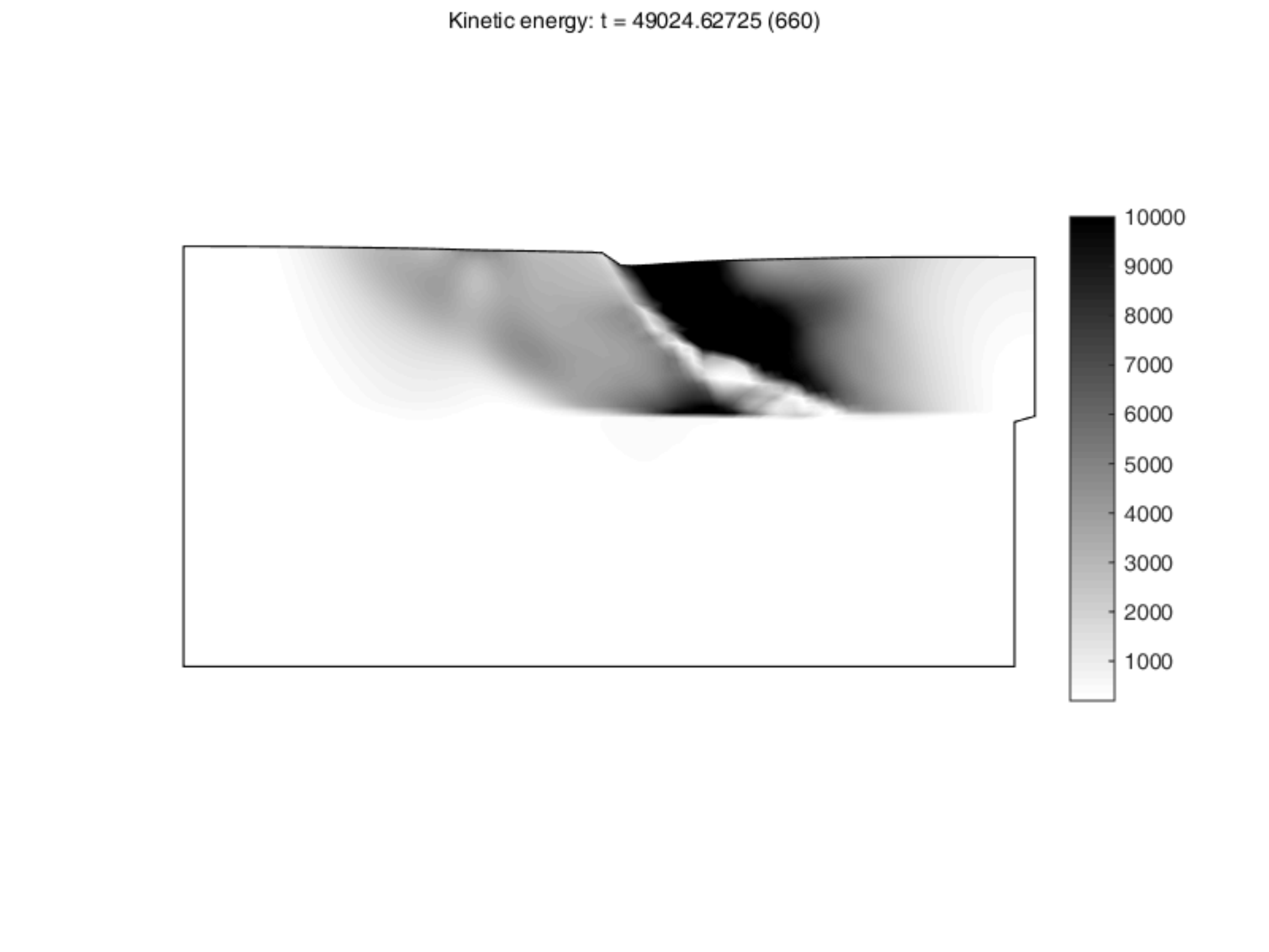}%\hspace*{.5em}
\includegraphics[width=0.32\textwidth,bb=100 170 660 450,clip=true]{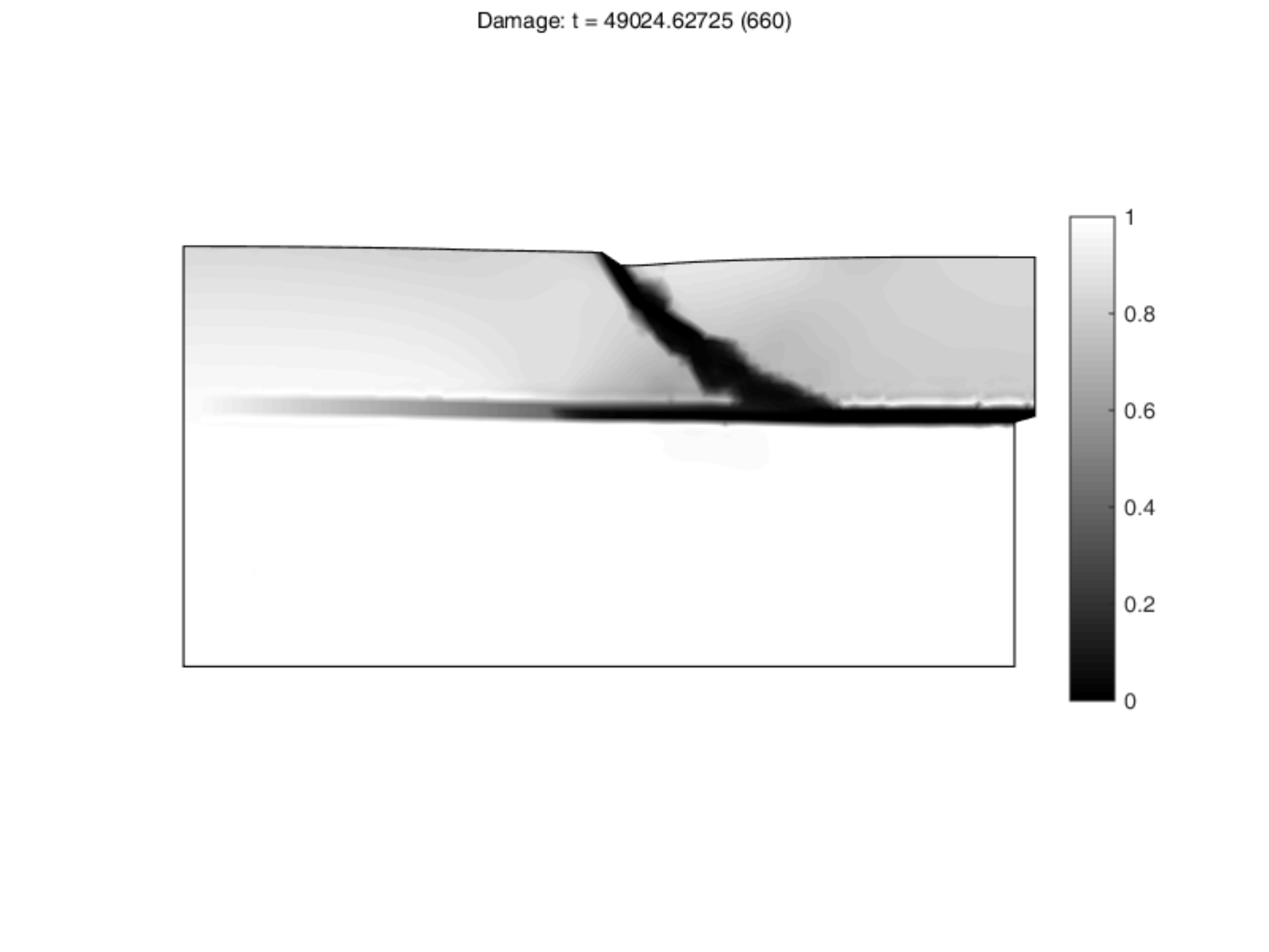}%\hspace*{.5em}
\includegraphics[width=0.32\textwidth,bb=100 170 660 450,clip=true]{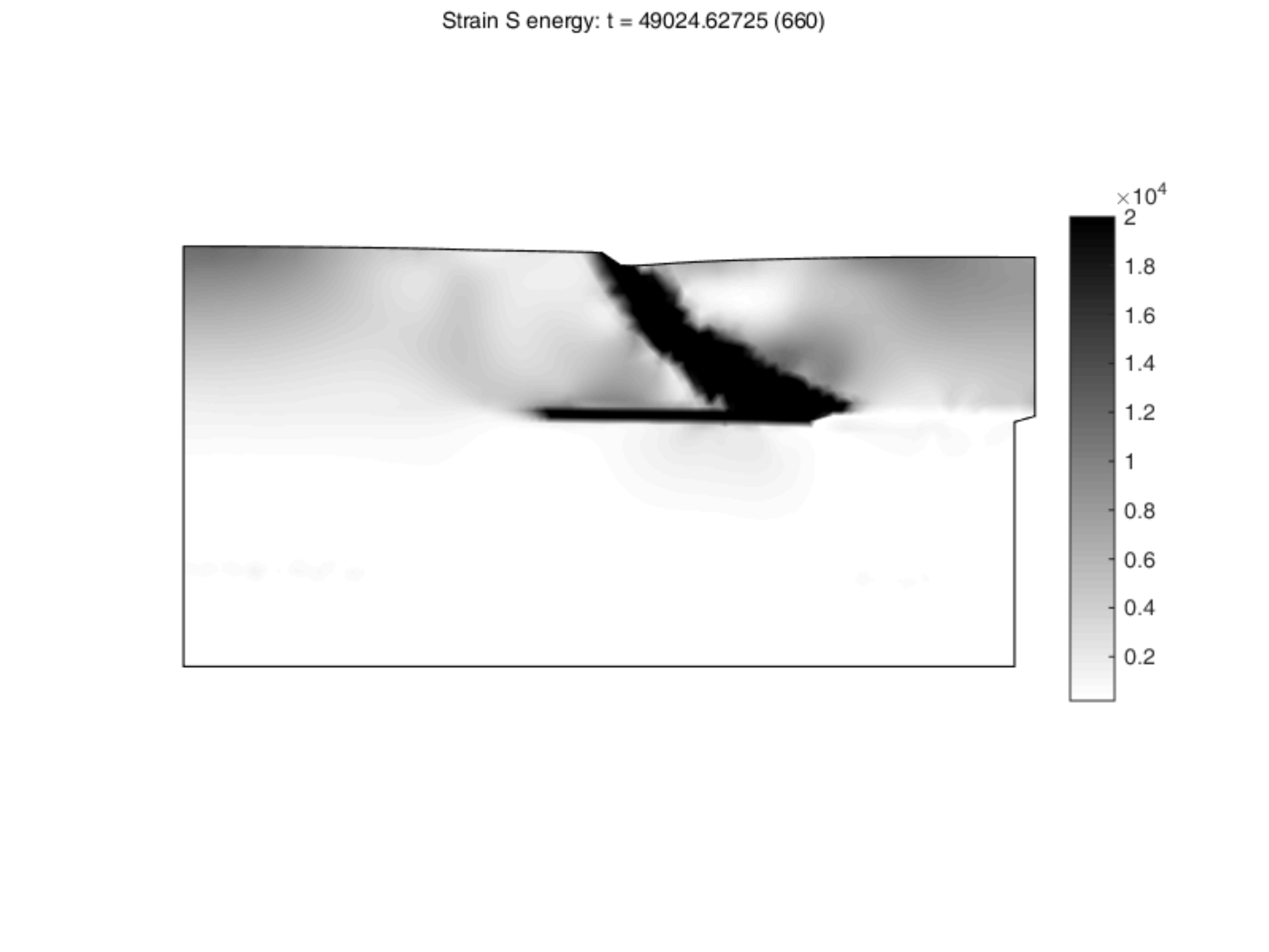}
\\
\rotatebox[origin=lt]{90}{\parbox{2cm}{\centering\COL{$t_5=t_1+20.44$\,s}}}\hspace*{-.2em}
\includegraphics[width=0.32\textwidth,bb=100 170 660 450,clip=true]{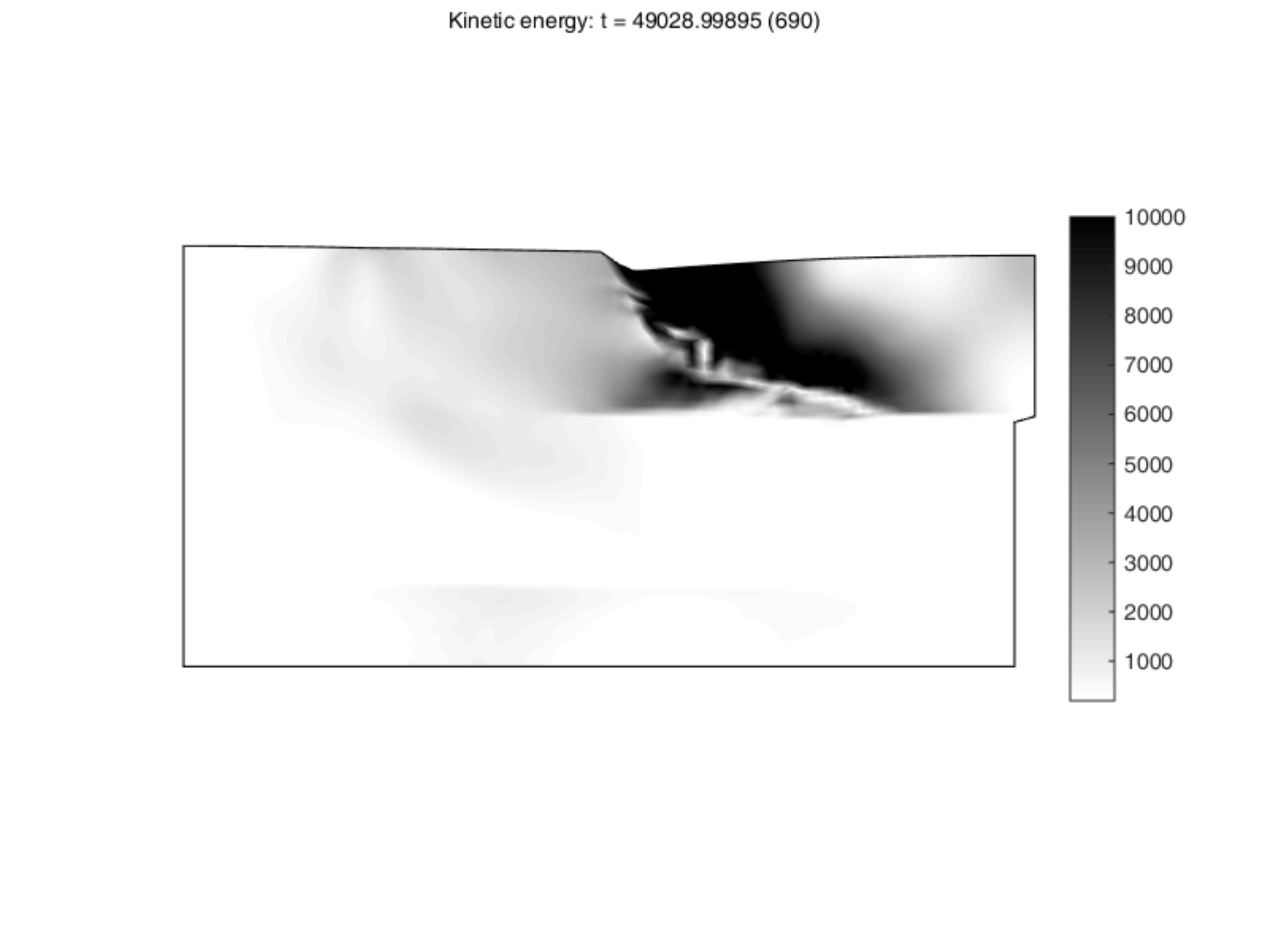}%\hspace*{.5em}
\includegraphics[width=0.32\textwidth,bb=100 170 660 450,clip=true]{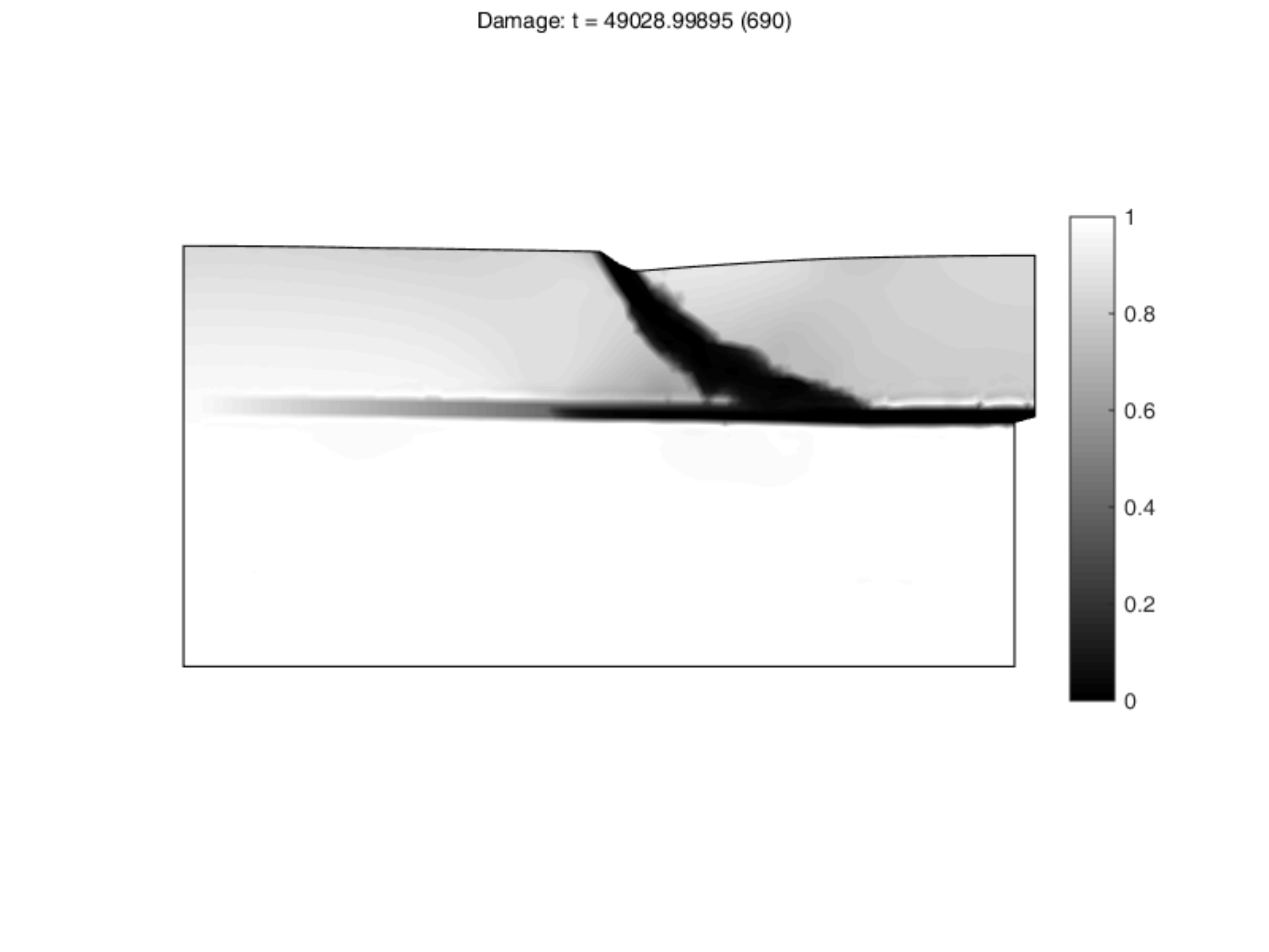}%\hspace*{.5em}
\includegraphics[width=0.32\textwidth,bb=100 170 660 450,clip=true]{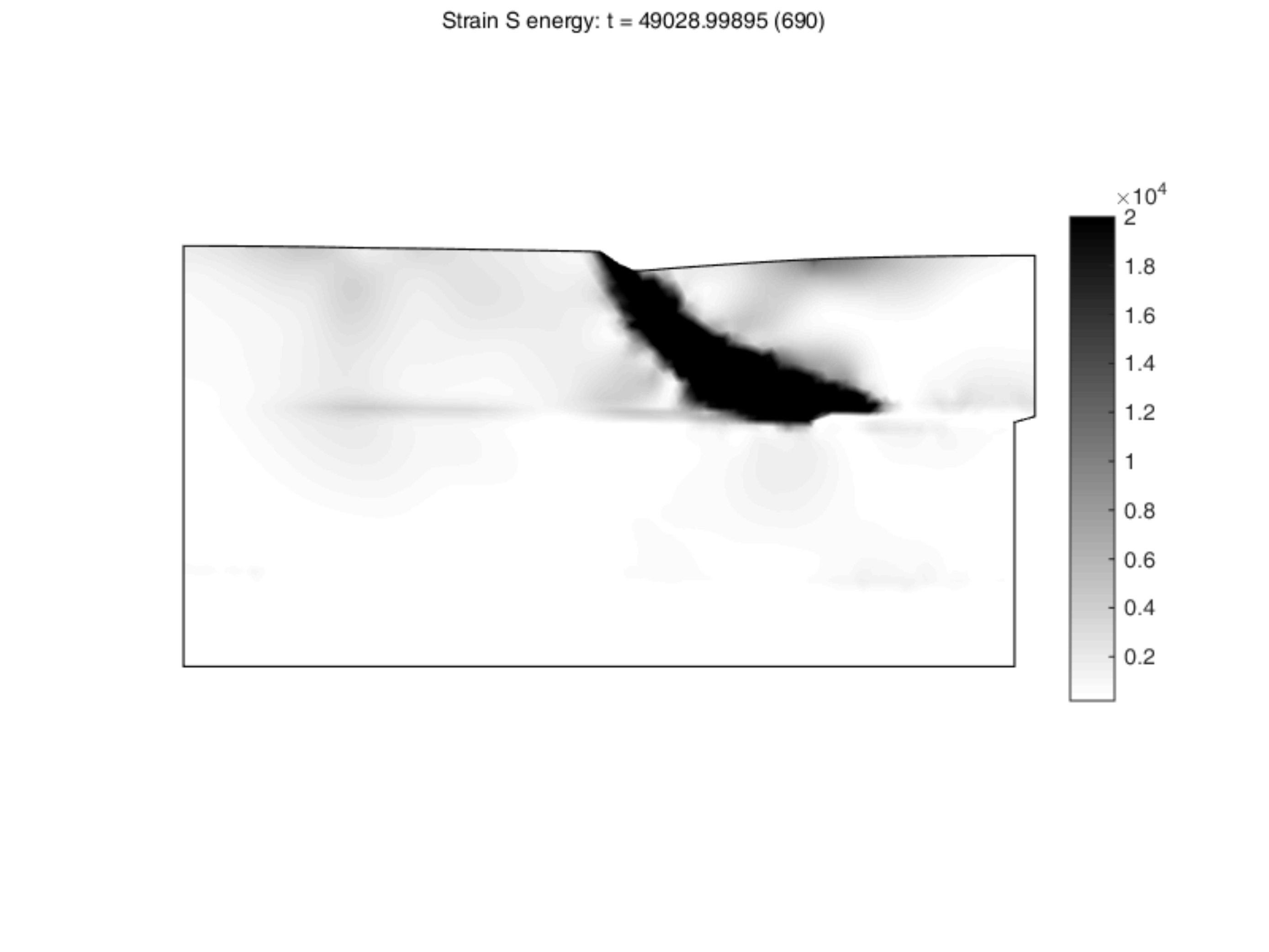}
\\
\rotatebox[origin=lt]{90}{\parbox{2cm}{\centering$t_6=t_1+24.82$\,s}}\hspace*{-.2em}
\includegraphics[width=0.32\textwidth,bb=100 170 660 450,clip=true]{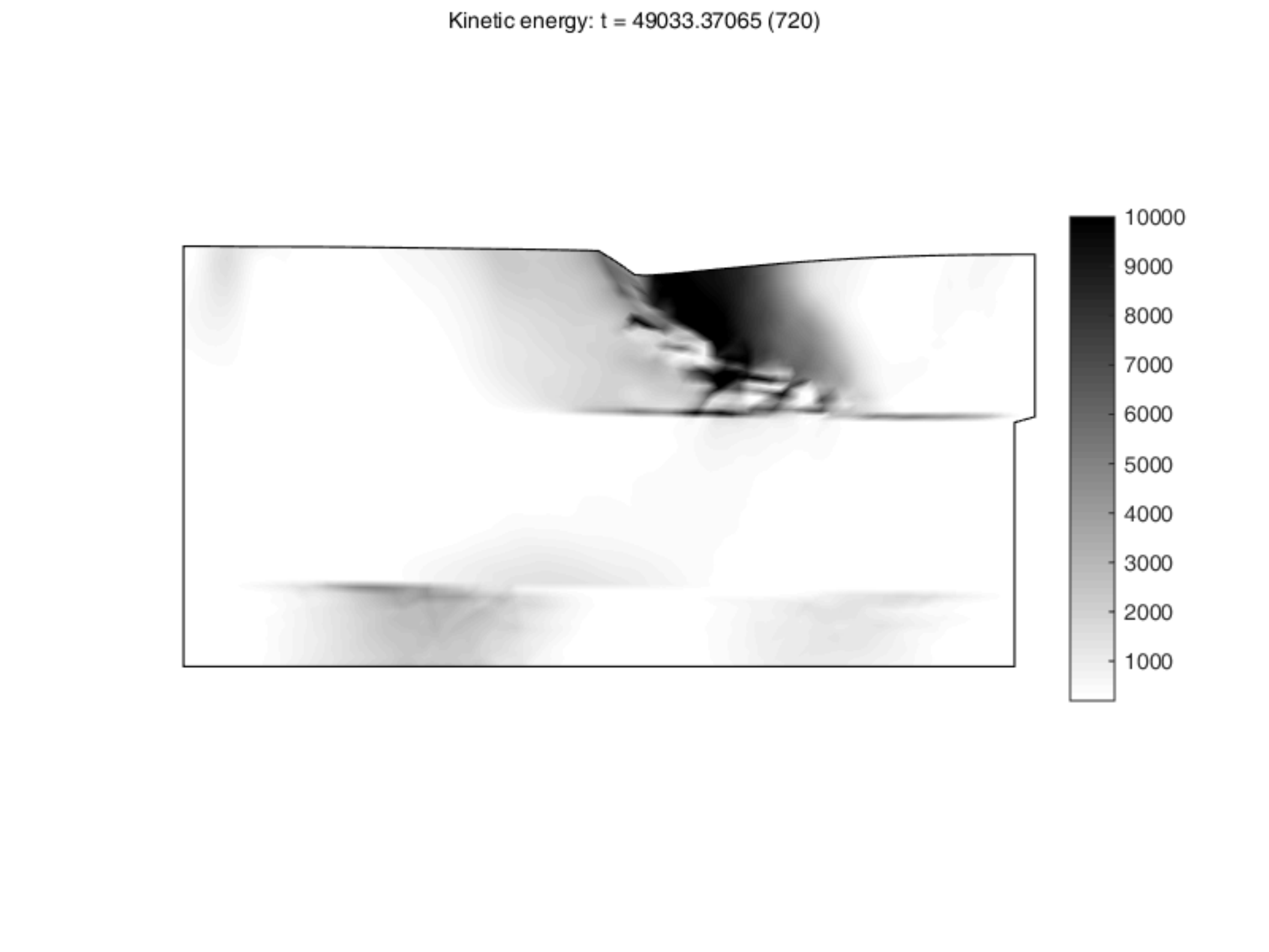}%\hspace*{.5em}
\includegraphics[width=0.32\textwidth,bb=100 170 660 450,clip=true]{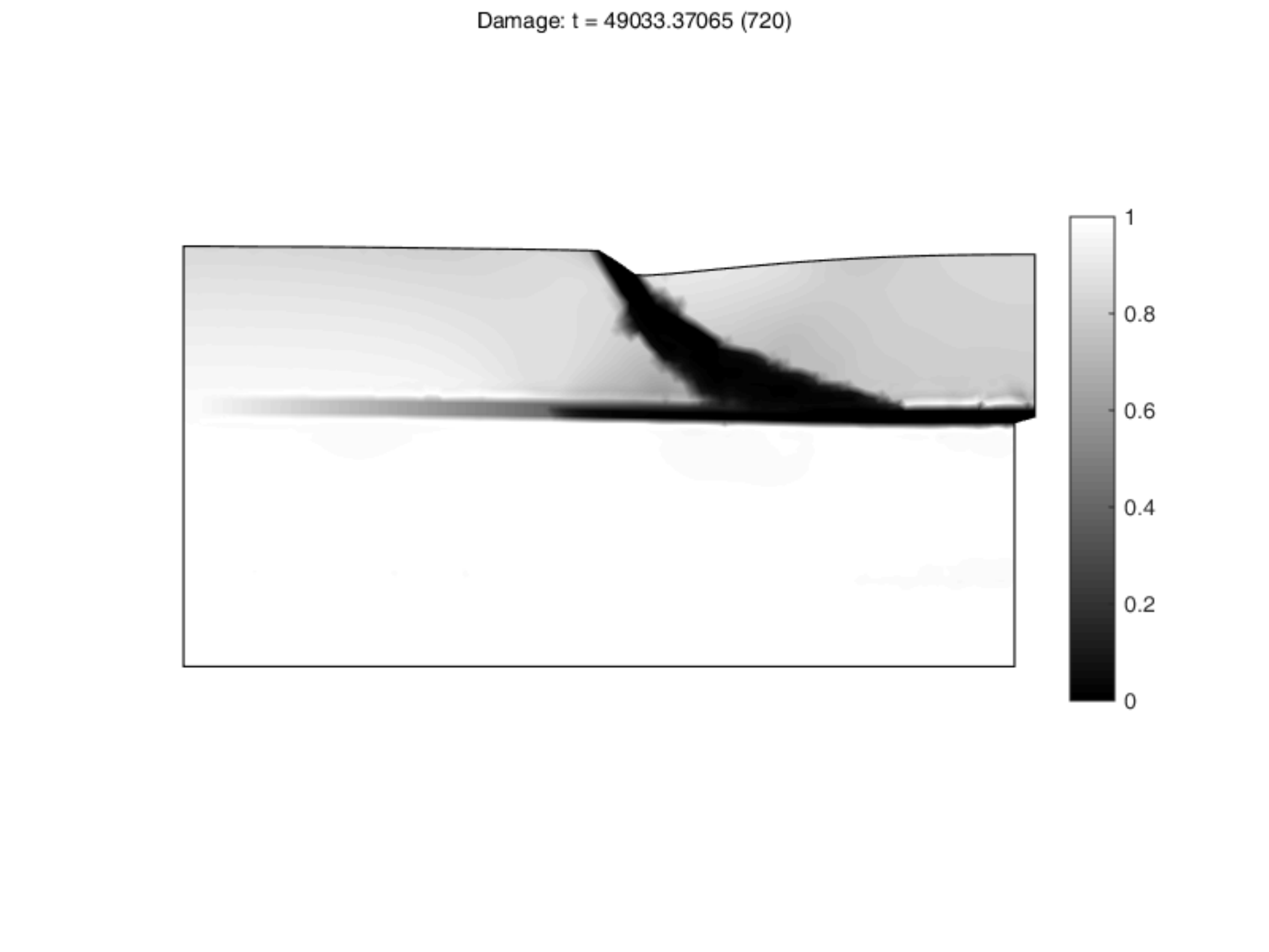}%\hspace*{.5em}
\includegraphics[width=0.32\textwidth,bb=100 170 660 450,clip=true]{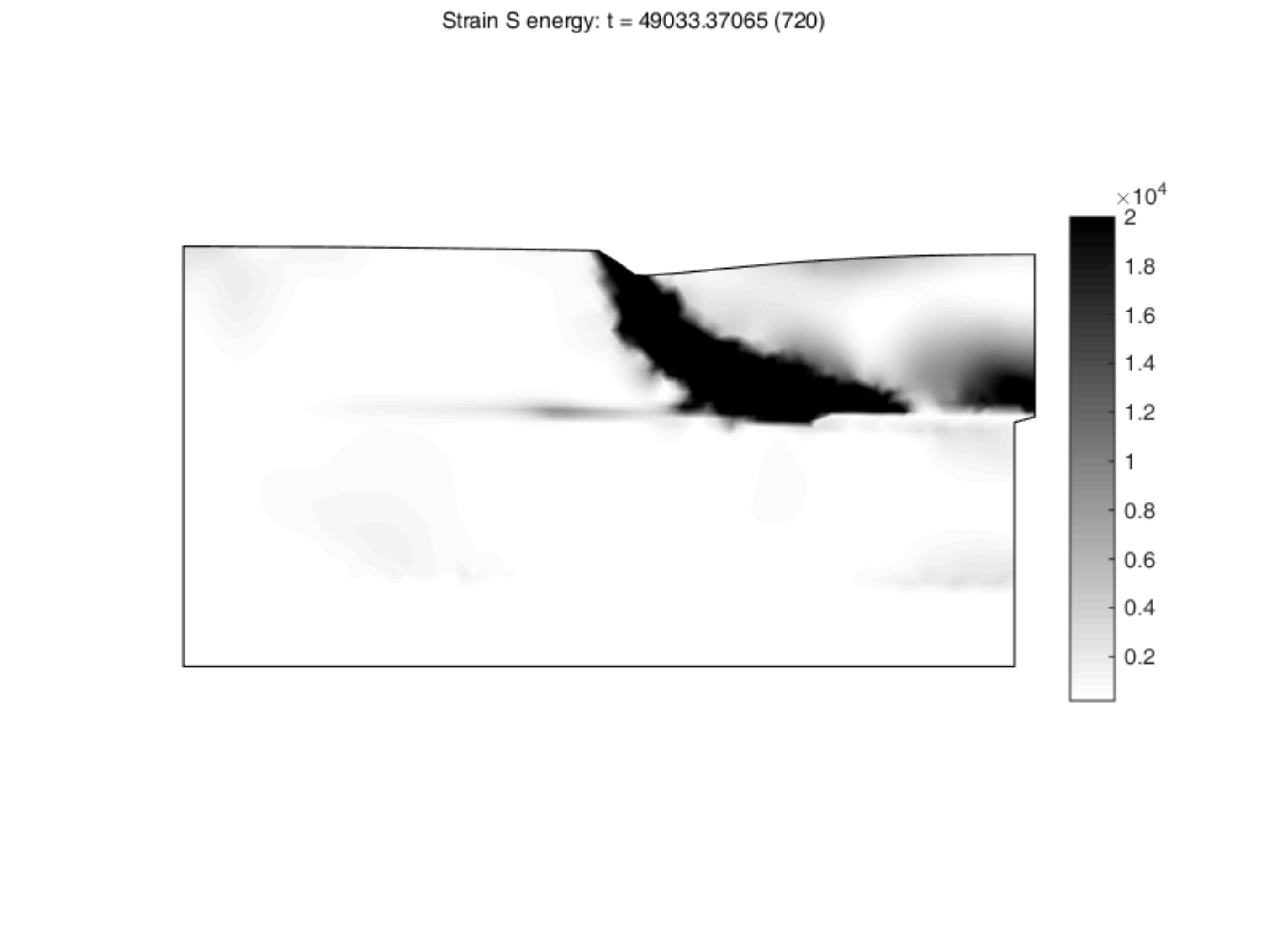}
\\
\rotatebox[origin=lt]{90}{\parbox{2cm}{\centering\COL{$t_7=t_1+29.2$\,s}}}\hspace*{-.2em}
\includegraphics[width=0.32\textwidth,bb=100 170 660 450,clip=true]{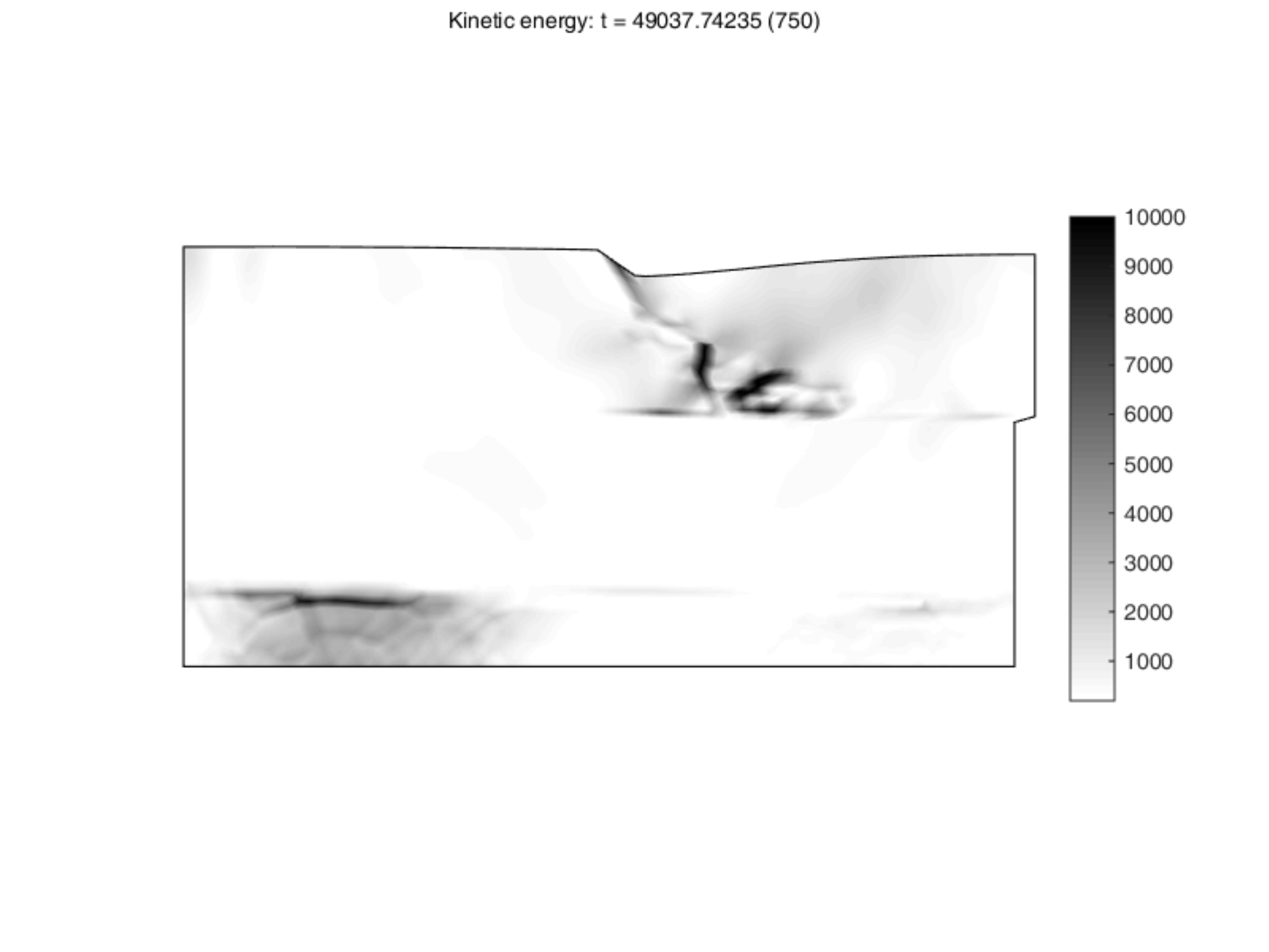}%\hspace*{.5em}
\includegraphics[width=0.32\textwidth,bb=100 170 660 450,clip=true]{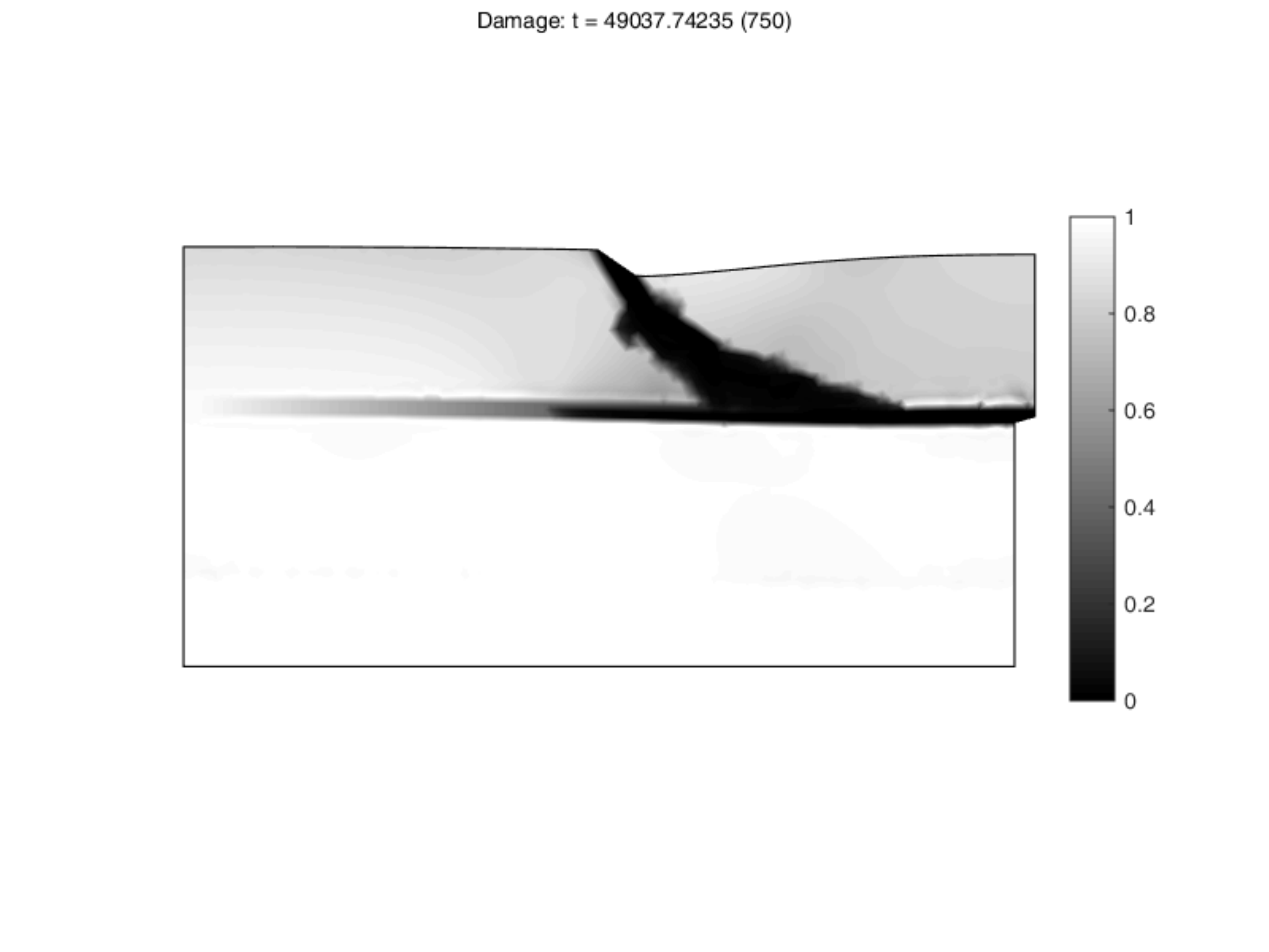}%\hspace*{.5em}
\includegraphics[width=0.32\textwidth,bb=100 170 660 450,clip=true]{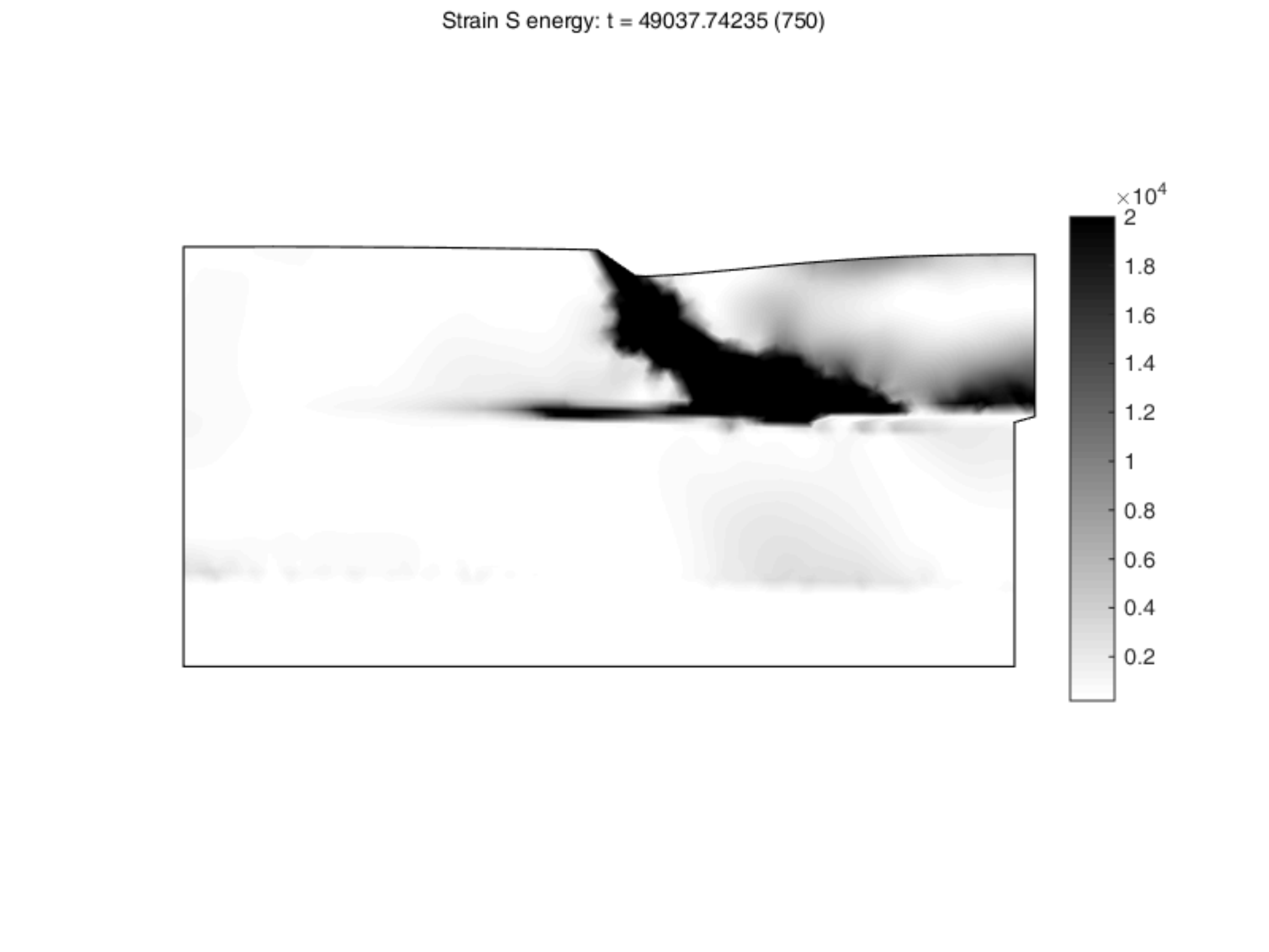}
\caption{\sl Simulations of a birth of a new fault in a normal position
and a depression on the Earth surface, together with emission of a seismic 
(mainly S-) wave during completion of the rupture, its propagation and creation 
of a rather strong P-wave in the fluidic domain. The displacement magnified 
100$\times$.}
\label{fig-EQ+}
\end{figure*}

The results of this example are depicted in Figure~\ref{fig-EQ+} in seven 
selected snapshots of spatial distribution similar to the previous 
example. When the upper plate is stretched enough, it starts rupturing
on an a-priori not pre-defined place. The energetics of the model dictates
that the new damaging area is a (relatively) narrow plane which is 
positioned at about 45 
%?????????????????????????? MELO BY BYT SPISE 60 !!!!! 
degrees with respect to the existing fault. Such position, together with the 
slip orientation, is referred to as a {\it normal fault}, in contrast to 
reverse (thrust) faults, or strike or vertical faults.
%One can observe that
In fact, the new fault plane slightly curves,
the dip being steeper near the surface while shallower with increased depth. 
This is referred to as a {\it listric normal fault}.
%, cf.\ \cite{BBDM81LNF} or e.g.\ \cite[Figs.6-7]{PDGL17FSCO}. 
%\COL{...HERE A REF. TO an (Italien) listric faults.............}

The rupture starts on the existing fault due to a slight stress concentration
and then propagates towards the Earth's surface with an increasing speed,
emitting a seismic wave. This wave propagates mainly through the 
upper plate (\COL{as an S-wave which can be observed in the plots of kinetic and shear stored energy for the time instants $t_4$ and $t_5$)} but partly penetrated through the existing horizontal fault into the
lower plate and then even towards the fluidic layer where it creates a relatively
strong P-wave, cf.\ Figure~\ref{fig-EQ+}-left \COL{the time instant $t_7$}.
%.....HERE DESCRIPTION OF THE RESULTS  

\COL{
\subsection{A new listric reverse fault birth followed by a 
%strike-slip 
horizontal fault rupture}\label{experinent3}
%           ~~~~~~~~~~~~~~~~~~~~~~~~~~~~~~~~

Another interesting situation, which may occur together with a dip-slip fault 
as shown in the previous example, is arising of a 
%strike-slip 
reverse (thrust) fault. 
The geometry of the 2-dimensional computation region together with the 
boundary conditions imposing the increasing displacement at both lateral 
faces of the top solid layer  but in opposite direction
is depicted in Figure~\ref{fig-reverse-geom+}.
\begin{figure}
\centering
\includegraphics[width=27em]{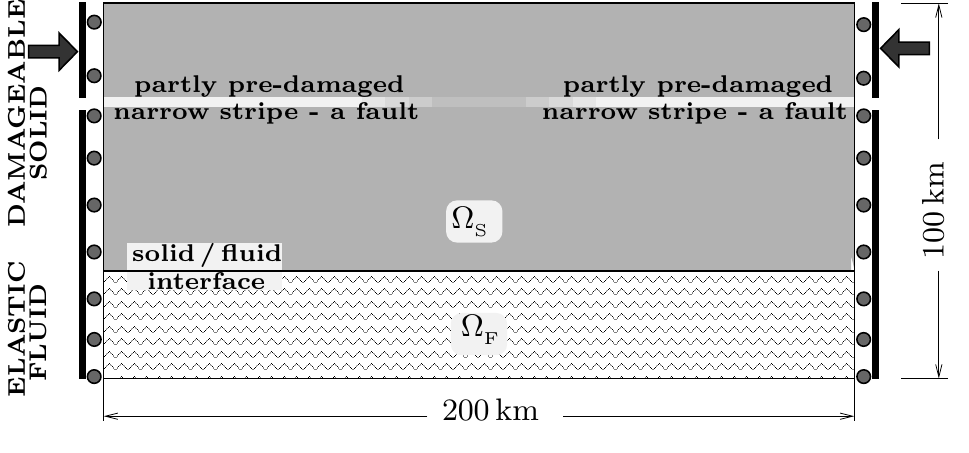}
%\\\definecolor{light}{rgb}{0.8,0.8,1}
%\fcolorbox{black}{light}{\Huge TO BE MODIFIED}
\caption{\sl A computational 2-dimensional domain and boundary conditions 
for the experiment in  Sect.\,\ref{experinent3}. 
At time $t=0$, the fault is ``compact'' at the central part only while 
both the left and the right parts are 
%partially 
substantially damaged.
}
\label{fig-reverse-geom+}
\end{figure}
It differs from the previous example also by the initial damage conditions
where both ends of the fault are largely damaged. Therefore, the upper plate 
easily slides on the lower plate at its 
%extremes
side parts while the central part 
%of the plates are
is well connected with the lower plate and allow for stress concentration.

The results of this example are shown in Figure~\ref{fig-reverse+} in eight 
selected snapshots of spatial distribution similarly as in the previous 
examples.
When the upper plate is substantially compressed, it starts rupturing
on an a-priori not pre-defined place. The energetics of the model dictates
that the new damaging area is a (relatively) narrow plane which is 
positioned in about 60 degrees with respect to the existing fault. Such 
position, together with the 
slip orientation, is referred to as a {\it reverse (thrust) fault}, here 
again a bit curved (listric) like in the previous normal-fault case.
%The emitted energy then cause another fault in the weakened central part 
%so that the predefined damaged faults connect.
During this first rupture, the newly born thrust fault 
%Therefore, 
starts on the existing horizontal fault 
%due to a stress concentration the fault 
and then propagates towards the Earth surface with an increasing speed, 
emitting a seismic wave. This wave propagates mainly through the upper 
plate but partly penetrates through the existing horizontal fault into the
lower plate and then even towards the fluidic layer where it creates a 
relatively strong P-wave, cf.\ Figure~\ref{fig-reverse+}-left.

After a certain time (determined rather by evolving boundary conditions), it 
causes the 
%strike-slip
horizontal fault rupture with emission of another S-waves which 
propagates/reflects in a similar way as in the first example in 
Sect.~\ref{sec-old-fault}.
%whose reflections 
%in the lower solid plate can be observed in the snapshots.
\begin{figure*}
\centering
{\footnotesize\bf \hspace{2em}KINETIC\ \,ENERGY\hspace{11.5em}DAMAGE \hspace{11em} SHEAR\ \,STORED\ \,ENERGY}
\\  
\rotatebox[origin=lt]{90}{\parbox{2cm}{\centering\COL{$t_1=$27.3\,ks}}}\hspace*{-.2em}
\includegraphics[width=0.30\textwidth,bb=130 130 880 520,clip=true]{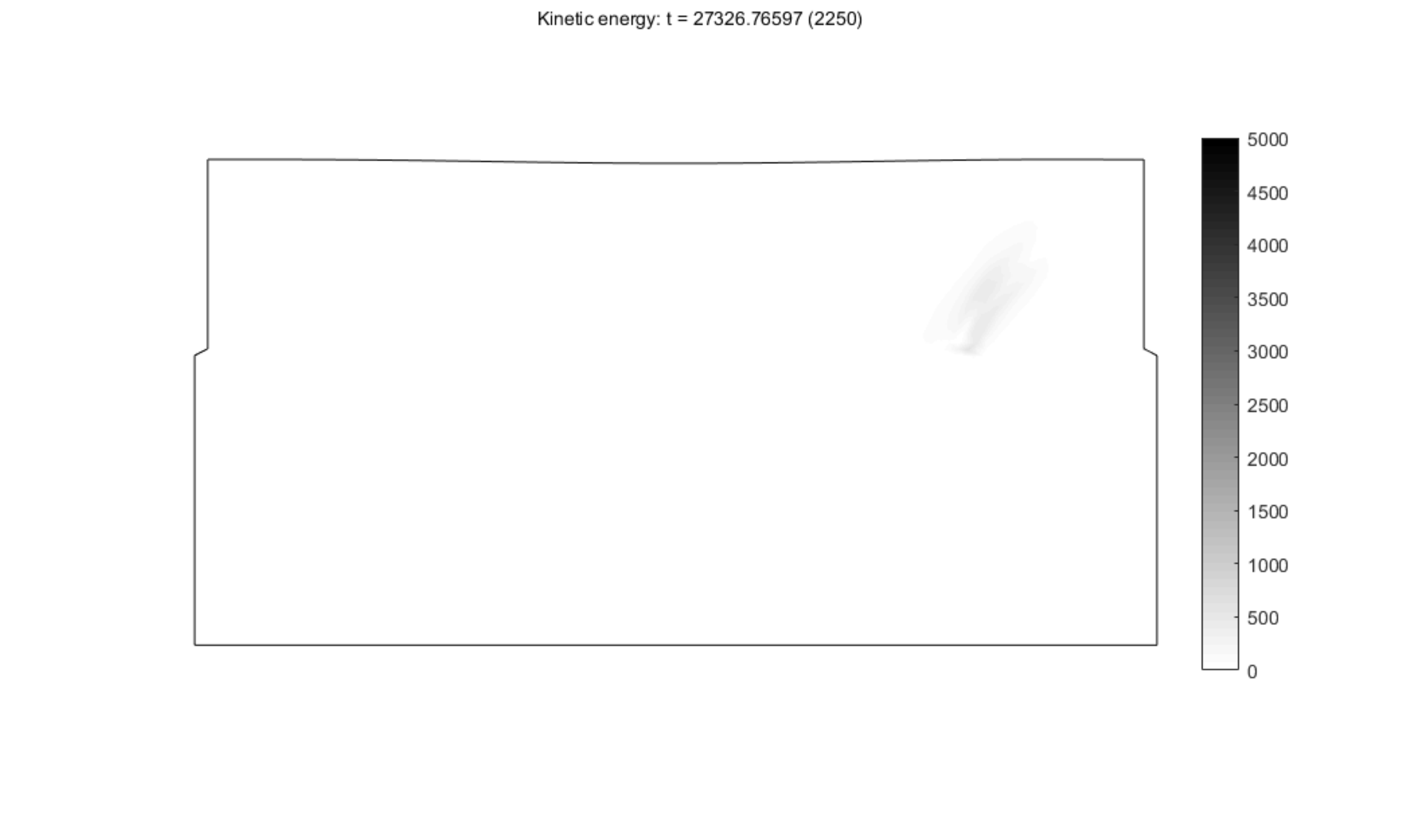}%\hspace*{.5em}
\includegraphics[width=0.30\textwidth,bb=130 130 880 520,clip=true]{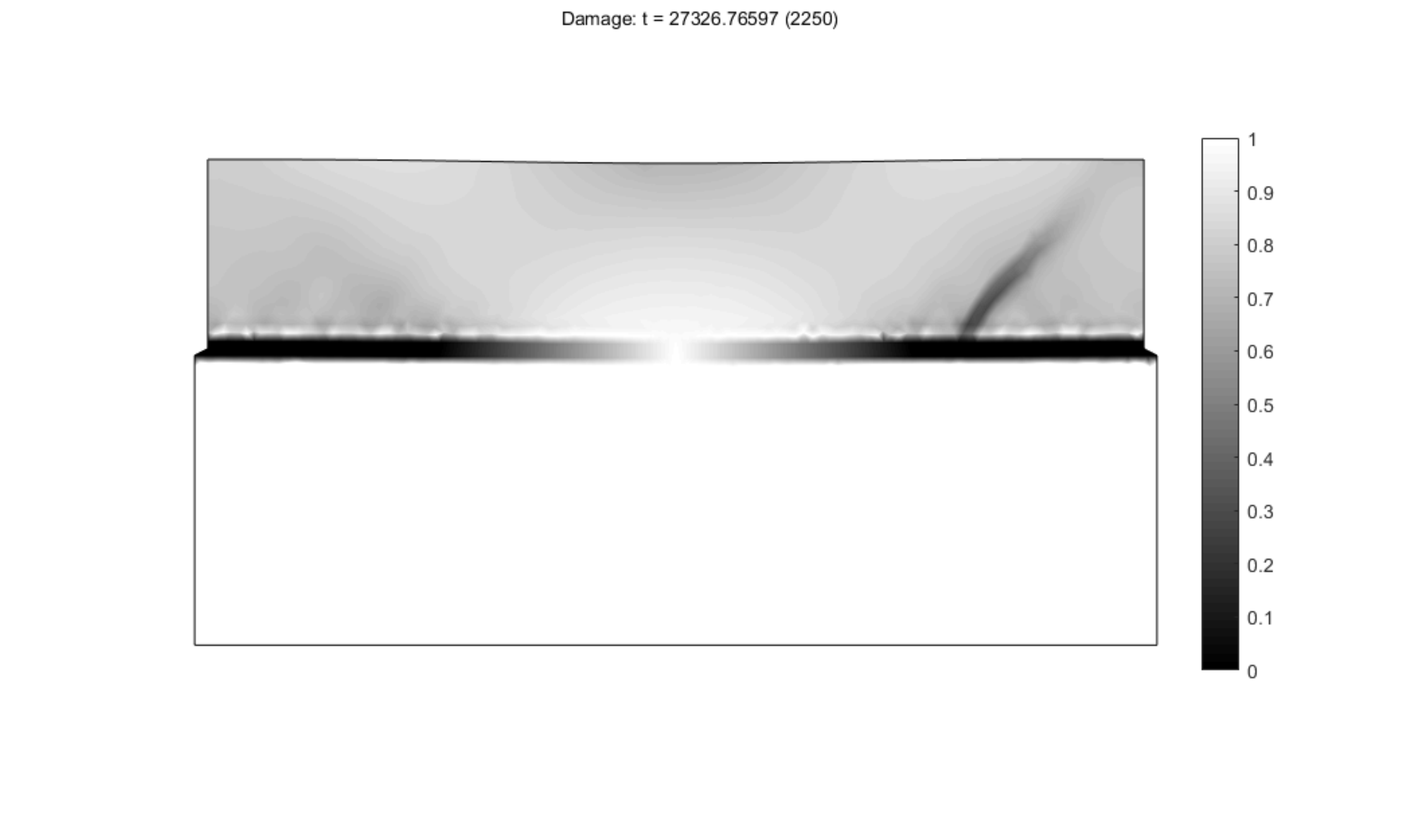}%\hspace*{.5em}
\includegraphics[width=0.30\textwidth,bb=130 130 880 520,clip=true]{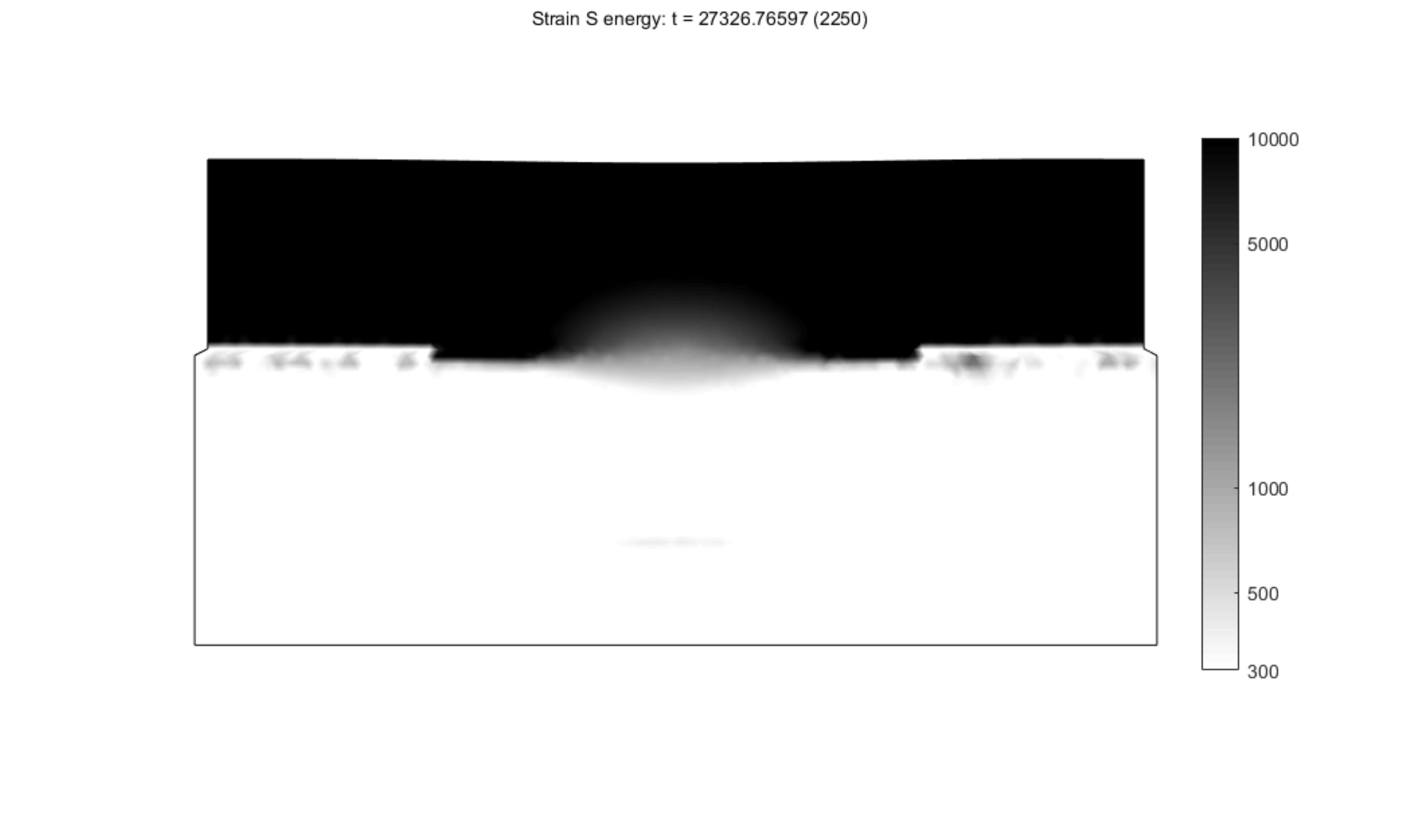}
\\
\rotatebox[origin=lt]{90}{\parbox{2cm}{\centering\COL{$t_2=t_1+7.30$\,s}}}\hspace*{-.2em}
\includegraphics[width=0.30\textwidth,bb=130 130 880 520,clip=true]{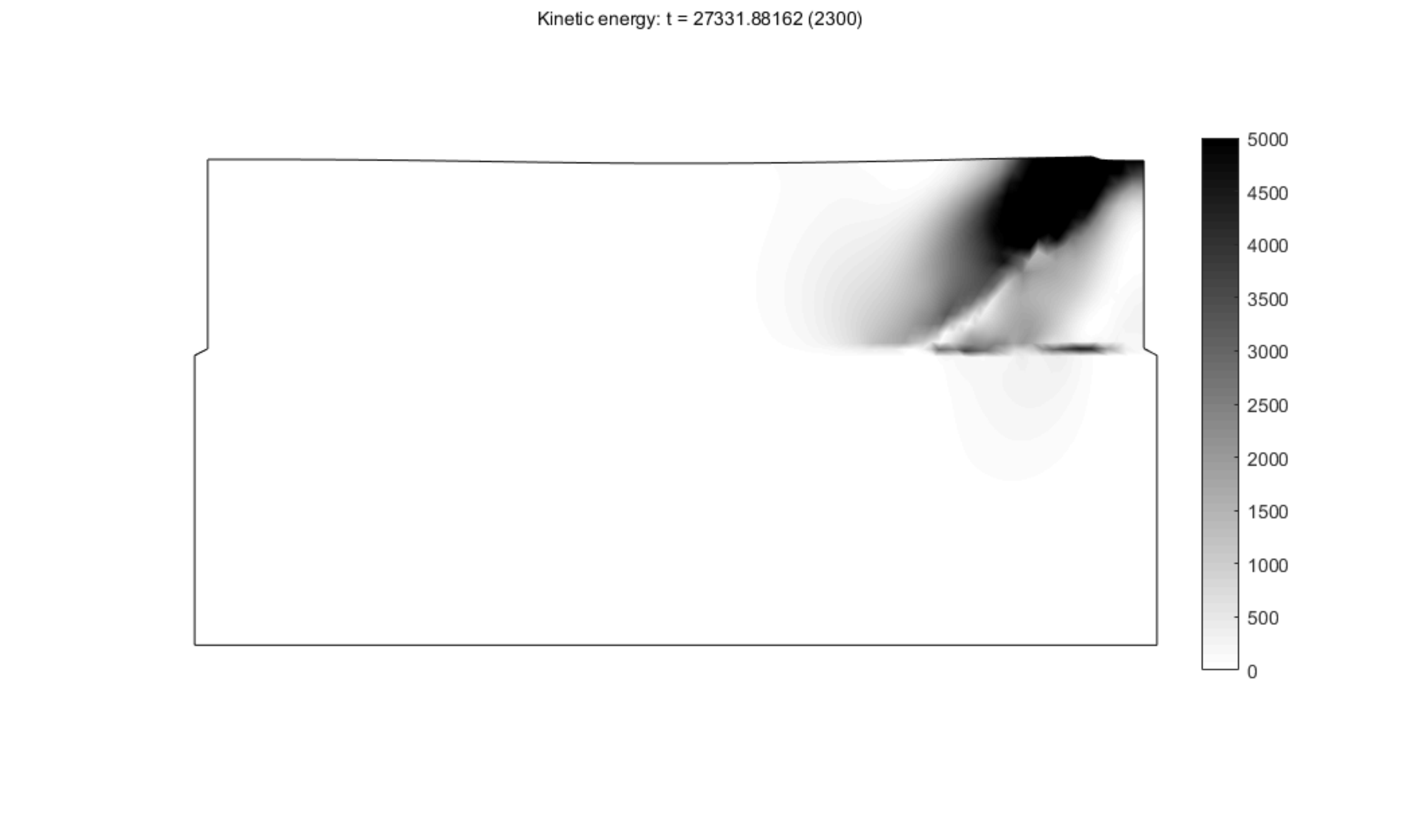}%\hspace*{.5em}
\includegraphics[width=0.30\textwidth,bb=130 130 880 520,clip=true]{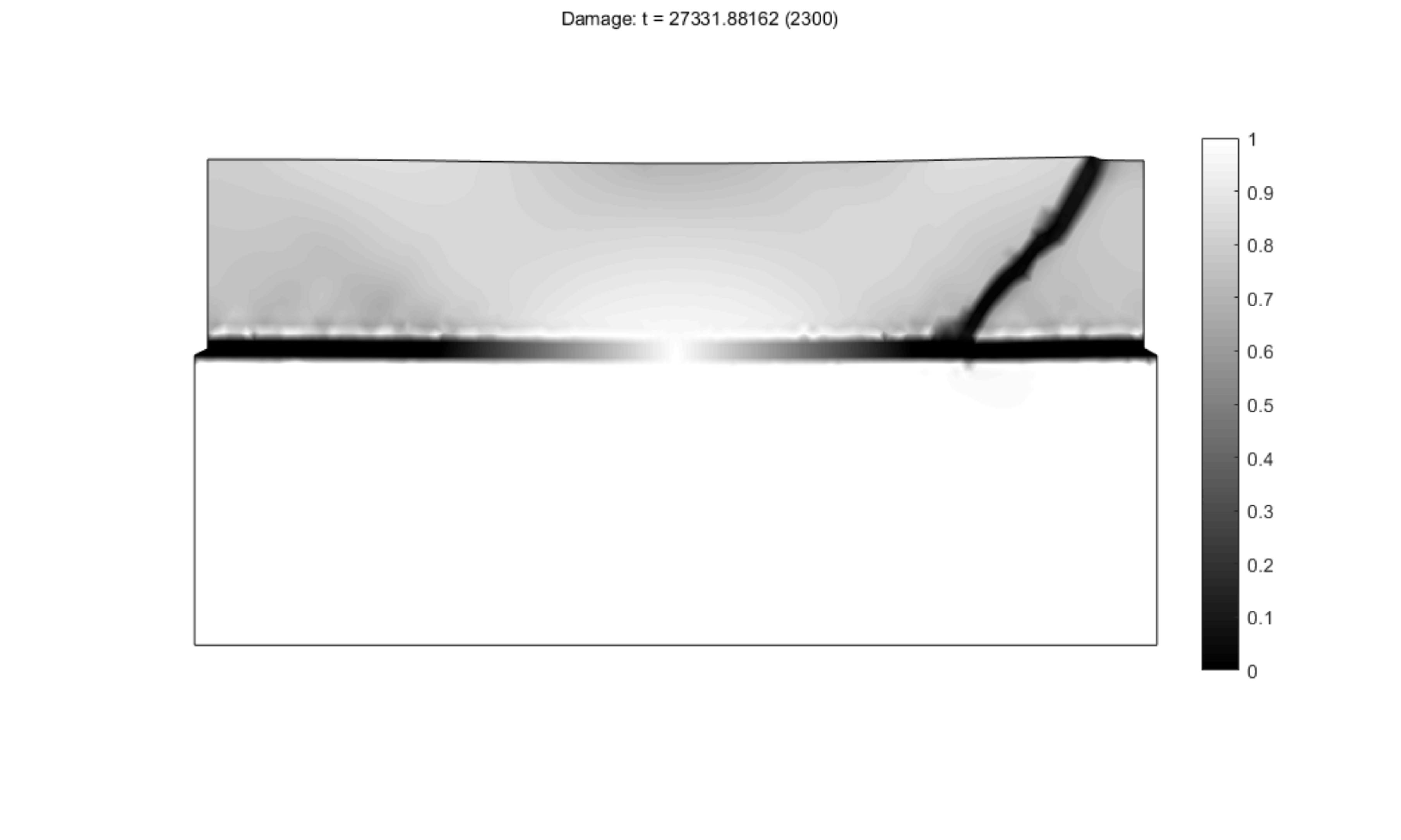}%\hspace*{.5em}
\includegraphics[width=0.30\textwidth,bb=130 130 880 520,clip=true]{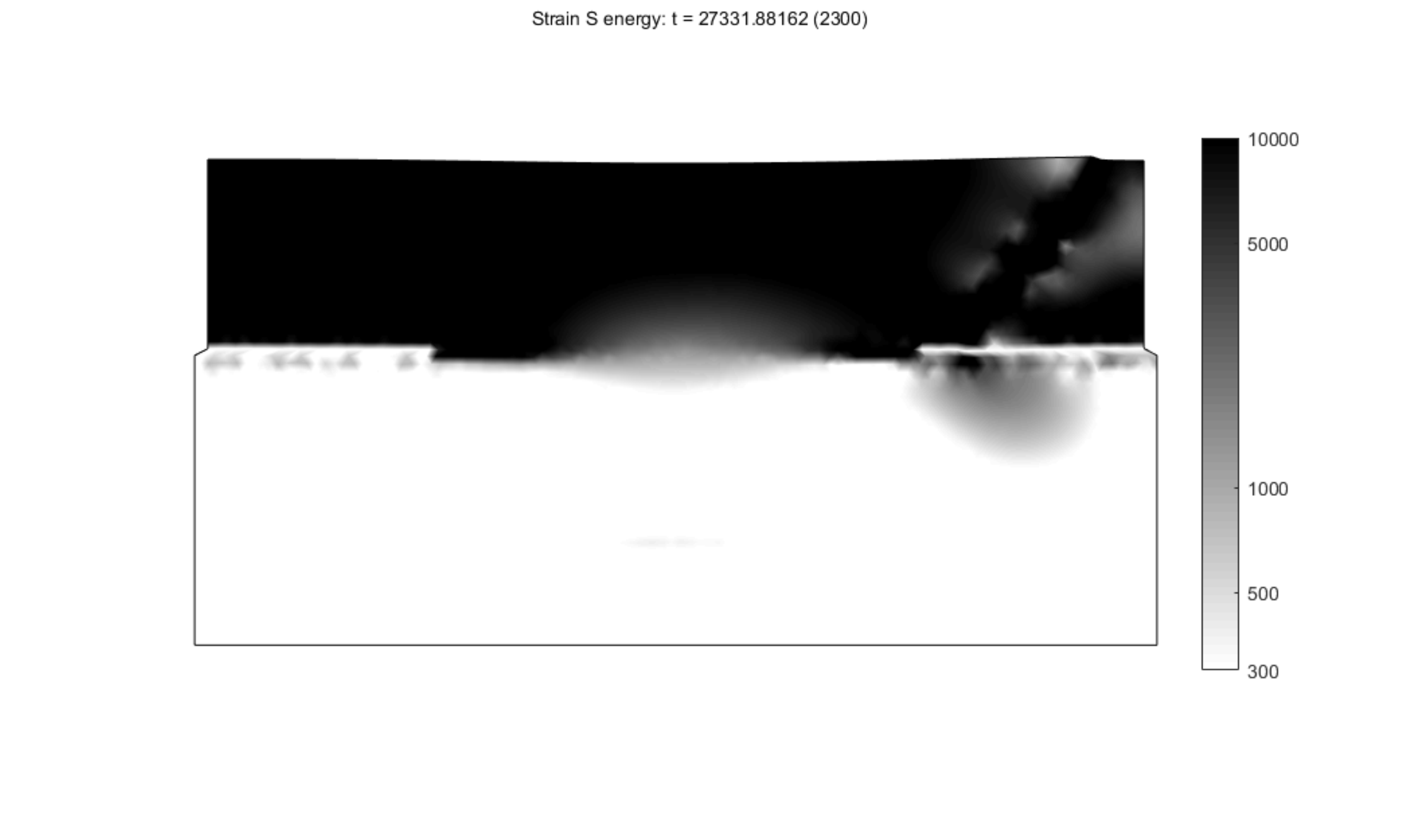}
\\
\rotatebox[origin=lt]{90}{\parbox{2cm}{\centering\COL{$t_3=t_1+24.82$\,s}}}\hspace*{-.2em}
\includegraphics[width=0.30\textwidth,bb=130 130 880 520,clip=true]{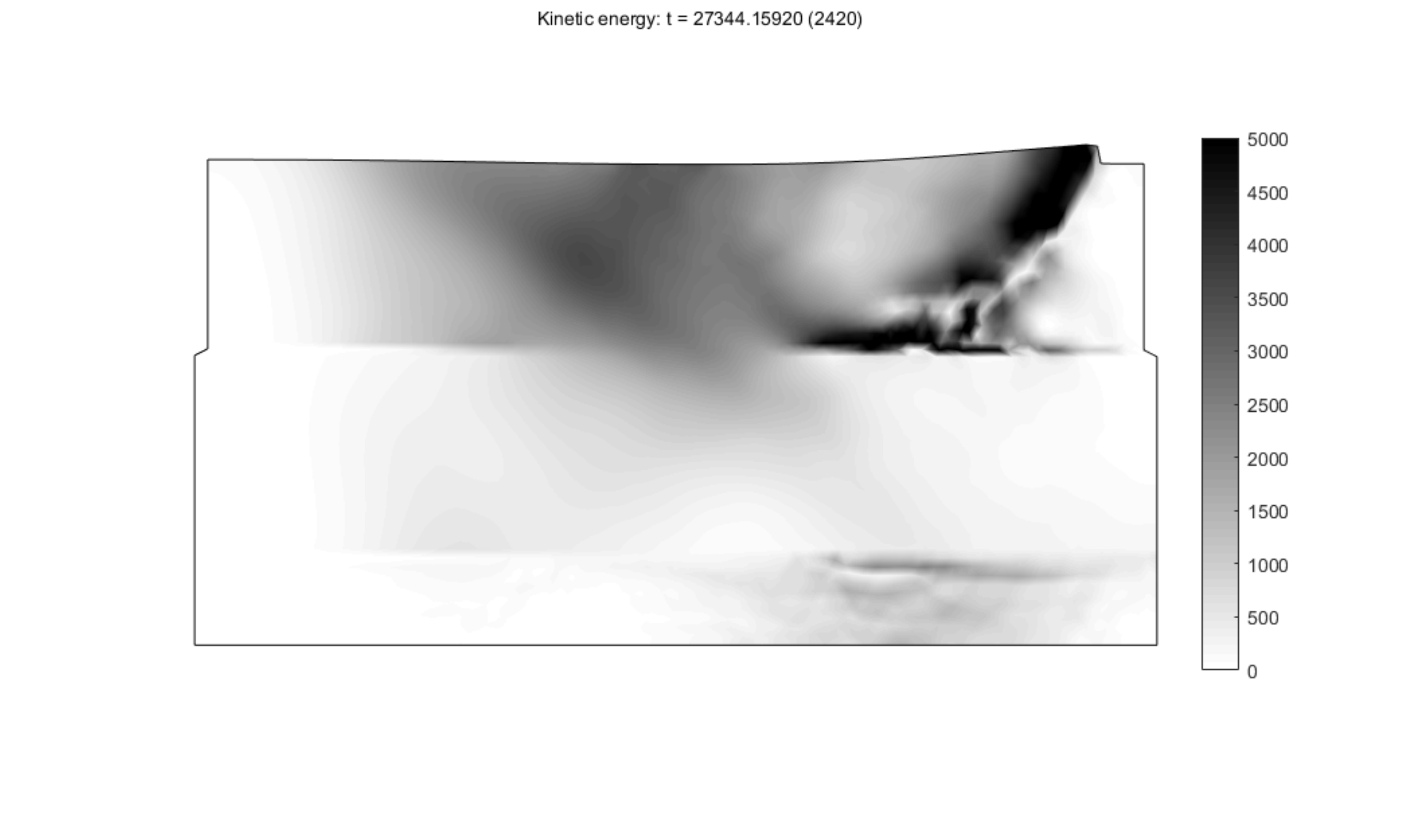}%\hspace*{.5em}
\includegraphics[width=0.30\textwidth,bb=130 130 880 520,clip=true]{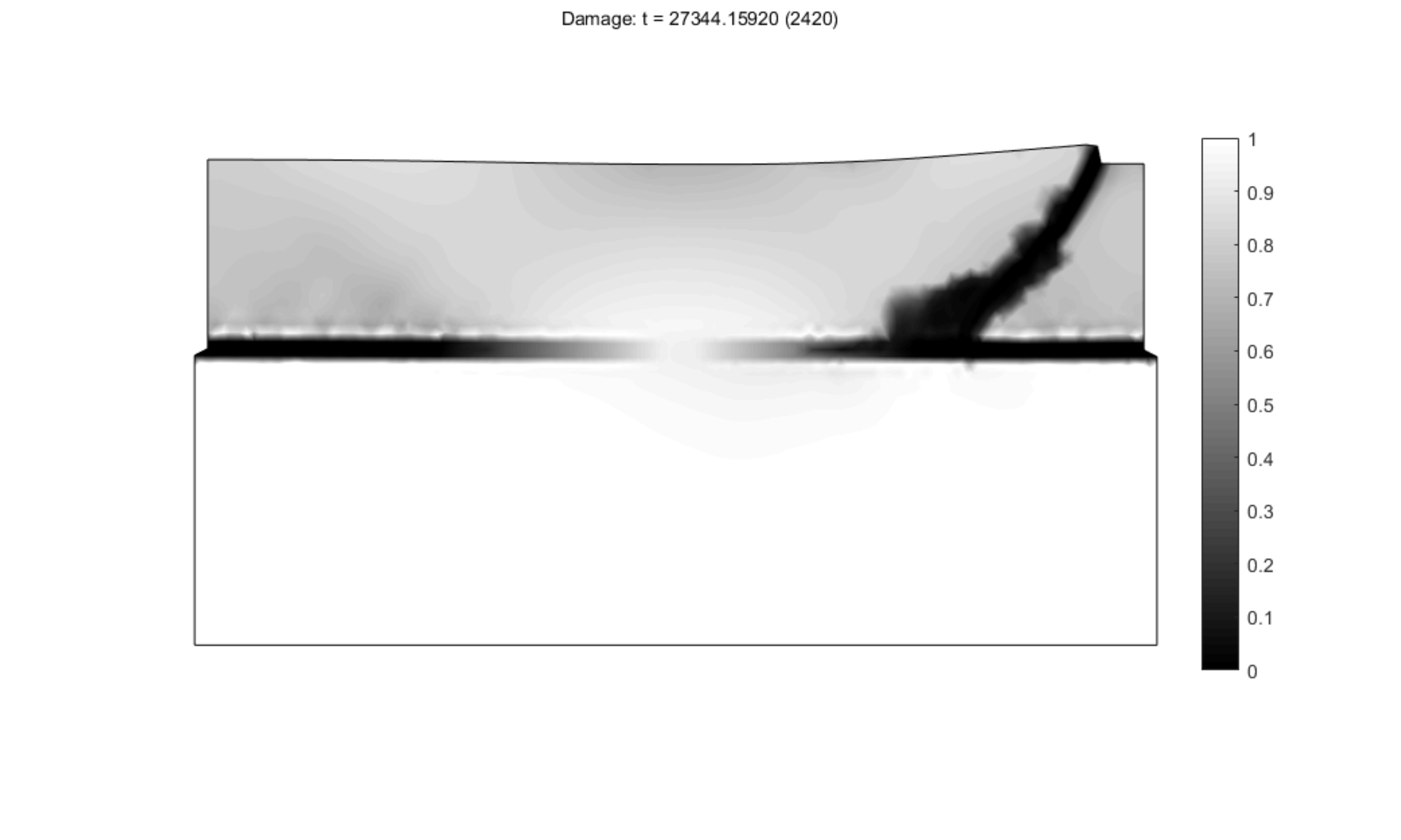}%\hspace*{.5em}
\includegraphics[width=0.30\textwidth,bb=130 130 880 520,clip=true]{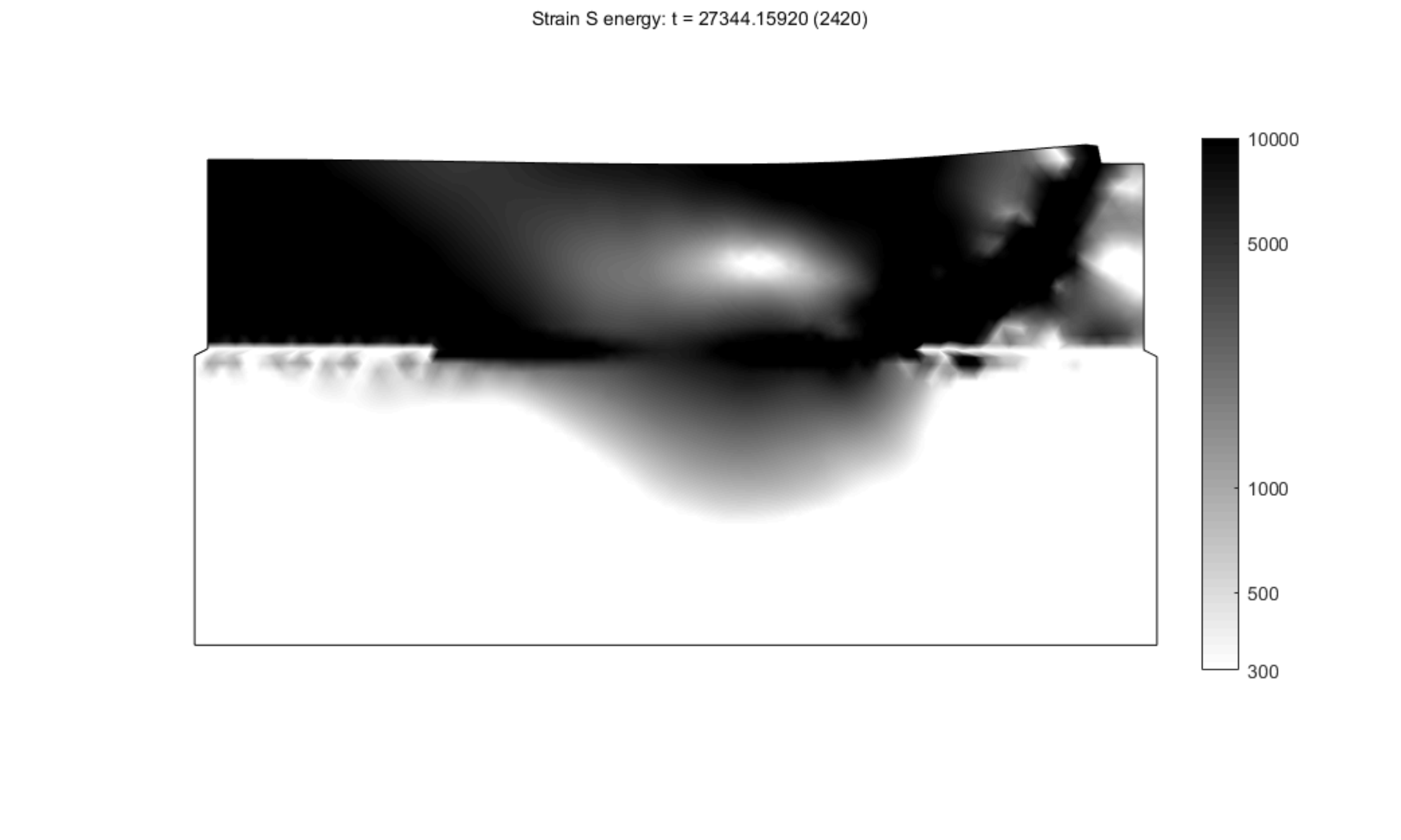}
\\
\rotatebox[origin=lt]{90}{\parbox{2cm}{\centering\COL{$t_4=t_1+33.58$\,s}}}\hspace*{-.2em}
\includegraphics[width=0.30\textwidth,bb=130 130 880 520,clip=true]{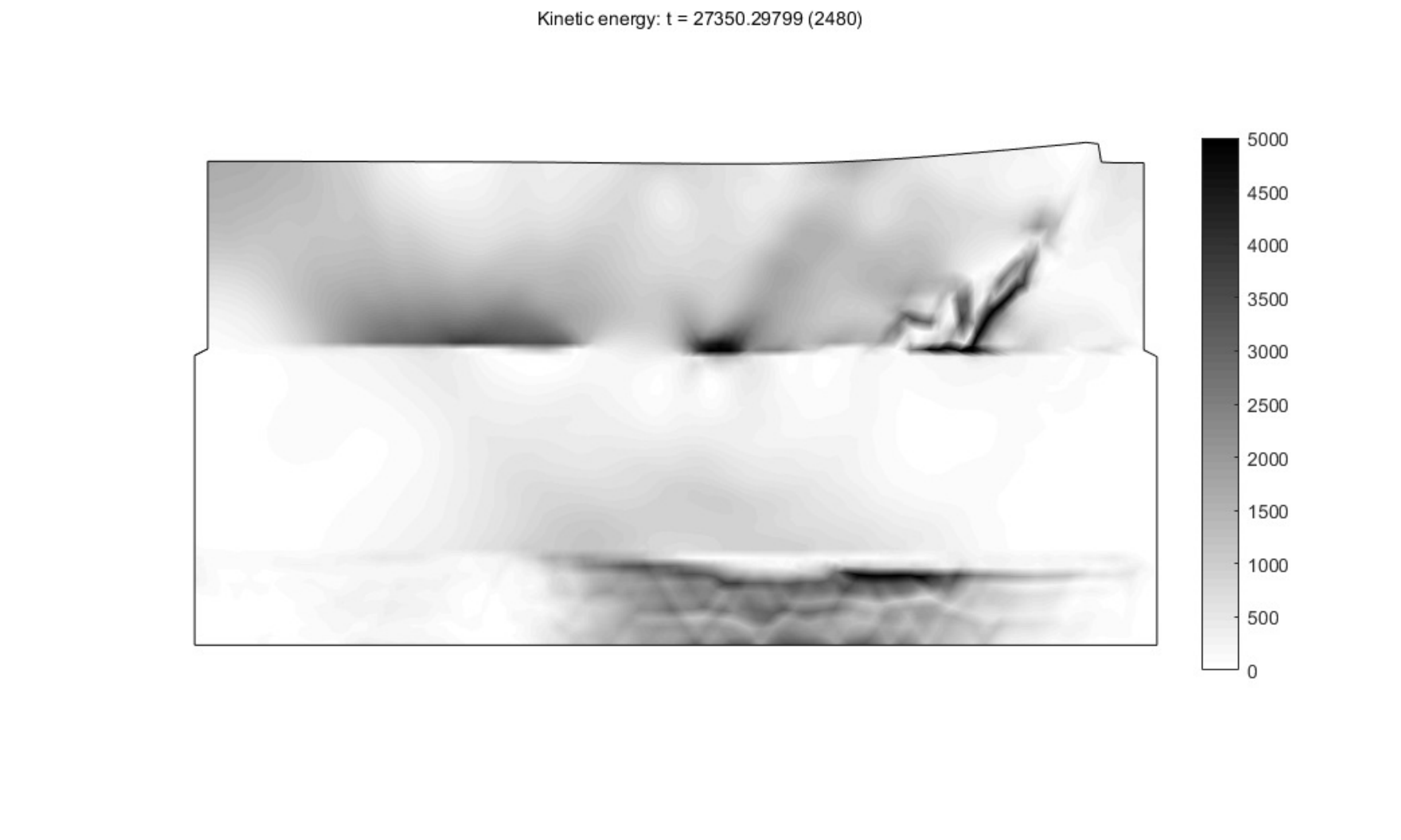}%\hspace*{.5em}
\includegraphics[width=0.30\textwidth,bb=130 130 880 520,clip=true]{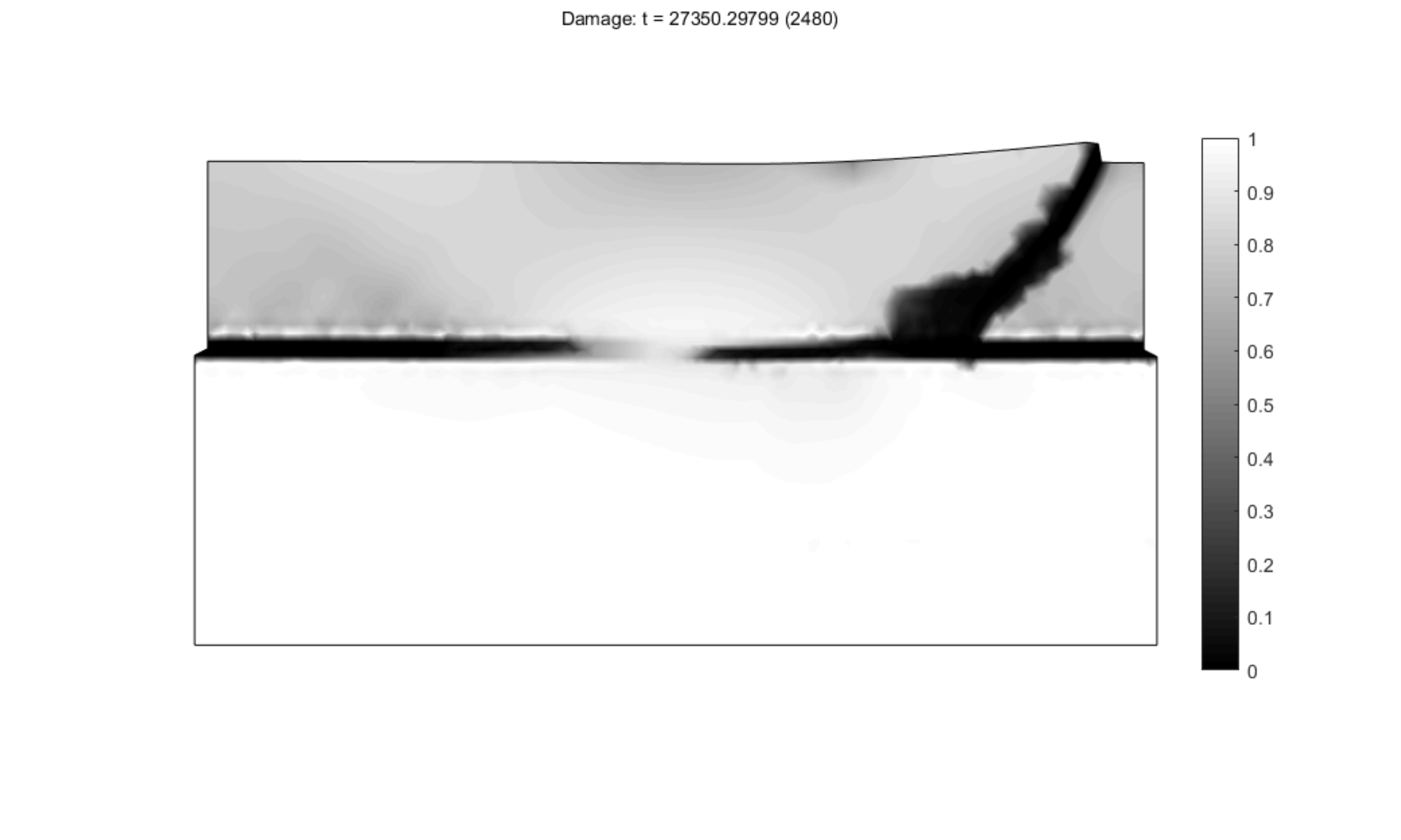}%\hspace*{.5em}
\includegraphics[width=0.30\textwidth,bb=130 130 880 520,clip=true]{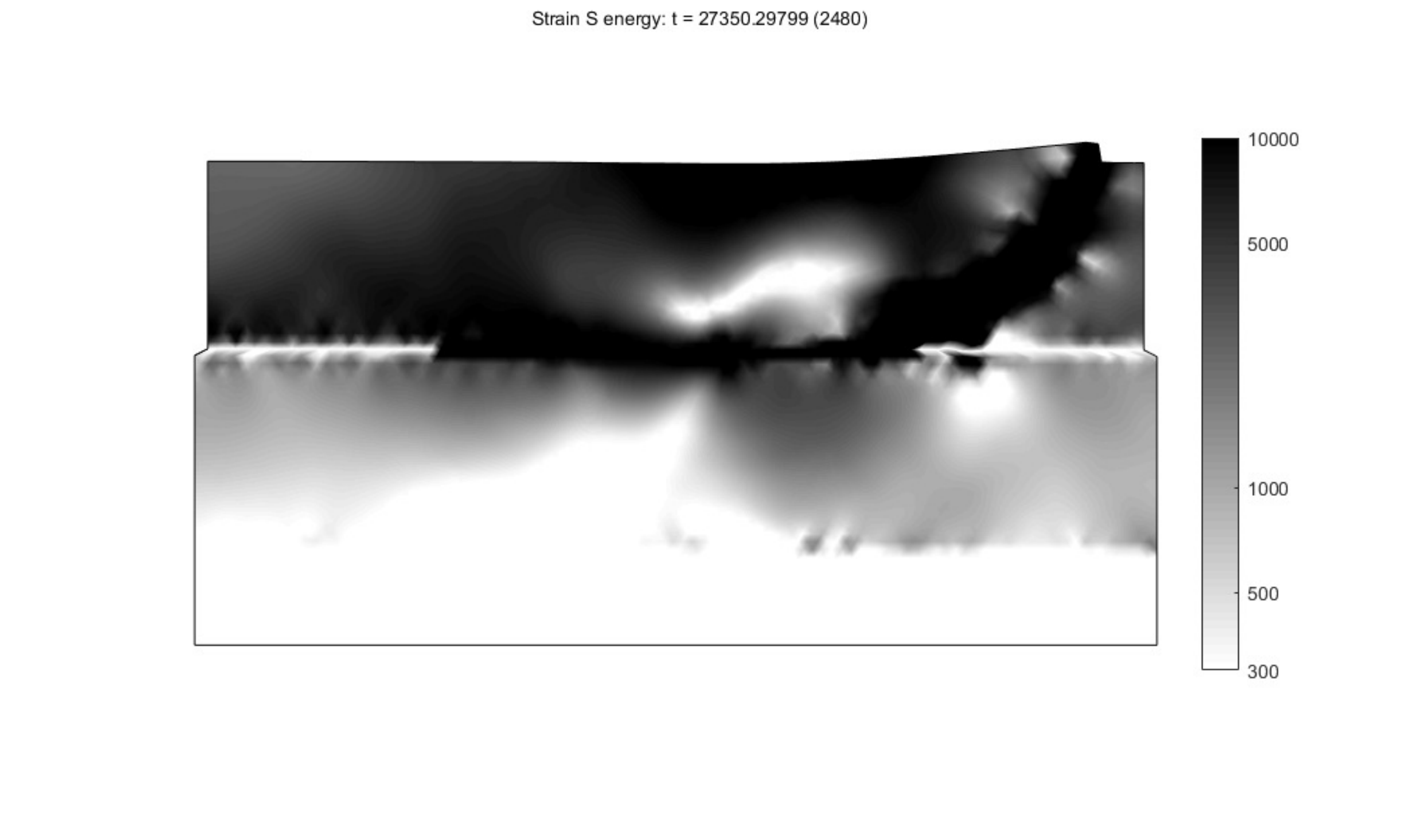}
\\
\rotatebox[origin=lt]{90}{\parbox{2cm}{\centering\COL{$t_5=t_1+37.96$\,s}}}\hspace*{-.2em}
\includegraphics[width=0.30\textwidth,bb=130 130 880 520,clip=true]{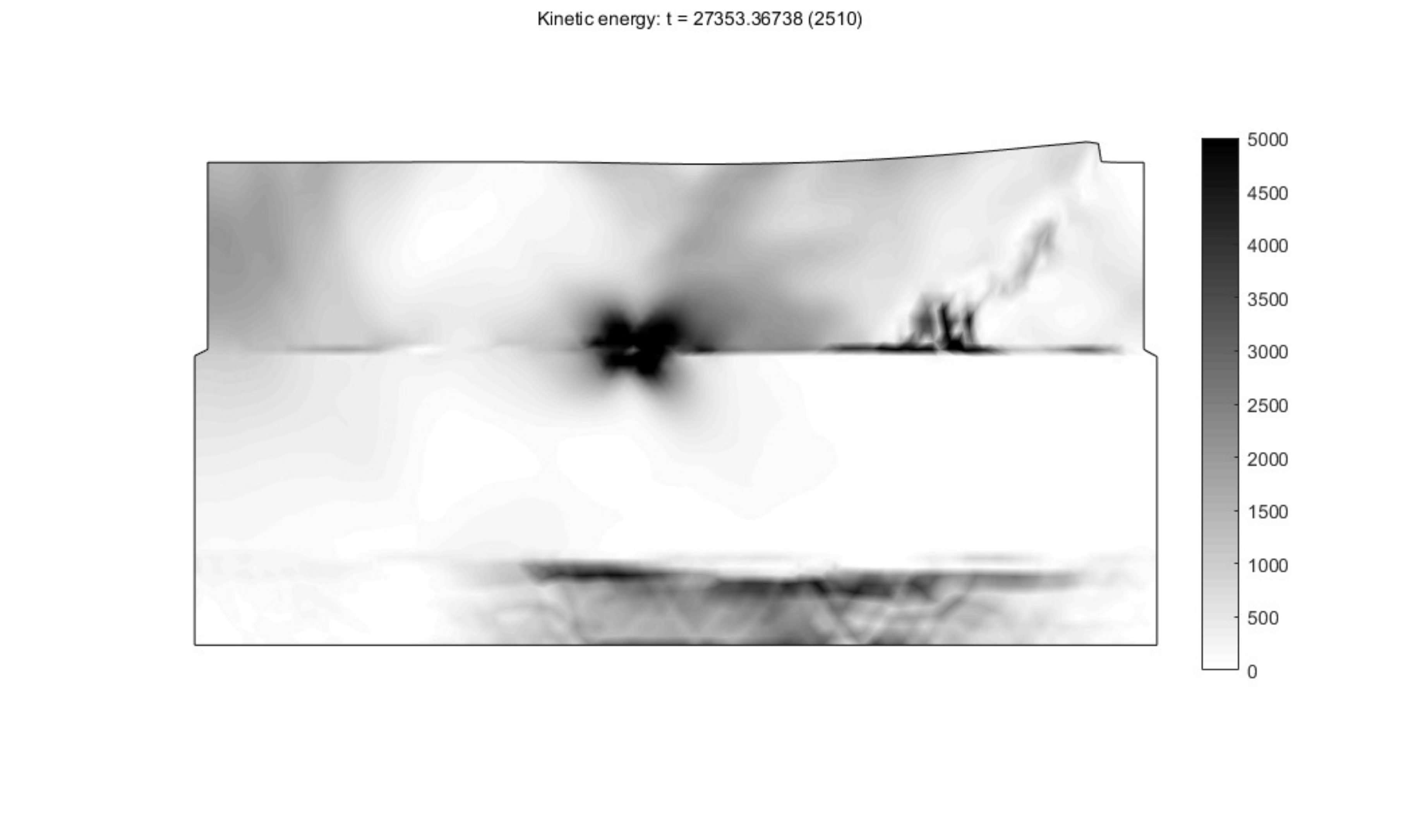}%\hspace*{.5em}
\includegraphics[width=0.30\textwidth,bb=130 130 880 520,clip=true]{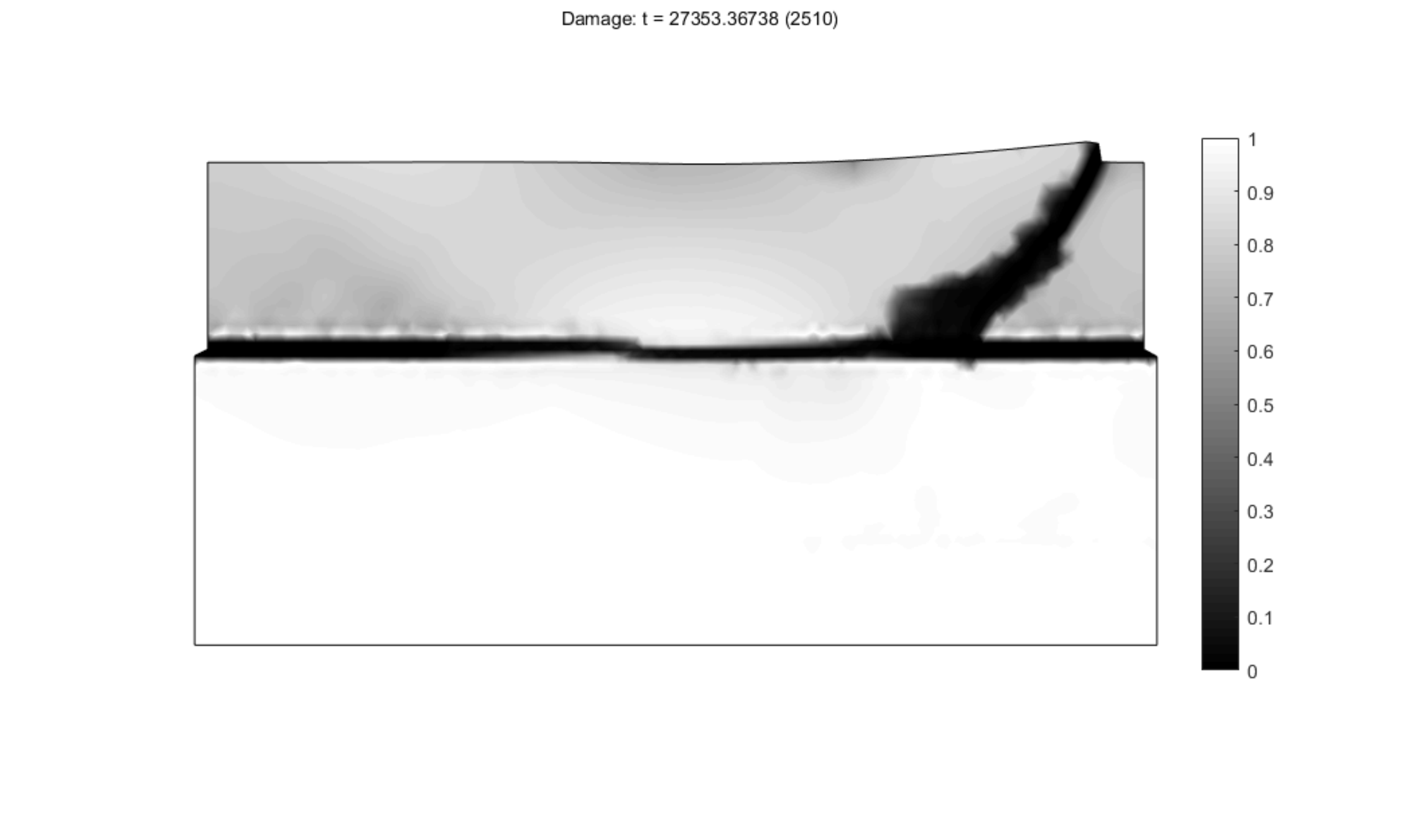}%\hspace*{.5em}
\includegraphics[width=0.30\textwidth,bb=130 130 880 520,clip=true]{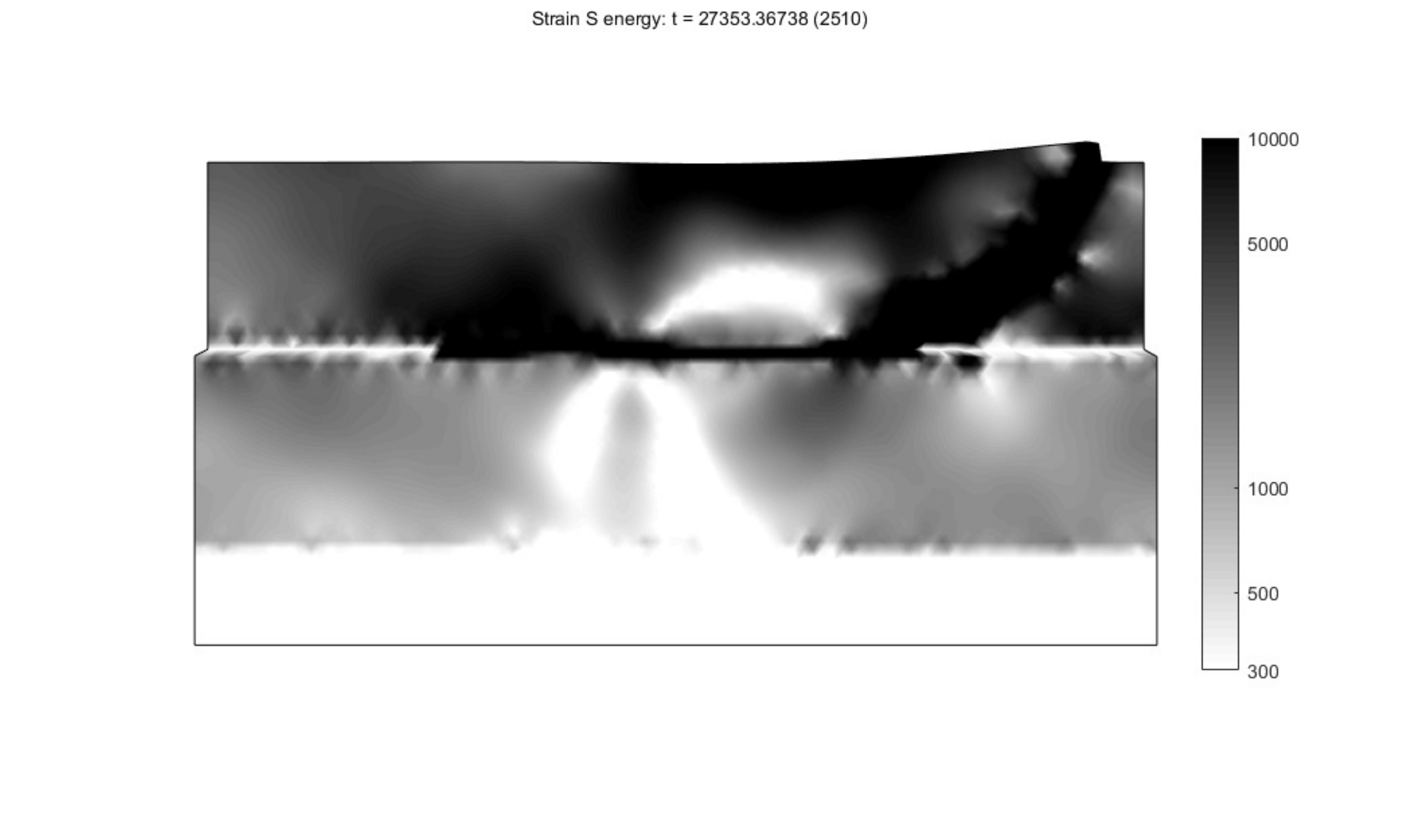}
\\
\rotatebox[origin=lt]{90}{\parbox{2cm}{\centering$t_6=t_1+40.88$\,s}}\hspace*{-.2em}
\includegraphics[width=0.30\textwidth,bb=130 130 880 520,clip=true]{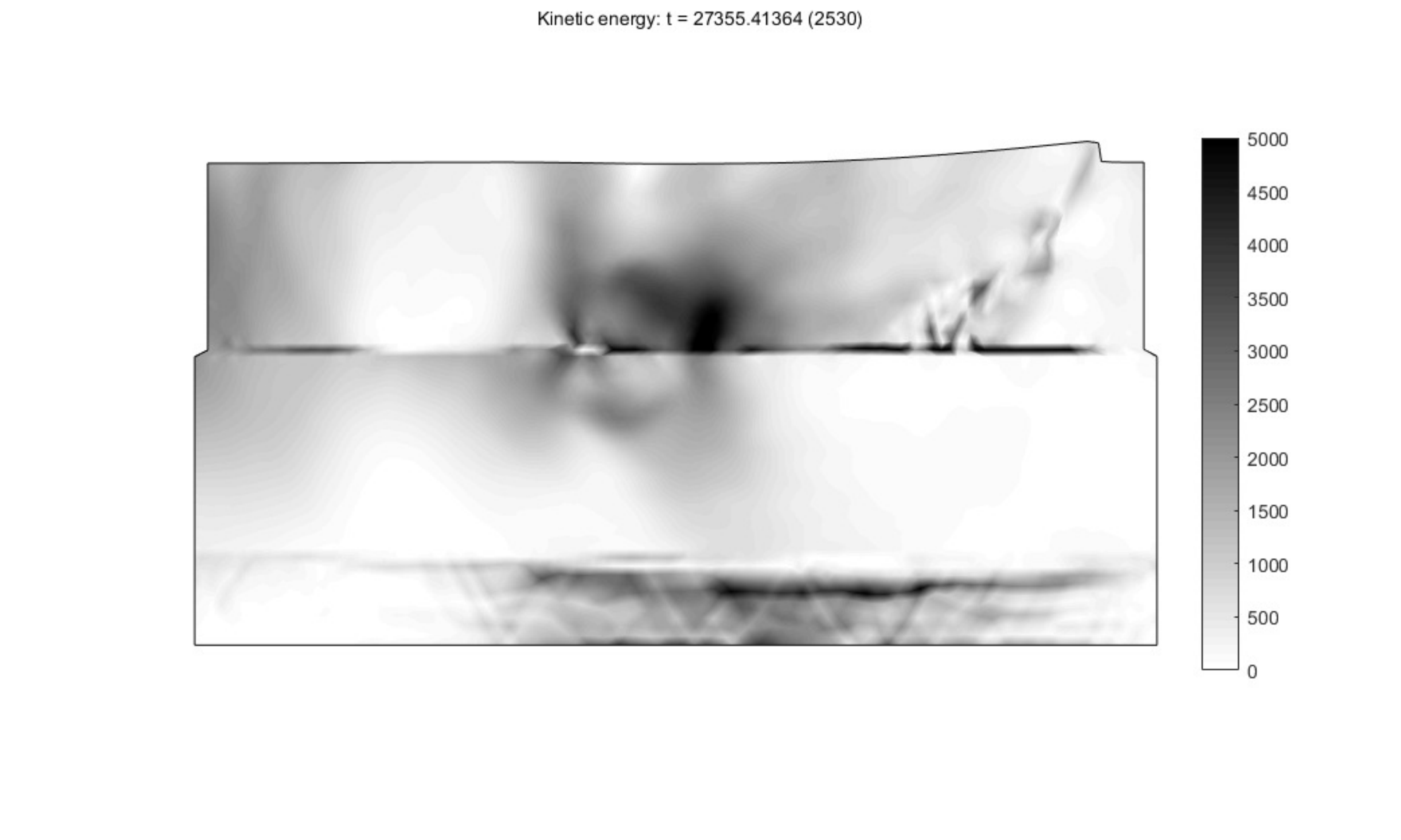}%\hspace*{.5em}
\includegraphics[width=0.30\textwidth,bb=130 130 880 520,clip=true]{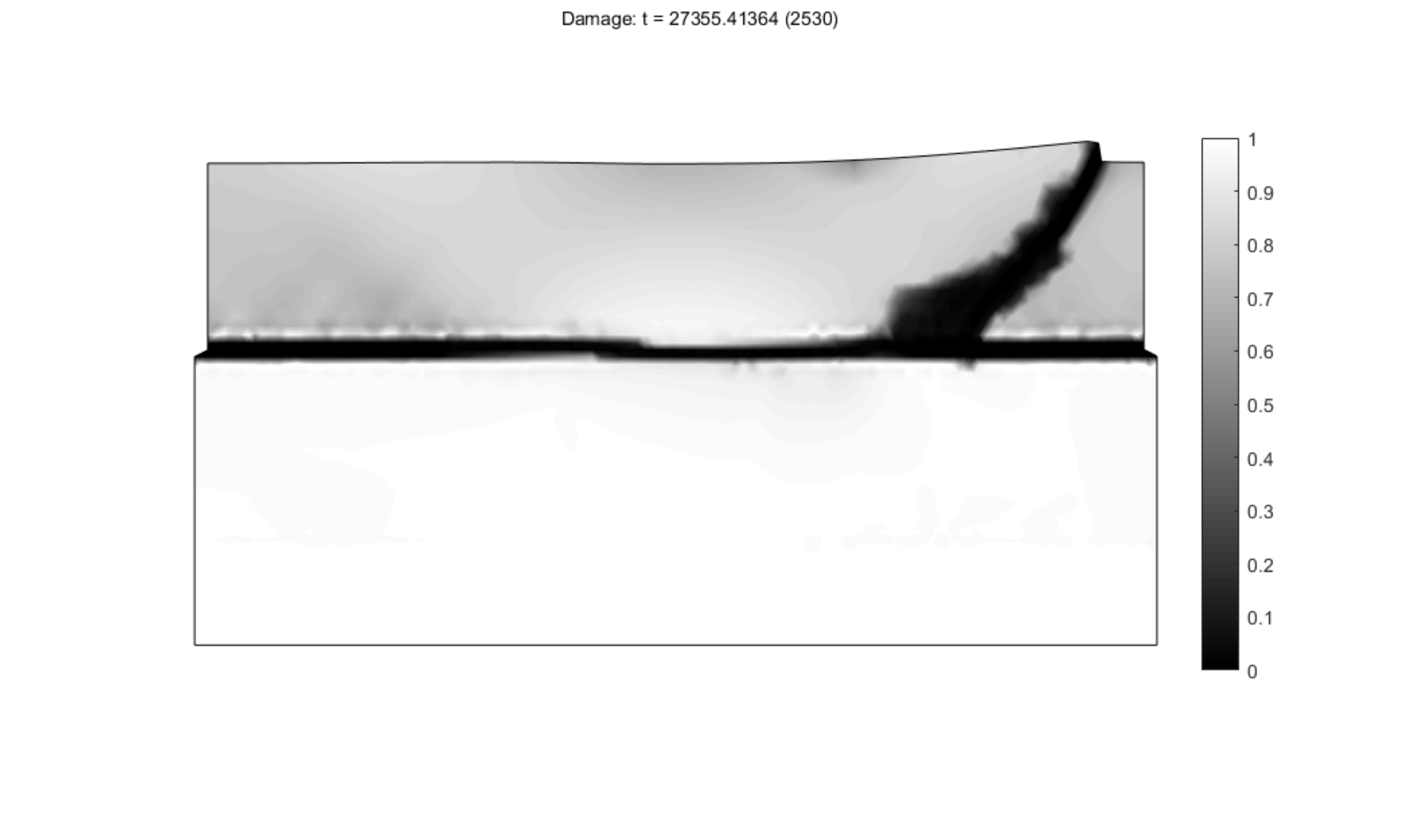}%\hspace*{.5em}
\includegraphics[width=0.30\textwidth,bb=130 130 880 520,clip=true]{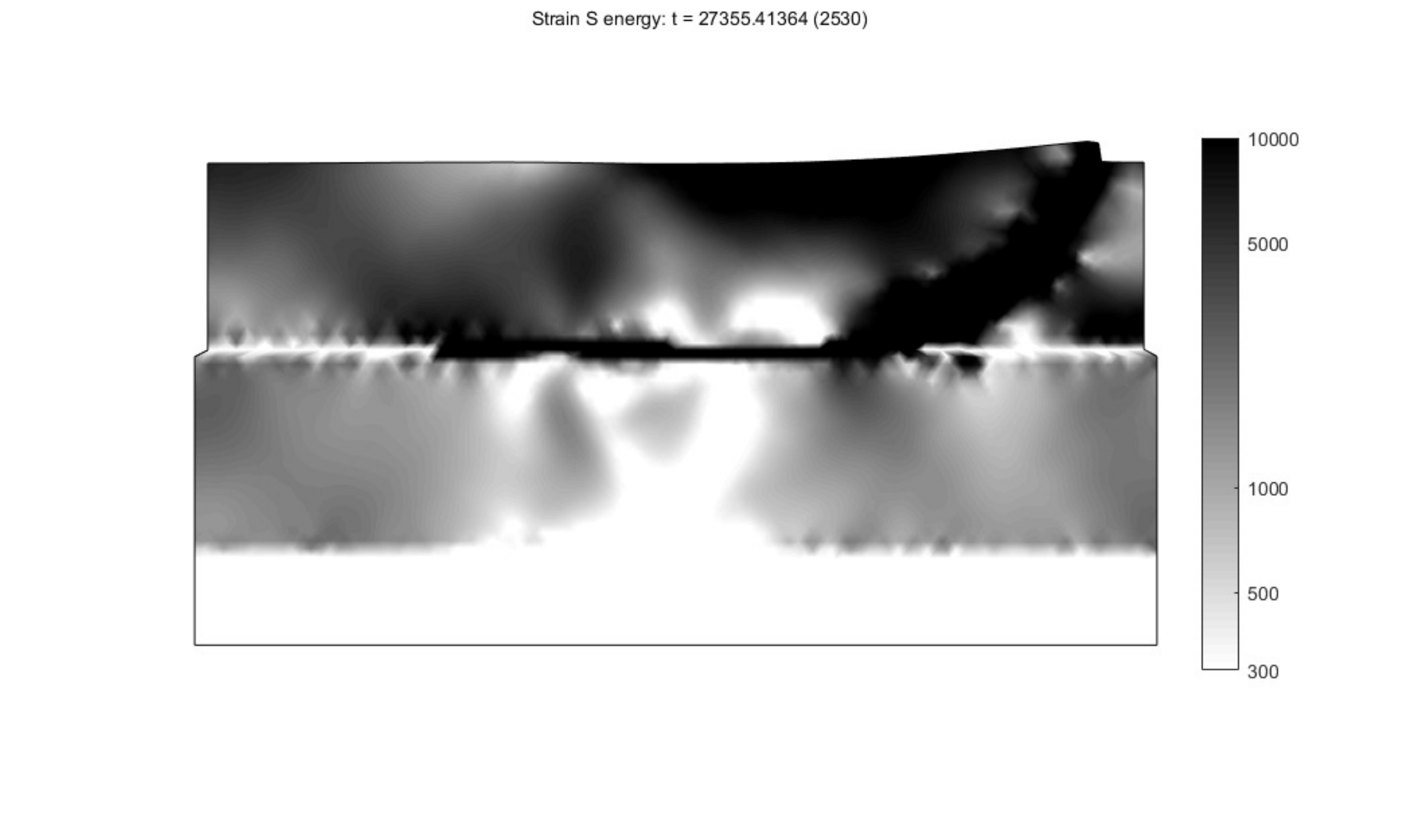}
\\
\rotatebox[origin=lt]{90}{\parbox{2cm}{\centering\COL{$t_7=t_1+43.80$\,s}}}\hspace*{-.2em}
\includegraphics[width=0.30\textwidth,bb=130 130 880 520,clip=true]{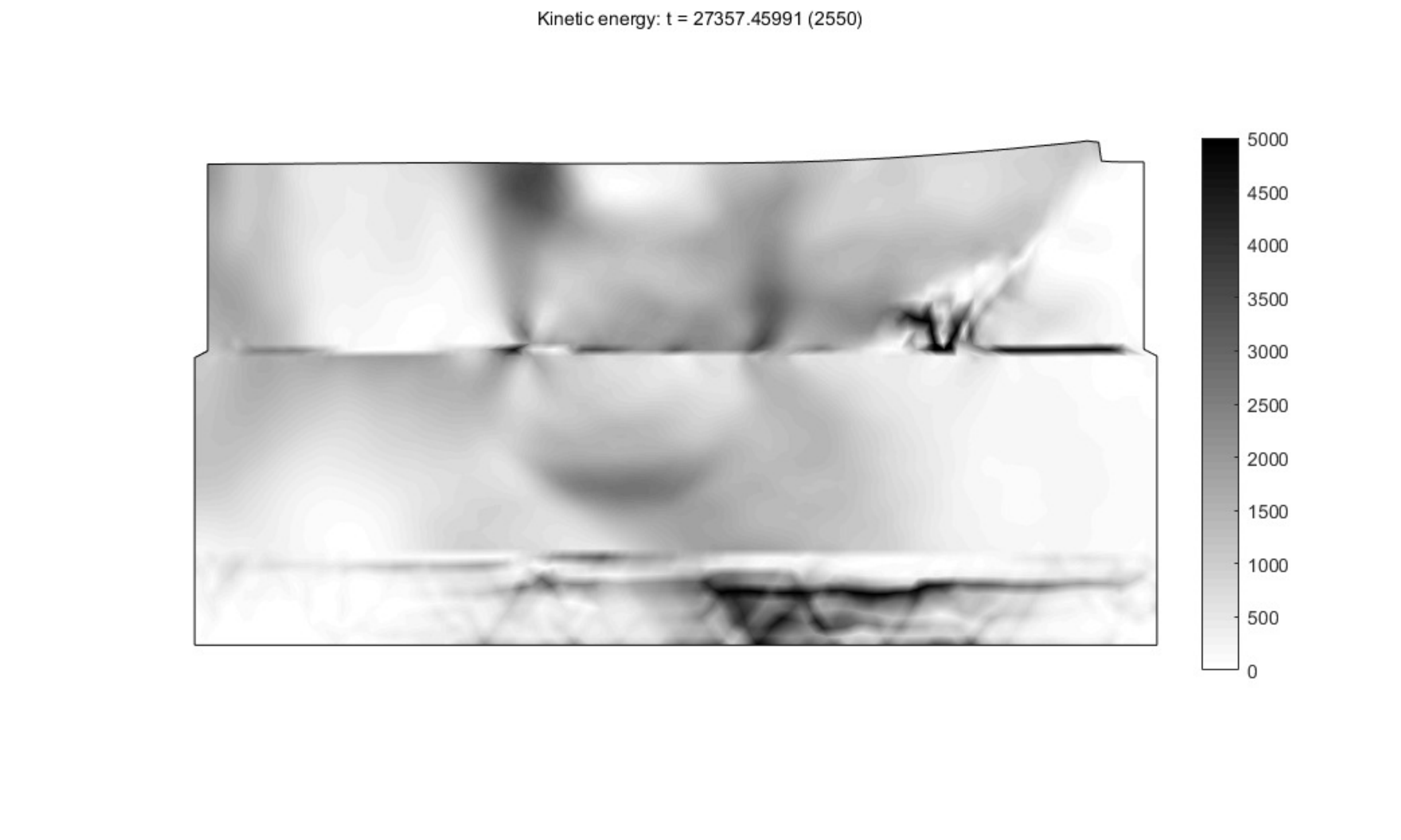}%\hspace*{.5em}
\includegraphics[width=0.30\textwidth,bb=130 130 880 520,clip=true]{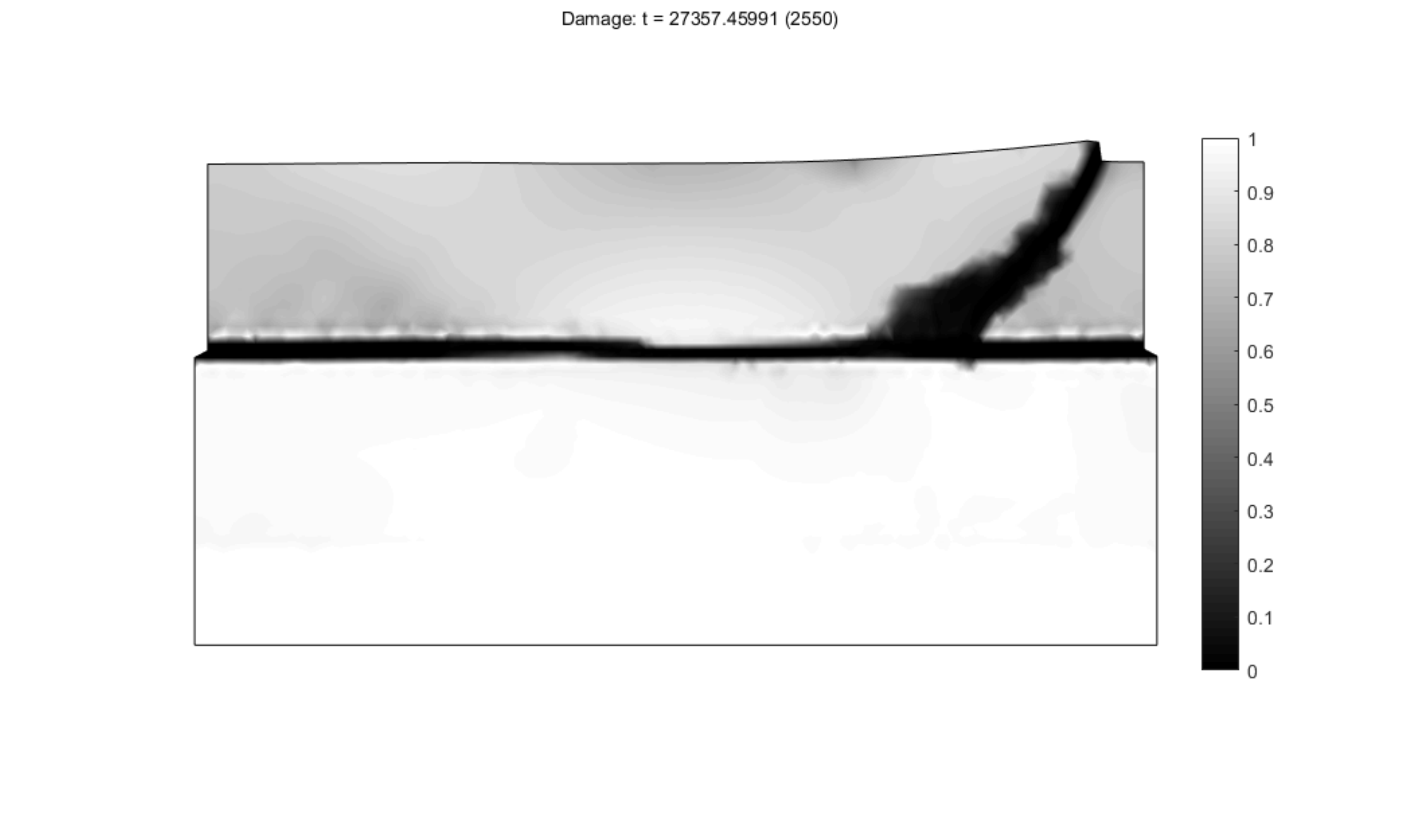}%\hspace*{.5em}
\includegraphics[width=0.30\textwidth,bb=130 130 880 520,clip=true]{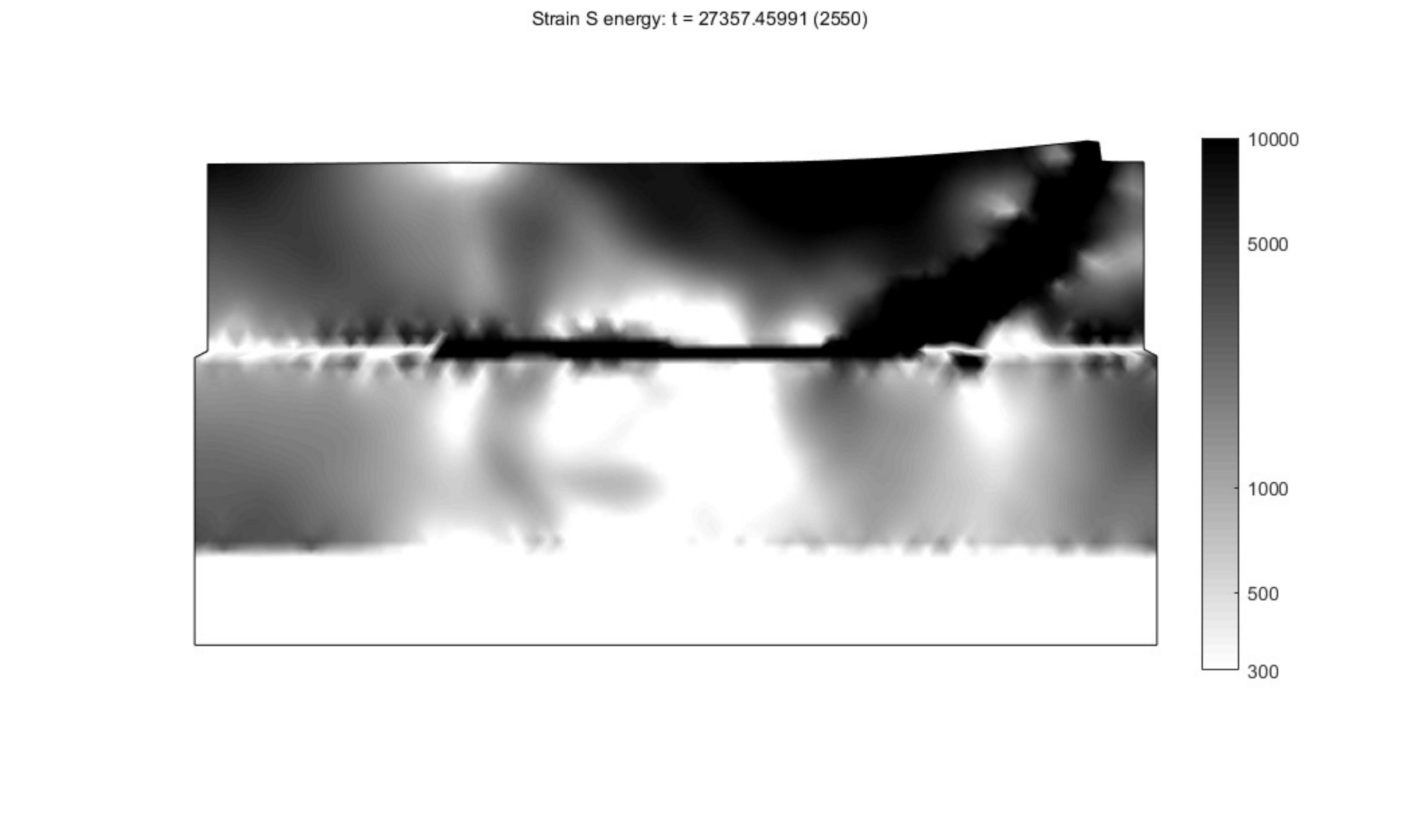}
\\
\rotatebox[origin=lt]{90}{\parbox{2cm}{\centering\COL{$t_8=t_1+49.64$\,s}}}\hspace*{-.2em}
\includegraphics[width=0.30\textwidth,bb=130 130 880 520,clip=true]{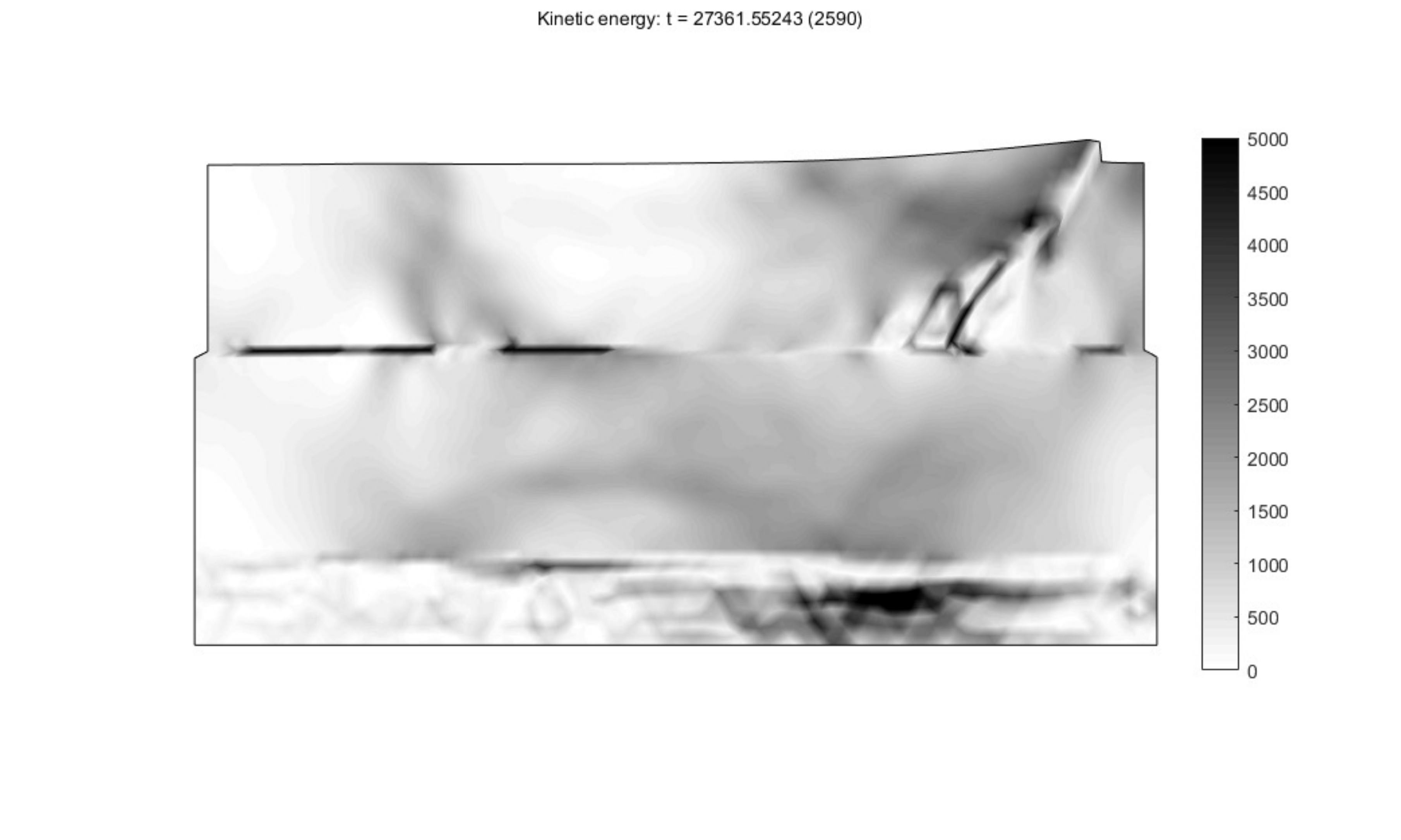}%\hspace*{.5em}
\includegraphics[width=0.30\textwidth,bb=130 130 880 520,clip=true]{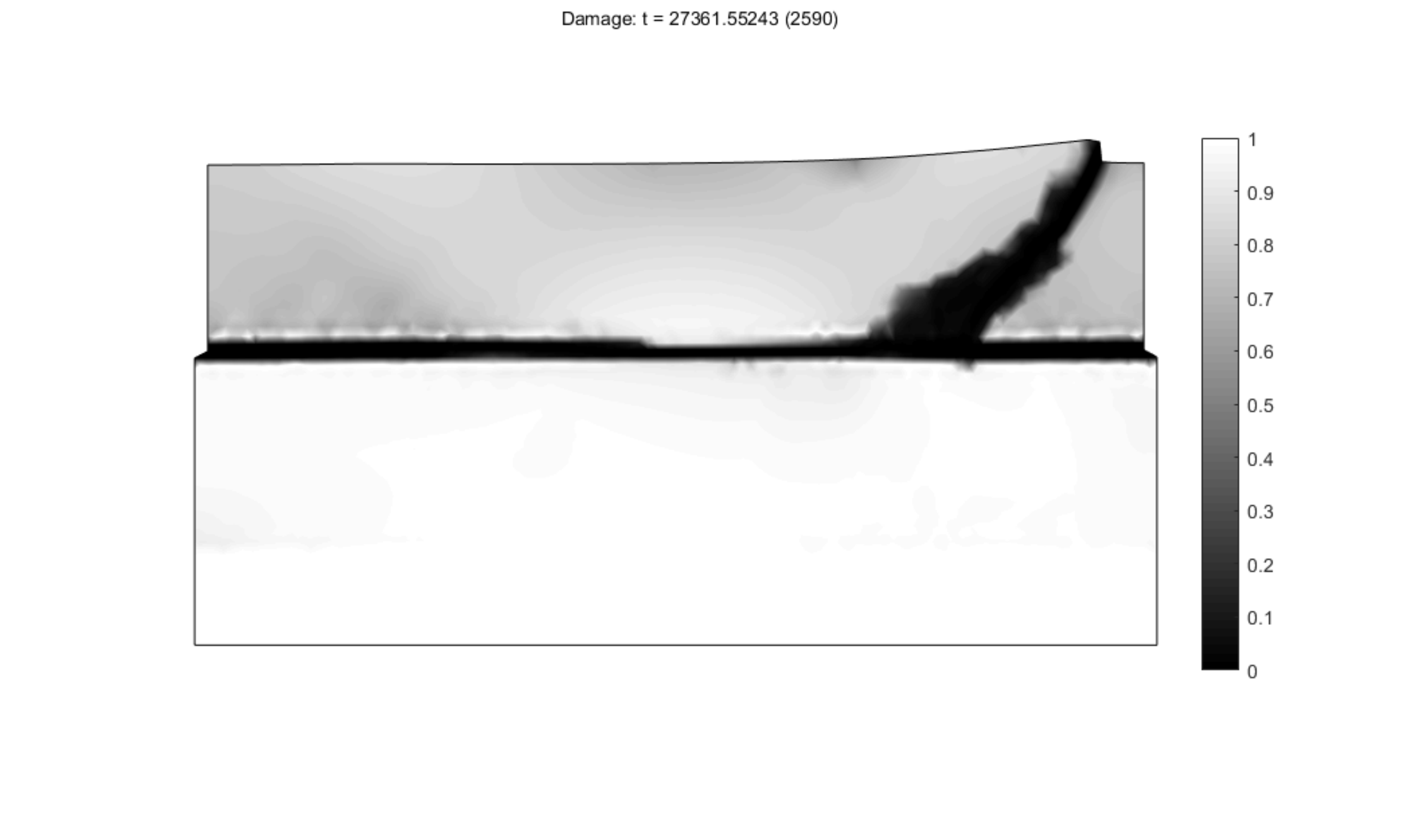}%\hspace*{.5em}
\includegraphics[width=0.30\textwidth,bb=130 130 880 520,clip=true]{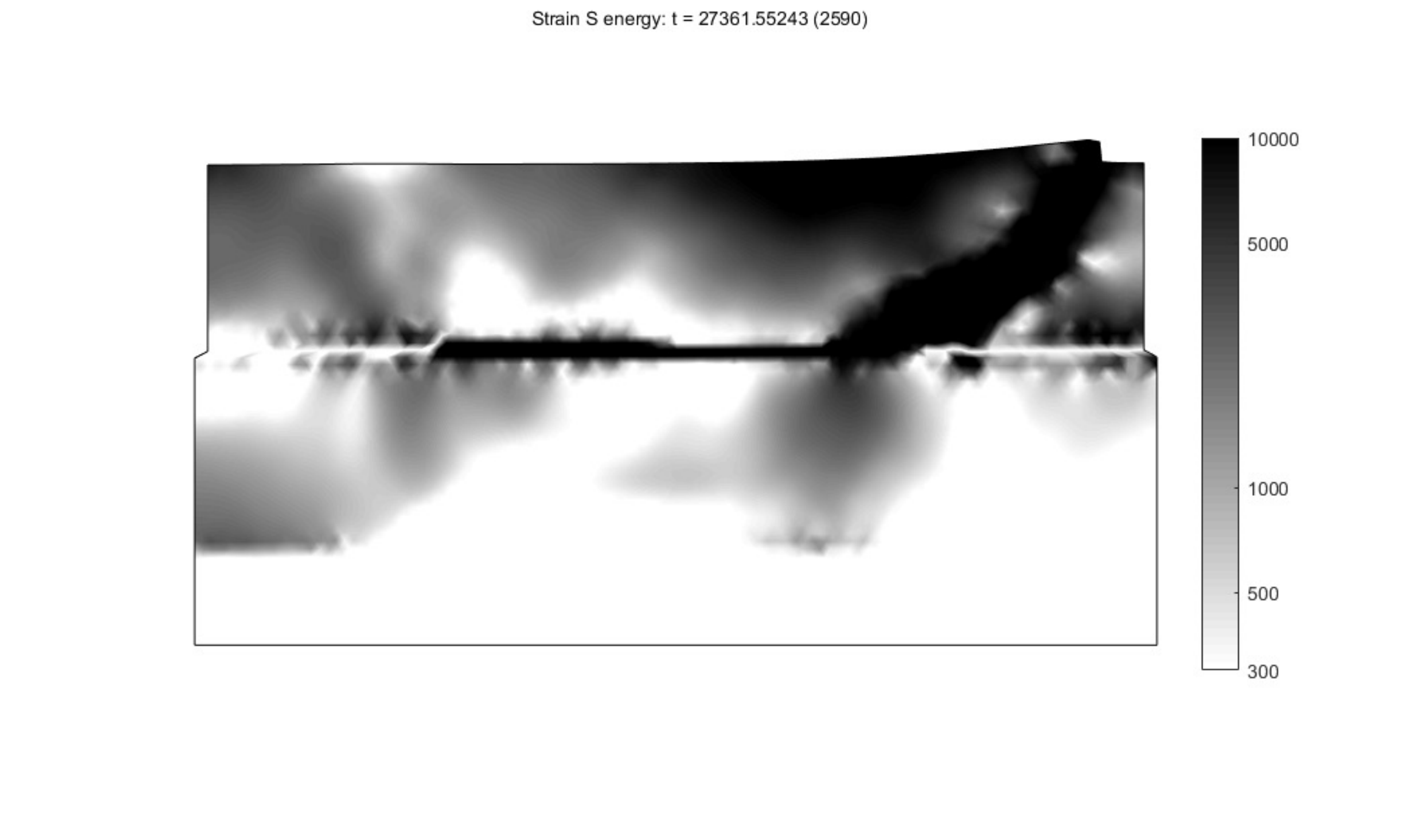}
\caption{\sl\COL{Simulations of a birth of a new reverse (thrust) 
fault %in a reverse position 
in combination with a
%nother 
pre-existing strike-slip fault 
%in strike-slip position
rupturing with a certain delay, together with emission of seismic 
(mainly S-) waves during completion of these two subsequenct ruptures, 
their propagation and creation of rather strong P-waves in the fluidic domain. 
The displacement again magnified 100$\times$ and 
the new mountain build during the reverse fault birth is visible.}}
\label{fig-reverse+}
\end{figure*}
}

\section{Modifications or extensions}\label{sec-ext}
%      ~~~~~~~~~~~~~~~~~~~~~~~~~~~~~

The mentioned academical character of the presented, rather simple model
can be suppressed \COL{when enhancing it } by various ways.

\subsection{Towards easier earthquake nucleation}
%           ~~~~~~~~~~~~~~~~~~~~~~~~~~~~~~~~~~~~~
It is a well recognized effect that the Griffith fracture model, 
which is approximated by the ansatz \eqref{AT-ansatz} if $\eps\to0$, 
is realistic as far as fracture (cracks) propagation but has difficulties
with nucleation of cracks. 

This effect can be seen when calculating the damage driving force 
$\varphi_\alpha'$ used in \eqref{AT-ansatz}, %for $\alpha=1$, 
obtaining $G_{_{\rm E}}'(\alpha)|e_{\rm dev}|^2+\gamma_{_{\rm DAM}}'(\alpha)=
2(1{-}\alpha)G_{_{\rm E0}}|e_{\rm dev}|^2+G_{\rm c}
%(1{-}\alpha)
\alpha/\eps$. Counting with the shear stress $\sigma_{\rm dev}:=
{\rm dev}\,\varphi_e'=2G_{_{\rm E}}(\alpha)e_{\rm dev}$, one can see that 
the stress needed for triggering damage (rupture) indeed growths like 
$1/\sqrt\eps$ for $\eps\to0$, which causes the mentioned drawback of
the limit Griffith-type model. 

One way to get rid of this drawback is to consider,  
instead of $G_{_{\rm E}}(\alpha)=({\eps^2}/{\eps_0^2}{+}(1{-}\alpha)^2)G_{_{\rm E1}}$
used in \eqref{AT-ansatz}, a more general convex decreasing 
nonlinearity $G_{_{\rm E}}:[0,1]\to(0,\infty)$ depending on 
$\eps$ used in the crack surface density in \eqref{AT-ansatz}, cf.\ e.g.\ 
%Bourdin at al.\ 
\cite{BMMS14MPCC}. When requiring $G_{_{\rm E}}'(0)$ to blow up like 
$1/\eps$, one can still consider $\eps>0$ small to imitate the 
narrow crack zones while keeping both nucleation and propagation
under control through two mutually independent parameters $G_{_{\rm E}}'(0)$ 
and $G_{\rm c}$. Such {\it softening} may lead to localization of damage 
even without considering the Ambrosio-Tortorelli model, as shown 
by \cite{BazJir96SIDL}.

Another worthy modification which facilitates nucleation of rupture 
is to make $e_{\rm el}=e{-}\pi\mapsto\varphi(e,\pi,\alpha)$ nonconvex 
if $\alpha$ is close to 1, reflecting the lost of stability of the
rock when damage is sufficiently developed. The 2-homogeneous ansatz 
suggested by \cite{LyaMya84BECS} and used often in geophysical models is 
\begin{align}\label{Vladimir}
\varphi(e,\pi,\alpha)=\frac{\lambda_{_{\rm E}}}2|{\rm sph}\,e|^2+G_{_{\rm E}}(\alpha)|e{-}\pi|^2
-\alpha\gamma_{_{\rm R}}|e{-}\pi|\,{\rm sph}\,e
\end{align}
with $\lambda_{_{\rm E}}>0$ the first Lam\'e constant (fixed)
%($K=\lambda+\frac23G$),
and with the elastic shear modulus \COL{ given by an affine relation $G_{_{\rm E}}(\alpha)=G_0{-}\alpha G_{_{\rm R}}$ with undamaged shear modulus $G_0$ like in~\eqref{AT-ansatz} and $0<G_{_{\rm R}}<G_0$ sensitivity of the shear modulus to damage,}
while $\gamma_{_{\rm R}}>0$ makes it non-Hookean if the damage develops enough, 
cf.\ e.g.\ \cite{HaLyAg04CEDP,LHAB09NDRW,LyHaBZ11NLVE,LyaBZ14DBRM} and 
references therein. This leads 
to a nonlinear ill-posed hyperbolic problem which does not need to have 
any global solution in whatever sense. Its rigorous justification has 
been done in \cite{Roub17GMHF} only when one slightly modifies the ansatz 
\eqref{Vladimir} to reduce the fall of $\varphi$ to $-\infty$ and augment it, 
beside $\nabla\alpha$ and $\nabla\pi$
as in \eqref{AT-ansatz}, also by a strain-gradient term, 
referring sometimes as a concept of non-simple materials.
This may be related with an anomalous wave dispersion.  

In both mentioned modifications that can be even combined, $\varphi$ is no 
longer component-wise quadratic. Therefore, in contrast to \eqref{AT-ansatz}, 
the staggered scheme in Sect.\,\ref{sec-staggered} leads to non-quadratic 
minimization problems for which iterative solvers have to be used.
\COL{Cf.\ also \cite{Roub??MDDP} for a brief survey of various options 
to improve the previous model.}

\subsection{Towards other geophysical phenomena}\label{sec-more-geo}
%           ~~~~~~~~~~~~~~~~~~~~~~~~~~~~~~~~~~~
The set of internal variables $(\pi,\alpha)$ from Sect.\,\ref{sec-coupling}
is a very minimal scenario to hit basic phenomena in their simplest variant.
Actual modelling towards a geophysically relevant event should include more
internal variables.

Putting the model into full thermodynamical context involving 
temperature and heat transfer would allow to include, e.g., flash
heating within huge earthquakes as well as the popular Dieterich-Ruina-type
rate-and-state friction model on the fault, cf.\ \cite{Roub14NRSF}.
The energy-based ansatz \eqref{Biot} allows for 
%an extension for unisothermal situations which is simultaneously 
thermodynamical consistency, i.e.\ the total-energy conservation, 
nonnegativity of temperature, and the Clausius-Duhem entropy inequality.
In particular, the dissipation rate $\Xi=\Xi(q,\DT q)$ from \eqref{def-of-Xi}
should be non-negative even locally for subsystems. In addition,
$\Xi$ may depend also on temperature.

Another expansion might consider a water flow in the 
porous rock as in \cite{HaLyAg04CEDP}. It is well known that 
the water content influences both the rupture process as well as the
attenuation of the seismic waves. Following Biot's theory, 
the stored energy \eqref{def-of-phi} is to be augmented as
\begin{align}%\nonumber
&\varphi(e,\pi,\alpha,c)
%=\frac 32K_{_{\rm E}}|{\rm sph}\,e|^2
%+G_{_{\rm E}}(\alpha)|e_{\rm dev}|^2+\gamma_{_{\rm DAM}}(\alpha)
=\frac 32K_{_{\rm E}}|{\rm sph}\,e|^2
+G_{_{\rm E}}(\alpha)|{\rm dev}\,e{-}\pi|^2
%\\&\qquad\qquad
\label{def-of-phi+}
+\gamma_{_{\rm DAM}}(\alpha)+\frac12M|\beta(\,{\rm sph}\,e){:}\mathbb I
-(c{-}\phi)|^2,
\end{align}
where $c$ denotes the water content, and $M$ and $\beta$ 
are so-called Biot's modulus and Biot's coefficient, respectively,
%${\rm tr}\,e={\rm sph}\,e....$
while $\phi$ denotes the porosity. Then the flow is governed by the 
{\it Darcy law}:
\begin{align}
%\nonumber
\DT c={\rm div}\big(m\nabla p)\ 
\ \ \text{ with }\ \ \ p=M((c{-}\phi)-\beta(\,{\rm sph}\,e){:}\mathbb I)
\end{align}
where $p$ is the {\it pore pressure} and $m$ is the so-called hydraulic 
conductivity; note that $p=\varphi_c'$ is in the position of a 
{\it chemical potential} and $m$ is the mobility coefficient. 
Also the dissipation potential \eqref{zeta-split} can be considered 
$c$-dependent. A staggered energy-conserving time-discretisation scheme 
for this complex poro-thermodynamic model has been proposed 
in \cite{Roub17ECTD}. Moreover, the porosity itself does not need to be 
considered fixed but may be subjected to an evolution rule similar to damage, cf.\ \cite{HaLyAg04CEDP}. 

%Cf.\ also \cite{Roub14NRSF,RoSoVo13MRLF} for models of 
%rupturing lithospheric faults and, in particular, a relation with 
%the popular Dieterich-Ruina rate-and-state friction model.

Eventually, the gravitational force $g$ in \eqref{system-u++} need not be 
considered a-priori given but rather as $g=-\varrho\nabla\phi$, resulting 
from a gradient of the gravitational potential $\phi$ satisfying %the equation  
\begin{align}\label{grav-force}\Delta\phi&=
4\pi\mathfrak{g}\big(\varrho-{\rm div}(\varrho u)\big)
\end{align}
considered on the whole universe with the ``boundary'' condition 
$\phi|_{|x|=\infty}^{}=0$ and with gravitational constant 
$\mathfrak{g}\doteq 6.674\times10^{-11}$m$^3$kg$^{-1}$s$^{-2}$.
This model of {\it self-gravitating planet} is relevant in ultra low frequency
vibrations, and can be completed with considering also 
centrifugal and Coriolis forces. Thus the overall bulk force is
\begin{align}\label{Coriolis-force}
g=-\varrho\big(\nabla\phi+2\omega{\times}\DT u
+\omega{\times}(\omega{\times}(x{+}u))\big)
\end{align}
with a given angular velocity $\omega\in\R^3$ and the vectorial cross 
product ``$\times$''.

\bbb

\subsection{Towards bridging the scales}
%           ~~~~~~~~~~~~~~~~~~~~~~~~~~~
Some applications (and in particular those in geophysics of solid Earth)
are heavily multi-scaled as far ar both time and space concerns. 
Coping with this phenomenon is, numerically, very demanding and no universally
applicable hints exist. Our staggered implicit scheme is well compatible with
local mesh refinement which may handle spatial scales, e.g., on (and near) the
fracturing surfaces (faults). Also, it is well compatible with the time-stepping
refinement in (and near) the moments of ruptures and subsequent wave
propagation. On the other hand, the implicit schemes ultimately require
solving large systems of algebraic equations, which brings computational
restrictions. In particular, when waves are propagating through large
distances, one needs a fine mesh (or high-order finite elements) everywhere.
For this reason, rather explicit time-discretisations are used, which
ultimately, however, require the mentioned CFL condition even during periods when only
slow-loading processes are on effect without any rupturing and wave emission
and which makes usage of such methods limited. Here, one may think about an
adaptive combination of the implicit discretisation using rather large
time-steps and admitting local mesh refinement with an explicit
disretisation using short time steps but only during
relatively short events (typically minutes during earthquakes versus
years or thousands of years in between them). If the 2nd-order system
is implemented in a staggered way as a 1st-order system, varying of time step
is easy. A rigorous combination of explicit methods with implicit discretisation
of damage and Maxwellian viscoelastic rheology or poroelasticity
model is not trivial and has recently be devised in \cite{RoPaTs19ETDE}.
The spatial scales would require still several meshes: a fine for
resolving waves by explicit discretisation (possibly with special
so-called spectral finite elements
\cite{KaLaAm08SEMS,KomTro02SESG,KomTro02SESG2}) while an only locally refined
mesh for solving systems of equations arising from the gradient-damage 
model.

\subsection{Towards very long time scales}
%           ~~~~~~~~~~~~~~~~~~~~~~~~~~~~~

Even solid geophysical materials (lithospheric rocks in the Earth mantle)
become rather fluidic on very long time scales (millions of years).
Displacements and plastic deformation can then be large even if the elastic
strain is small. A compromising model still based on small strains then
involves Korteweg-like stress contribution arising from the transport of
internal variables, and it has been formulated in \cite{LyaBZ14DBRM} and later
analysed in \cite{Roub17GMHF}. The full thermodynamically consistent model
should be formulated at large strains, cf.\ \cite{RouSte18TEPR}, but
its numerical implementation is very cumbersome.

\eee

\section{Conclusion}
%        ~~~~~~~~~~

A coupled model for wave source due to fracture in elastic solid
(approximated by a phase-field approach) and propagation of these
waves in the layered solid and fluidic
continua has been presented and tested computationally on 2-dimensional
examples. The interpretation as seismic sources by tectonic earthquakes
generating seismic waves and for their propagation was suggested.
The ability of this model to capture basic phenomena occurring 
during earthquakes on pre-existing faults and during the creation of new faults
has been demonstrated even when using rather basic damage-type
models for the rupture.

Other, more ``local'' and engineering, usages might be for various destructive
processes in rocks or concrete constructions adjacent to water or oil reservoirs,
or in hydraulic fracture merged with natural faults (like e.g.\ in 
\cite{KhVaHi16MIFD} in quasistatic variant).

The algorithmically efficient staggered time discretisation scheme has been 
used. This was combined with the simplest P1 finite elements for the
space discretisation. Yet, it is well known that this 
discretisation is not optimal for elasticity and more sophisticated 
spectral elements 
\cite{KaLaAm08SEMS,KasDum06AHOD,KDPI07AHOD,KomTro02SESG,KomTro02SESG2}
or discontinous P2-elements \cite{TCVE12ADGM} are implemented in this 
context.

%It should be however remarked that special techniques would be needed
%to implement this monolithic model in the context of global seismicity,
%where rather explicit schemes are used for discretisation in time
%and more sophisticated (spectral) element method for discretisation in time,
%cf.\ e.g.\ 
%\cite{KaLaAm08SEMS,KasDum06AHOD,KDPI07AHOD,KomTro02SESG,KomTro02SESG2}.

Eventually, various expansions of this basic model have been outlined.
In fact, Sections~\ref{sec-energetic}--\ref{sec-staggered} already introduce 
another internal variable (the inelastic strain) that should be considered 
for long-time processes as creep or as healing during long lasting periods 
in between earthquakes, although it was not implemented in the presented
computational simulations in Sect.~\ref{sec-comput}.

\bigskip

%\baselineskip=11pt\centerline{\it Acknowledgments.} {\small

%\medskip

%\noindent
%The support from the grants 16-03823S 
%%``Homogenization and multi-scale computational modelling of flow and 
%%nonlinear interactions in porous smart structures''
%and 17-04301S
%%``Advanced mathematical methods for dissipative evolutionary systems''
%(Czech Sci.\ Foundation),
%VEGA 1/0078/16 (Ministry of Education, Science, Research and Sport of the Slovak Republic) and the institutional support RVO:61388998 (\v CR) 
%are acknowledged. }


%\bibliographystyle{plain}
%  \bibliography{trseismo6}   % name your BibTeX data base

\end{sloppypar}
\end{document}